\documentclass[12pt,article]{iopart}

\usepackage[utf8]{inputenc}
\usepackage[dvipsnames]{xcolor}
\usepackage{epsfig}
\usepackage{graphicx}
\usepackage{dsfont}
\usepackage{xcolor}
\usepackage{bbold}
\usepackage{float}
\usepackage{dutchcal} 
\usepackage{soul}
\usepackage{amssymb}
\usepackage{amsbsy}

\providecommand{\keywords}[1]
{
  \small	
  \textbf{\textit{Keywords---}} #1
}

\usepackage{graphicx}

\usepackage{braket}

\usepackage{xcolor}

\renewcommand{\vec}[1]{\mathbf{#1}}  

\newcommand{\init}[1]{{#1}^{(\textrm{i})}}
\newcommand{\fin}[1]{{#1}^{(\textrm{f})}}

\newcommand{\vecfin}[1]{\fin{\vec{#1}}}
\newcommand{\vecinit}[1]{\init{\vec{#1}}}

\newcommand{\cop}[1]{\hat{b}_{#1}^\dagger}
\newcommand{\anop}[1]{\hat{b}_{#1}}

\newcommand{\tE}{{t_{\mathrm{E}}}}    
\newcommand{\tEsp}{{t^{\rm sp}_E}}  
\newcommand{\tH}{{t_{\mathrm{H}}}}    
\newcommand{\tD}{{t_{\mathrm{D}}}} 

\newcommand{\heff}{{\hbar_{\mathrm{eff}}}} 
\newcommand{\lsp}{{\lambda_{\mathrm{sp}}}} 
\newcommand{\rhosp}{{\rho_{\mathrm{sp}}}} 

\begin{document}

\title{Semiclassical roots of universality \\
in many-body quantum chaos
}

\author{Klaus Richter}
\address{Institut f\"ur Theoretische Physik, Universit\"at Regensburg, 93040 Regensburg, Germany}

\author{Juan Diego Urbina}
\address{Institut f\"ur Theoretische Physik, Universit\"at Regensburg, 93040 Regensburg, Germany}

\author{Steven Tomsovic}
\address{Department of Physics and Astronomy, Washington State University, Pullman, WA~99164-2814, USA}

\begin{abstract}
Quantum chaos of many-body systems has been swiftly developing into a vibrant research area at the interface between various disciplines, ranging from statistical physics to condensed matter to quantum information and to cosmology. In quantum systems with a classical limit, advanced semiclassical methods provide the crucial link between classically chaotic dynamics and corresponding universal features at the quantum level. Recently, single-particle techniques dealing with ergodic wave interference in the usual semiclassical limit $\hbar \rightarrow 0$ have begun to be transformed into the field theoretical domain of $N$-particle systems in the analogous semiclassical limit $\heff = 1/N \rightarrow 0$, thereby accounting for genuine many-body quantum interference.  This semiclassical many-body theory provides a unified framework for understanding random-matrix correlations of both single-particle and many-body quantum chaotic systems. Certain {\it braided bundles} of classical orbits, and of mean field modes, govern interference, respectively, and provide the key to the foundation of universality.  Case studies presented include a many-body version of Gutzwiller's trace formula for the spectral density and out-of-time-order correlators along with brief remarks on where further progress may be forthcoming.

\end{abstract}

\centerline{ \today }
\vspace{1cm}


\submitto{Special Issue: \jpa }

\maketitle

\tableofcontents

\section{Introduction}
\label{sec:Introduction}

An integral part of Fritz Haake's scientific life was dedicated to the study of quantum chaos or, as he used to call it, {\em quantum signatures of chaos}~\cite{Haake18}. Among his many contributions to various aspects of quantum chaos, he and his co-workers' achievements towards a semiclassical understanding of random matrix universality for quantum-chaotic single-particle dynamics are particularly striking.  Based on a short review of such accomplishments and those of many others, we summarize more recent work showing how these earlier semiclassical single-particle (SP) methods engender a semiclassical theory of many-body (MB) quantum dynamics. 

\subsection{Facets of quantum chaos}
\label{sec:pillars}

One main branch of this field originated many years ago in the physics of strongly interacting nuclear MB systems.
There, Bohr’s compound nucleus model~\cite{Bohr36} may be viewed as the first quantum chaotic system, although at that time there was no concrete association with classically chaotic dynamics. Instead, Wigner's foundational work on random matrix ensembles~\cite{Wigner55, Wigner58} and subsequent contributions by several others~\cite{PorterBook} allowed for understanding nuclear statistical  spectral and scattering properties such as level repulsion~\cite{Brody81, Bohigas83}, the Porter-Thomas distribution~\cite{Porter56}, and Ericson fluctuations~\cite{Ericson60, Verbaarschot85}. 
Much later, random matrix theory (RMT) with a broadened focus,  evolved into one of the methodological pillars of quantum chaos~\cite{Haake18, Bohigas88, Guhr98, Beenakker97RMP, Verbaarschot00, StockmannBook, Mehta04}.

Given that the notion of classically chaotic dynamics is absent from RMT, in a seminal series of papers~\cite{Gutzwiller71} starting from Feynman's path integral, Gutzwiller derived a semiclassical trace formula expressing a SP quantum spectrum as a sum over unstable classical periodic orbits. Fifty years ago, he thereby set the cornerstones of the bridge connecting the classical and quantum mechanics of non-integrable systems. Subsequently, the semiclassical mechanics of chaotic dynamical systems became a second central pillar of quantum chaos studies~\cite{Haake18,StockmannBook,Gutzwiller90,Brack03}. 

Finally, in a parallel development universal features in spectral and quantum transport properties of disordered conductors had been predicted~\cite{Altshuler80,Lee85a} and observed~\cite{Washburn86}. Afterwards, this research line, comprising localization phenomena, criticality, and universality in predominantly non-interacting disordered systems, has evolved into its own field~\cite{Imry02, Akkermans07, EfetovBook, Evers08} and can be considered as representing a third methodological foundation of quantum chaos studies.

\begin{figure}
  \centering
  \includegraphics[width=0.8\linewidth]{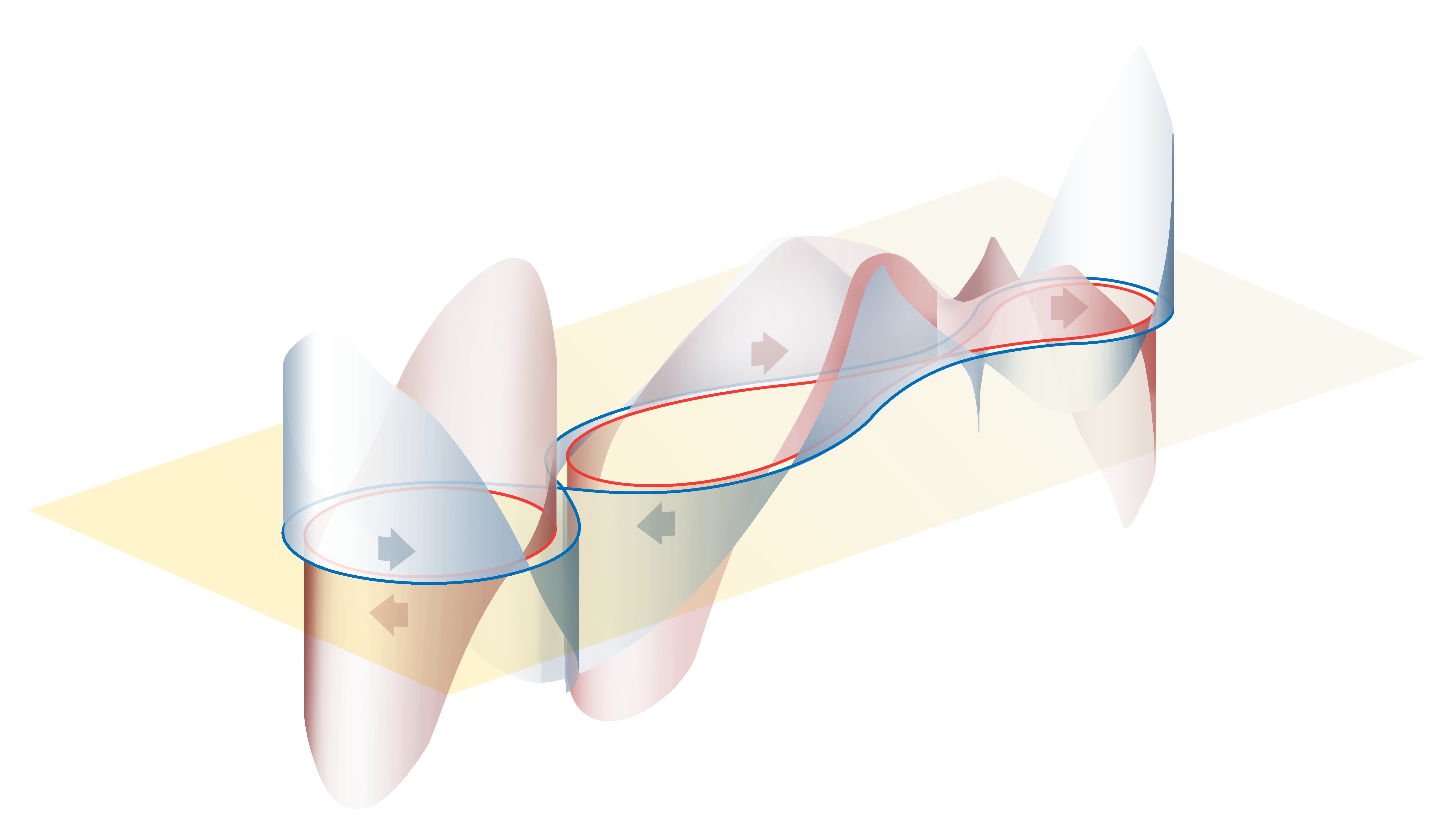}
    \caption{
   {\bf
   Correlated periodic orbits} | Phase-space sketch of classical periodic single-particle orbits that are nearly identical up to encounter regions, located at self-crossings in a two-dimensional configuration space (yellow $xy$-plane). Vertical components indicate the respective momenta in $y$-direction  
  (taken from Ref.~\cite{Haake11} with permission). \label{fig:po-pair-HR}}
\end{figure}

These three pillars, RMT, semiclassical theory, and the theory of disordered systems, initially developed rather independently, and only much later were their deep mutual links recognized and revealed.  Of particular interest is the fundamental relation between the complementary approaches underlying semiclassical and random matrix theories.  The assumptions for using RMT had originally been justified by invoking complexity of interacting MB dynamics, but the famous conjecture of Bohigas, Gianonni, and Schmit (BGS)~\cite{Bohigas84} represented a paradigm shift, namely that the deeper rationale for the justification and applicability of RMT was fully and exponentially unstable dynamics.  It was a profound unifying concept that strongly interacting MB systems and conceptually simple quantum-chaotic SP systems exhibit to a large extent common statistical spectral properties, i.e.~a prime example of universality in quantum physics.  Based on earlier works by Hannay and Ozorio~\cite{Hannay84}, Berry~\cite{Berry85}, and Sieber and one of the authors~\cite{Sieber01}, the group of Fritz Haake and Petr Braun, to whom this review is dedicated, contributed significantly towards a proof of the BGS conjecture~\cite{Heusler07} with regard to SP dynamics. Classical correlations between periodic orbits (for an example see Fig.~\ref{fig:po-pair-HR}) turned out to be the key to understanding random matrix-type spectral universality. Extending these approaches to many interacting particles involves further challenges as shown ahead.

Although quantum MB physics has a long tradition and the foundations of statistical mechanics were laid together with those of quantum mechanics, these subjects have witnessed a recent rebirth due to advances in atomic, molecular, optical and condensed matter physics that allow for building, controlling and monitoring synthetic MB systems with strongly interacting quantum degrees of freedom. Studying their dynamics~\cite{Polkovnikov11,Eisert15,Ueda20} has allowed for identifying particular classes of states that fail to quantum thermalize~\cite{Nandkishore15,Altman18,Sacha17,Turner18}. Studying their evolution towards equilibrium is especially important because equilibration is associated with chaos, which underlies scrambling of quantum correlations across MB systems' many degrees of freedom. In particular, after the proposals on out-of-time-order correlators (OTOCs)~\cite{Shenker14} and a universal limit of their growth rates, {\em i.e.}~quantum 'bounds on chaos'~\cite{Maldacena16},
such aspects related to MB chaos, ergodicity, and ergodicity breaking have recently received a great deal of attention. During the last decade corresponding activities, ranging from MB quantum dynamics via statistical physics to quantum gravity, have merged into a swiftly expanding field in theoretical physics that may be subsumed under the topic {\em many-body quantum chaos}. Corresponding research is
dramatically redirecting towards quantum MB dynamics, harking back to its origins in nuclear MB physics.

\subsection{Semiclassical regimes of quantum many-body dynamics}
\label{subsec:limits}

Referring to chaos, an inherently classical concept, requires properly defined notions of classical and semiclassical limits in MB physics.  Although there generally exists a variety of meanings for the term `{\em semiclassical}', depending on the respective field~\cite{Richter22}, here this term is being used in the original sense, just as in quantum chaos, referring to physics in the crossover regime between the classical and quantum worlds.  Semiclassical theory may then be formally based on asympototic (effective) $\hbar$ expansions of quantum mechanical (MB) Feynman propagators. The resulting semiclassical expressions, although based on classical quantities for input, nevertheless fully account for quantum (or wave) interference as an integral part of the theory. 

Large classes of MB quantum chaotic systems possess a classical limit and reside at such a semiclassical interface between non-integrable MB quantum and classical dynamics. In fact, this occurs in a two-fold way.  First, far out-of-equilibrium quantum dynamics are associated with high-energy excitations and thereby with the usual short-wavelength limit, {\em i.e.}\ small $\hbar$.  Alternatively, the  limit of large particle numbers $N$ can also be regarded as semiclassical, governed by an effective Planck constant $\heff = 1/N$. Whereas this analogy can be made rigorous in the case of non-dilute systems where $N/L \gg 1$, with $L$ the number of possible SP microstates, in which case it simply corresponds with the notion of a thermodynamic limit with large enough densities,  there is some evidence of its validity for dilute regimes characteristic of fermionic systems.  We therefore consider MB chaotic quantum systems in the limits where either $\hbar$ or $\heff$ is small but nonzero.  Both types of quantum-classical transitions are singular implying disruptive changes and complexity for the broad class of quantum systems residing at the edge of classical MB chaos. Typically, these systems require exceedingly difficult numerical quantum simulations due to vastly growing Hilbert space dimensions. Thus, there has been a quest for MB methods specifically devised for these complementary crossover regimes. In the following, the underlying concepts and challenges of a corresponding semiclassical MB theory are indicated.


\begin{figure}
    \centering
  \includegraphics[width=0.8\linewidth]{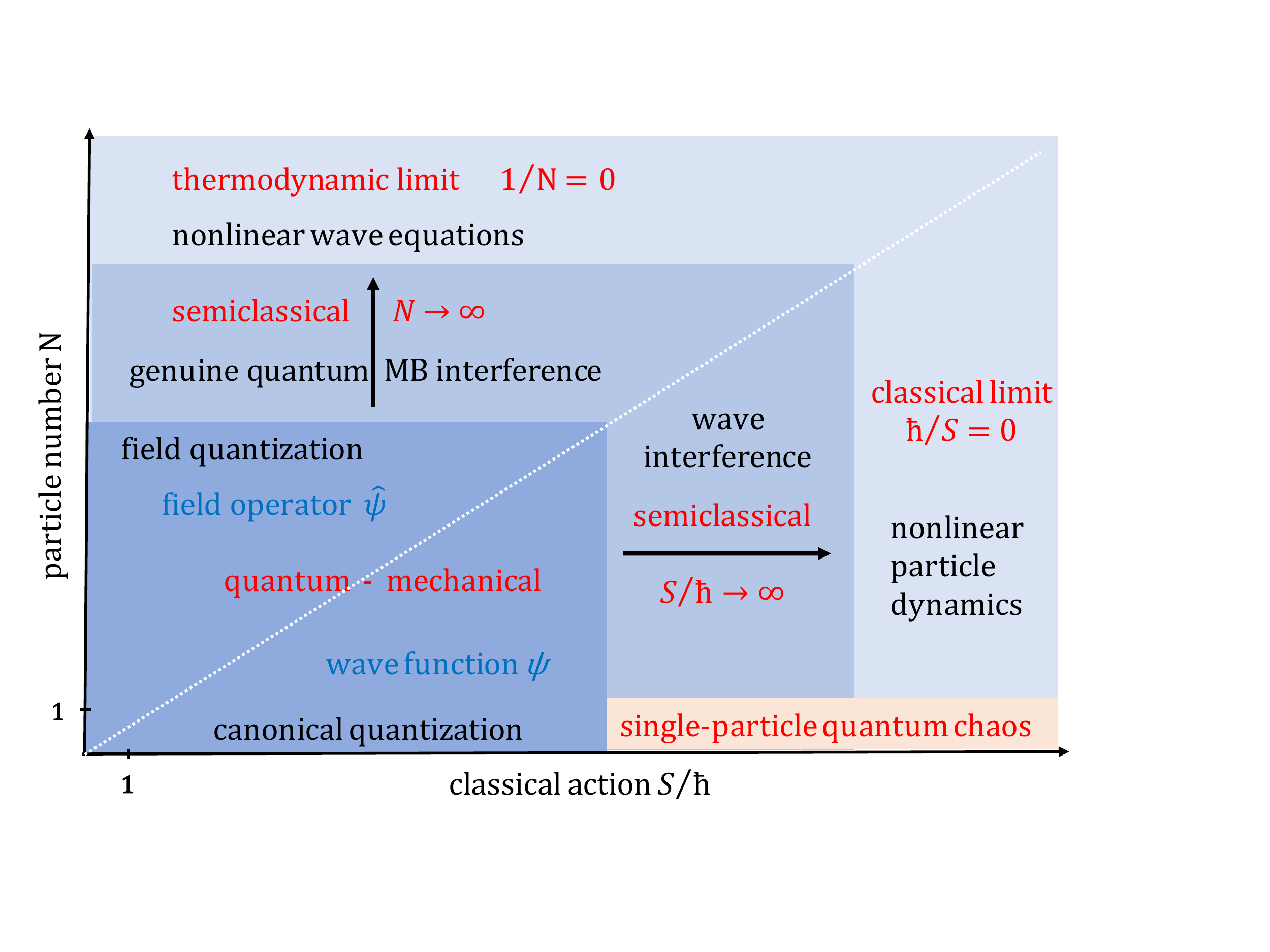}
    \caption{\label{fig:sc-limits}
   {\bf
   Semiclassical regimes and limits of quantum MB dynamics} | At fixed particle number $N$, the usual semiclassical limit $S/\hbar \rightarrow \infty$, often referred to as $\hbar \rightarrow 0$,  corresponds to a (horizontal) transition from quantum mechanical waves to classical particles, nearly always involving nonlinear dynamics. Semiclassical theory within the field of quantum chaos has traditionally addressed single-particle systems, i.e.~the lower right zone.  From the perspective of quantum field theory, 
   the vertical direction of increasing particle number $N$, consistent with what is usually considered as thermodynamic for non-dilute systems, formally corresponds to a complementary semiclassical regime with effective $\heff= 1/N \rightarrow 0$. In that limit quantum fields pass into nonlinear waves.
   }
\end{figure}

\subsubsection{The usual limit of high excitations}
\label{subsec:limit1}

Consider the familiar case of a SP quantum system with an existing classical limit that is approached in the limit $\hbar \rightarrow 0$.  More precisely, the semiclassical limit is one in which the dimensionless ratio $\hbar/S \ll 1$ with $S=\int {\bf p} d{\bf q}$, a typical classical action of the particle with momentum $p$. This is the standard limit of short wave lengths $\lambda$ in view of the relation $\lambda = h/p$.
In the schematic Fig.~\ref{fig:sc-limits} with horizontal scale $S/\hbar$ and vertical scale denoting the particle number $N$, this semiclassical limit corresponds to the horizontal crossover (for $N=1$) from the deep quantum regime at $S/\hbar \sim 1$ into the semiclassical range where SP wave mechanics approaches classical mechanics, {\em i.e.} classical particles most frequently possessing nonlinear, possibly chaotic dynamics.

Since Gutzwiller's and Berry's early works~\cite{Gutzwiller71, Berry76}, semiclassical approaches in quantum chaos have nearly exclusively been focused on the case $N=1$, {\em i.e.}~in the lower right region of the $S$-$N$-landscape of Fig.~\ref{fig:sc-limits}.  This region is the subject of several textbooks~\cite{Haake18, StockmannBook, Gutzwiller90,Brack03, Reichl21}.  However, the limit $\hbar/S \rightarrow 0$  also formally applies to systems with more than one particle in $D$ dimensions by considering semiclassical methods applied in a respective $2D\!\cdot\! N$-dimensional phase space. In Fig.~\ref{fig:sc-limits} this corresponds to moving vertically upwards in the right semiclassical regime and considering the limit $\hbar \rightarrow 0$ for given $N$.
In this case, the MB density of states of $N$ interacting confined particles with MB energies $E_n^{(N)}$ 
is conveniently decomposed into a smooth and an oscillatory part,
\begin{equation}
    \rho(E,N) = \sum_n \delta(E-E_n^{(N)}) = 
    \bar{\rho}(E,N) + \rho^{osc}(E,N) \, .
\label{eq:SCDOS}
\end{equation}
However, extending such semiclassical approaches from one to $N$ particles is accompanied by a variety of notable challenges.  First of all, in practice the calculation, classification, and visualization of classical dynamics in high-dimensional phase spaces quickly reaches its limits. For instance, the implementation of Gutzwiller's trace formula for $\rho^{osc}(E,N)$ on the basis of $N$-particle periodic orbits seems practically impossible for many particles. Secondly, it is necessary to account for the symmetry character of MB states representing $N$ identical particles in quantum physics.  Finally, and perhaps most challenging, to be truly valuable interaction effects must be incorporated as an integral part of MB quantum chaos.

Consequently, corresponding attempts have been rare, even for (non-integrable) few-particle systems.  An early example is the successful semiclassical quantization of the correlated Coulomb dynamics of the electrons in the helium atom, a longstanding problem dating prior to Schrödinger and his equation, i.e.~to the `old quantum theory'~\cite{Bohr13a},
 see \cite{Tanner00a,Kragh12} for reviews of the history.  By applying a  cycle expansion to chaotic dynamics in 6-dimensional phase space ground and exciting states of helium could be semiclassically computed with high precision~\cite{Ezra91}. 
 In Ref.~\cite{Primack98} the accuracy of the semiclassical trace formula for systems with more than two degrees of freedom was considered and numerically studied for the three-dimensional Sinai-billiard.
 More recently, the dynamics of quantum maps with up to 8-dimensional phase spaces has been visualized and thoroughly investigated~\cite{Richter14}. Furthermore, interesting collective MB dynamics were recently semiclassically identified in kicked spin-chains up to particle numbers of order $N\sim 20$~\cite{Akila17,Waltner17} making use of a remarkable particle number-time duality in the Ising spin chain~\cite{Akila16}.

There are only a few scattered examples for generalizations of the van Vleck-Gutzwiller propagator~\cite{Gutzwiller71} to truly many particles.  In~\cite{Weidenmueller93} Gutzwiller’s trace formula for the density of states $\rho^{osc}(E,N)$ was reconsidered for systems of non-interacting identical particles, in particular fermions.  However, being based on classical SP phase space, this construction of the MB density of states remains purely formal without a direct interpretation in terms of MB phase space.
With regard to semiclassical theory for many interacting fermions (for a review see~\cite{Ullmo08}), in~\cite{Ullmo98} it was shown that the orbital magnetic response can be greatly enhanced by the combined effects of interactions and finite size.  In the context of MB scattering~\cite{Urbina16} contains a semiclassical calculation of the transmission probabilities through mesoscopic cavities for systems of many non-interacting particles, in particular photons, thereby generalizing advanced semiclassical SP techniques~\cite{Richter02, Mueller09, Berkolaiko12} for scattering in chaotic conductors.  There the interplay between interference at the SP level and due to quantum indistinguishability leads to specific universal correlations in MB scattering properties with relevance for boson sampling~\cite{Aaronson10} and the Hong-Ou-Mandel effect~\cite{Hong87}.  

In a parallel development, remarkable progress has been achieved by Gutkin and co-authors providing certain classical foundations of many particle chaos based on models for coupled cat maps~\cite{Gutkin16,Gutkin21}. In view of correlations between partners within ``braided classical orbit bundles'', known to be relevant for universal spectral properties and to be reviewed below, they highlighted the existence of partner orbits specific to MB systems. For a sufficiently large particle number $N$, these new partners are considered as relevant for construction of a consistent MB semiclassical theory. Very recently, a chaotic scalar lattice field theory (in one dimension) has been proposed~\cite{Lakshminarayan11,Liang22}, complementary in spirit to Gutzwiller's periodic-orbit approach for low-dimensional chaotic dynamics~\cite{Gutzwiller1971} and to generalizations presented in Sec.~\ref{sec:SC-MB}.

The above works comprise various attempts to generalize semiclassical periodic-orbit theory for the fluctuating level density $\rho^{osc}(E,N)$ to spectra of systems with few to many particles. The MB phase space dimensions increase with $N$, and accordingly the spectral MB density of states grows immensely~\cite{Garcia17}.  In turn, the spacing between MB levels tends to zero and individual highly excited MB levels are with rare exceptions (such as slow neutron resonances) no longer resolvable.
Hence the smooth part $\bar{\rho}(E,N)$ of the spectral MB density in Eq.~(\ref{eq:SCDOS}), which is not sensitive to the nature of the dynamics, chaotic or regular, and often referred to as Weyl part of the spectrum~\cite{Weyl11}, gains particular importance. 
It plays for instance a central role in computing thermodynamic equilibrium and non-equilibrium properties. However, to compute even this smooth part quantum mechanically is numerically challenging since systems with fixed $N$ require elaborate MB techniques generating ground and low excited states.  They quickly reach their limits when increasing $N$ or the degree of excitations.  This has prompted the development of MB techniques specifically devised to directly compute $\bar{\rho}(E,N)$, thereby circumventing the intricate or often impossible calculation of individual excited MB levels, which requires detailed information that is afterwards smoothed out anyway.  For example, in the nuclear physics context, French and co-workers developed statistical spectroscopy based on moment methods~\cite{Chang71, Mon75, Brody81, KotaBook}. 

In the SP case, the Weyl expansion~\cite{Weyl11} provides a well defined semiclassical $1/\hbar$ expansion of the smooth part \cite{ Brack03,Baltes76}. In~\cite{Hummel14, Hummel17, Hummel19} the SP Weyl expansion has been generalized to MB systems of $N$ indistinguishable particles in $D$ dimensions.
Corresponding expressions for $\bar{\rho}(E,N)$ take the form of sums over clusters of particles moving freely around manifolds in configuration space invariant under permutations.
This approach contains the famous Bethe law \cite{Bethe36} for the mean fermionic spectral density as a limiting case
\footnote{Shell corrections to the Bethe law for the MB density of states were semiclassically considered in Ref.~ \cite{Leboeuf05} and are more generally reviewed in Ref.~\cite{Brack93}.}.
Furthermore, the correct emergence of the fermionic MB ground state is a consequence of a delicate cancellation effect of cluster contributions. Moreover, by including interaction effects in a non-perturbative way this MB Weyl appraoch has further been extended to systems of experimental relevance in cold atom physics, such as interacting bosons in traps,
demonstrating for instance that systems with very few up to many particles share the same underlying spectral features~\cite{Hummel19}.
We believe that such underlying MB scaling laws have much in common with related semiclassical scalings in recent generalizations of Thomas-Fermi theory~\cite{Okun21}.


\subsubsection{The thermodynamic limit of large particle number}
\label{subsec:limit2}

Besides the usual notion of $S/\hbar\rightarrow \infty$ discussed so far, in quantum field theory, where wave functions are replaced by field operators,  there is the complementary limit of large particle number $N$, but not necessarily small $\hbar$.  In the limiting case of $N=\infty$ and as long as the typical occupations of SP states are large enough, referred as the non-dilute case, the equations for the quantum fields pass into nonlinear wave equations characteristic of the thermodynamic limit.  From the viewpoint of quantum field theory, these wave equations appear as a kind of `classical' fluid dynamics. For instance, in the large-$N$ limit, systems of interacting bosons are described by the Gross-Pitaevskii equation.  Formally, the large-$N$-but-not-infinite regime, corresponding to the upper (left) region in Fig.~\ref{fig:sc-limits}, can be associated with an effective Planck constant $\heff =1/N \ll 1$ and hence also be considered semiclassical.

Wave interference is usually built into semiclassical propagators through coherent sums over classical paths with interfering amplitudes in configuration space, leading for instance to the van Vleck propagator~\cite{Vanvleck28} or the Gutzwiller trace formula~(\ref{eq:SCDOS})~\cite{Gutzwiller90} 
for $\rho^{osc}$ in terms of unstable periodic orbits of classical particles.  In the complementary limit, $\heff = 1/N \ll 1$, many-particle propagators and quantum observables derived from those can be formally described also by means of semiclassical sums over paths defined by classical field solutions, which have a completely different interpretation and meaning.  The summations are taken over collective modes of MB densities in a continuum version of high-dimensional MB Fock space~\cite{Engl14}, instead of particle trajectories in configuration space, as is outlined in Sec.~\ref{sec:SC-MB}. These Fock-space paths represent the various, in principle infinitely many, time-dependent solutions of the nonlinear wave equations in the classical limit $1/N = 0$ (upper region of Fig.~\ref{fig:sc-limits}).  Quantum MB interactions turn into nonlinearities in these wave equations and may result in unstable, possibly chaotic MB mode dynamics.  In this way chaos at the level of these classical-limit nonlinear waves implies {\em many-body quantum chaos} at the level of quantum fields\footnote{There are other conceivable routes to many-body quantum chaos not considered further in this contribution, but this issue is revisited for brief speculation in Sec.~\ref{sec:persp} (i).}.  This is entirely analogous to signatures of chaotic classical particle dynamics in wave functions at the Schrödinger equation level, i.e. quantum chaos in the limit $\hbar \rightarrow 0$.  In a sense, such an approach transports Gutzwiller's semiclassical theory of the time evolution operator from the level of ``first quantization'' to that of ``second quantization''.  Note that the classical quantities entering semiclassical path integrals have different meanings in the two complementary limits: for instance different Lyapunov exponents quantify the instability of particle trajectories and collective modes, respectively.
Remarkably, the semiclassical theory in the limit $\heff = 1/N \ll 1$ also applies to ground or low-lying excited MB states.

The classical paths in MB space, i.e.~the time-dependent solutions of the nonlinear wave equations, just represent mean-field solutions of the full MB problem.  This opens an interesting new perspective on the connections between chaotic mean-field dynamics, quantum correlations due to MB interactions, scrambling, and the generation of entanglement.  MB interaction effects beyond mean-field are commonly considered as correlation effects~\cite{Fulde95}. Hence, as will be explained in Sec.~\ref{sec:SC-MB}, the interpretation of Eq.~(\ref{eq:SCDOS}) as a coherent sum over different collective mean-field modes implies that massive MB interference between these chaotic modes describes or explains quantum correlations in the MB propagator. Hence MB quantum chaos and quantum correlation phenomena are intimately intertwined.  To highlight the difference between (SP) wave and MB quantum interference we coin the term for the latter case {\em genuine many-body quantum interference}.


\subsection{Outline of this review: 
Universality in many-body quantum chaos from a semiclassical perspective}
\label{subsec:outline}

The semiclassical approach reviewed below addresses these leading-order (in $\heff=1/N$) MB quantum mechanical contributions to the thermodynamic limit. The theory's strength is its capacity to apply broadly to dynamical MB systems that  are either fully chaotic, partially chaotic, or are even integrable as well. Hence, on the one hand it deals with systems not behaving in a universal manner and allows for addressing system specific individual, possibly quite atypical properties.  On the other hand, the MB semiclassical approach provides dynamical foundations for universal aspects of quantum chaotic MB systems, i.e.~in the statistical RMT-like sense. This review focuses on the latter issue, primarily based on the recent accomplishments in Refs.~\cite{Engl14,Engl15,Dubertrand16,Rammensee18}. 
Assuming fully chaotic MB mean-field dynamics, this branch of semiclassical MB theory follows the strategy of invoking ergodic properties and corresponding sum rules for the exponentially numerous classical paths, i.e.~collective modes entering into semiclassical trace formulas for various MB observables and correlation functions. Such assumptions often enable an analytical treatment of the arising multiple sums over chaotic MB modes. Generalizing corresponding SP techniques based on classical (periodic-)orbit correlations mentioned above provides the key to explaining aspects of RMT universality also in the many-particle context. 

The remainder of the review is structured as follows: in Sec.~\ref{sec:SC-SP} the earlier semiclassical theory providing the link between RMT-universality and SP chaos is summarized.  This includes the encounter calculus for special braided classical orbit bundles relevant for the evaluation of spectral correlation functions and, closely related, the treatment of phenomena at and beyond Ehrenfest time scales.  In Sec.~\ref{sec:SC-MB} the foundations of an advanced semiclassical theory of MB quantum fields and chaos are given for $\heff = 1/N \ll 1$.  After deriving a MB version of the van Vleck-Gutzwillerß propagator, a Gutzwiller-type trace formula for MB density of states is presented.  The resultant formulas provide the basis for deriving MB spectral correlators, response functions, and echo-type observables. With regard to the latter  the semiclassical theory of out-of-time-order-correlators (OTOCs)~\cite{Larkin69}, which have recently gained an enormous amount of attention~\cite{Maldacena16} in various fields of physics from condensed matter via cold atoms to cosmology, are sketched out.  This review is completed with perspectives and open questions discussed in Sec.~\ref{sec:persp}.


\section{Semiclassical theory of single-particle quantum chaos}
\label{sec:SC-SP}

Treating semiclassical limits of quantum theory by starting from Feynman's path integral and invoking corresponding stationary phase approximations naturally leads to expressing unitary quantum time evolution in terms of sums over phase-carrying classical paths. Quantum interference, as a direct consequence of the principle of quantum superposition, is then captured by the existence of multiple classical solutions and their coherent summation. Depending on the structure of the quantum observable to be considered, for instance spatial or spectral $n$-point correlation functions, a multitude of time evolution operators can be involved leading to a corresponding number of summations over classical paths. Modern semiclassical theory is concerned with the challenge of how such multiple summations can be carried out efficiently and appropriately while preserving the inherent underlying quantum interference mechanisms.  In this respect the Ehrenfest time $\tE$~\cite{Ehrenfest27} plays a key role. As will be discussed below, it has turned out to be of tremendous importance for the quantum dynamics of chaotic systems since it separates quantum propagation in the phase space around {\em one} dominant (unstable) classical trajectory at short time scales from subsequent times governed by strong wave interference, {\em i.e.}, involving propagation of amplitudes along {\em many} trajectories. 

Beyond $\tE$, semiclassical approaches that do not appropriately cope with such many-trajectory interferences break down. Due to the exponential sensitivity to initial conditions for chaotic dynamics, $\tE$ is a logarithmically short time scale as a function of $\hbar$, and hence absent the interference contributions the corresponding range of validity of such approaches is extremely limited.  In view of the fact that RMT-type spectral universality is reached in the limit where $\tE / \tH \rightarrow 0$, with 
\begin{equation}
    \tH = 2\pi \hbar \bar{\rho}(E)
\label{eq:tH}
\end{equation}
the Heisenberg time, the time dual to the mean level spacing $1/ \bar{\rho}(E)$ (with $ \bar{\rho}(E)$ the mean density of states), there has been the quest for devising advanced semiclassical methods to adequately treat post-Ehrenfest quantum dynamics. In fact, for a number of reasons, early on it was even thought that post-Ehrenfest quantum chaotic dynamics were beyond the range of any semiclassical approach. However, it was shown by the early $1990$'s that the validity of a complete semiclassical dynamics extended far, far beyond the logarithmic $\tE$ scale limit~\cite{Tomsovic91b, Oconnor92, Sepulveda92, Tomsovic93}.

In the following, a semiclassical theory is outlined that provides the link between chaos and RMT-universality for SP dynamics.  It is based on Gutzwiller's trace formula~\cite{Gutzwiller1971} for the SP density of states that is briefly introduced in Sec.~\ref{sec:SP-Gutzwiller}.  The theory further involves classical orbit correlations and the encounter calculus for braided orbit bundles (see Sec.~\ref{sec:SP-universality}). Intimately connected to that, it deals with interference phenomena at and beyond Ehrenfest time scales (see Sec.~\ref{sec:SP-Ehrenfest}).


\subsection{Single sums over paths: van Vleck propagator and Gutzwiller trace formula}
\label{sec:SP-Gutzwiller}

For a time-independent SP Hamiltonian $H$, the van Vleck propagator $K_{\rm sp}(t)$, and its refinement by Gutzwiller, is a  semiclassical approximation to the quantum time evolution operator $U(t) = \exp{(-(i/\hbar) H t)}$ in configuration space.  Here, the derivation of $K_{\rm sp}(t)$ and Gutzwiller's periodic orbit theory are skipped, as they can be found in various excellent text books~\cite{Haake18,StockmannBook,Gutzwiller90,Brack03,Reichl21}. The relevant expressions are directly introduced.

Evaluating the Feynman propagator $\langle {\bf r}_{f}| U(t)| {\bf r}_{\rm i}\rangle $ in a stationary phase approximation yields the 
van Vleck-Gutzwiller propagator for the evolution of a quantum particle  between initial and final coordinates ${\bf r}_{\rm i}$ and ${\bf r}_{\rm f}$ in $d$ dimensions:
\begin{equation}
\label{eq:vVG}
    K_{\rm sp} ({\bf r}_{\rm f},{\bf r}_{\rm i},t)
     = \sum_{\gamma} 
     \left(\frac{1}{(2\pi i \hbar)^d} \left| \frac{ \partial^2
     R_{\gamma}({\bf r}_{\rm f},{\bf r}_{\rm i},t)}{\partial {\bf r}_{\rm f} \partial {\bf r}_{\rm i}}
     \right|\right)^\frac{1}{2} 
     \
     {\rm e}^{i(R_{\gamma}({\bf r}_{\rm f},{\bf r}_{\rm i},t)/\hbar - \nu_\gamma \pi/2)} \, 
\end{equation}
with classical action
$ R_{\gamma}({\bf r}_{\rm f},{\bf r}_{\rm i},t) = \int_0^t L(\dot{\bf r}, {\bf r}, t) d t$ along the trajectory $\gamma$ connecting ${\bf r}_{\rm i}$ to ${\bf r}_{\rm f}$ in time $t$, and  topological index $\nu_\gamma$, which appropriately tracks the correct phase of the determinant's square root; see~\cite{Keller58}, for example.  For simplicity, all the various topological indices are referred to as Maslov indices.
This expression for $K_{\rm sp}(t)$ holds generally for either chaotic or integrable classical dynamics, and even for systems with coexisting stable and unstable phase space regions.
After computing the energy-dependent Green function via a Laplace transform of $K_{\rm sp}(t)$ and upon calculating the spatial trace integral by means of further stationary phase approximations, Gutzwiller derived the famous trace formula for the density of states $\rhosp(E)$ of a {\em classically chaotic} quantum SP system,
thereby laying the foundations of periodic-orbit theory in quantum chaos~\cite{Gutzwiller71}:
\begin{equation}
    \rhosp(E) \simeq \bar{\rho}_{\rm sp}(E) \ + \ \rho_{\rm sp}^{\rm (osc)}(E) =
    \bar{\rho}_{\rm sp}(E) \ + \frac{1}{\pi\hbar} {\rm Re}\left\{
    \sum_{\rm po} A_{\rm po}{\rm e}^{(i/\hbar) S_{\rm po}(E)} \right\} \, .
    \label{eq:SP-Gutzwiller}
\end{equation}
The Weyl term $\bar{\rho}_{\rm sp}(E)$ is a smooth function of energy $E$. It is obtained, to leading order in $\hbar$, by calculating the volume of the on-energy-shell classical phase space volume,
\begin{equation}
    \bar{\rho}_{\rm sp} (E)=\left(\frac{1}{2\pi\hbar}\right)^{d}\int \ d{\bf r} \ d{\bf p}\ \delta(E-H_{\rm sp} ({\bf r},{\bf p})) \, ,
    \label{eq:SP-Weyl}
\end{equation}
where $H_{\rm sp}$ is the classical Hamiltonian. In the above trace formula the remaining oscillatory part $\rho_{\rm sp}^{\rm (osc)}(E)$ of the density of states appears as a coherent sum over 
all {\it periodic orbits} (po) of the corresponding classical system at energy $E$, namely over the solutions $({\bf r}(t),{\bf p}(t))$ of Hamilton's equations with energy $E$ for which there exists a $T$ such that
\begin{equation*}
    ({\bf r}(t+nT),{\bf p}(t+nT))=
    ({\bf r}(t),{\bf p}(t))
\end{equation*}
for all integers $n$. The respective phases
\begin{equation}
    S_{\rm po}(E) = \int_{\rm po}  \ {\bf p} \cdot d {\bf q} -\hbar \mu_{\rm po}  \ \pi /2 \, .
    \label{eq:SP-action}
\end{equation}
contain their classical actions and Maslov indices  $\mu_{\rm po}$.
Note that the sum over all periodic orbits includes also all higher repetitions with periods $T_{\rm po}=nT_{\rm ppo}$ for $n\ge2$ of a given primitive periodic orbit (ppo) with period $T_{\rm ppo}$.
The amplitudes in Eq.~(\ref{eq:SP-Gutzwiller}) read
\begin{equation}
    A_{\rm po}(E) = \frac{T_{\rm ppo}(E)}{|{\rm d
    et}({\bf M}_{\rm po}(E) - {\bf I} |^{1/2}}  \, .
    \label{eq:SP-stab}
\end{equation}
The monodromy (or stability) matrix ${\bf M}_{\rm po}(E)$ takes into account, in a linearized way, the phase space structure in the vicinity of the periodic orbit. It characterizes its instability in terms of stability exponents (similar to Lyapunov exponents) for chaotic dynamical systems. 
 
 The trace formula, Eq.~(\ref{eq:SP-Gutzwiller}), decomposes the quantum spectrum in a Fourier-type hierachical way: whereas short periodic orbits contribute with long-energy-ranged cosine-like spectral modulations, accounting for contributions from longer and longer periodic orbits, in principle, generates an increasing spectral resolution~\footnote{There is an extensive literature about convergence properties of the trace formula and the challenges associated with semiclassically computing individual energy levels; see~\cite{Cvitanovic92}.}.  To resolve the quantum density of states at scales beyond the mean level spacing $1/\bar{\rho}(E)$ requires, in turn, to control semiclassical wave interference in the time domain on scales in the range of or longer than the Heisenberg time, $\tH$, the longest time scale involved. The challenge of coping with this {\em late-time behavior} leads to partially solved issues, but also many open questions that are addressed below in the context of spectral correlations.


\subsection{Multiple sums over paths: classical correlations and quantum universality}
\label{sec:SP-correlations}

\subsubsection{Spectral two-point correlation function}
\label{ref:sec-2point-cor}
In many circumstances, the quantities of interest are not the bare densities of states $\rhosp(E)$, but  rather the spectral $n$-point correlation functions. In particular, the normalized connected spectral two-point correlator
\begin{equation}
 C(\epsilon) = \frac{1}{\bar{\rho}_{\rm sp}(E)^2} 
 \left\langle 
 \rho_{\rm sp}^{\rm (osc)}
 \left(E+\frac{\epsilon}{2\bar{\rho}_{\rm sp}(E) }\right)\
  \rho_{\rm sp}^{\rm (osc)}
  \left(E-\frac{\epsilon}{2\bar{\rho}_{\rm sp}(E)} \right)
  \right\rangle_E 
    \label{eq:SP-2-point}
\end{equation}
with $\rho_{\rm sp}^{\rm (osc)}$ defined in 
Eq.~(\ref{eq:SP-Gutzwiller})
is a simple but fundamental measure of spectral correlations.
Here the angular brackets denote a running local average over energy $E$. The dimensionless variable $\epsilon$ stands for a spectral energy distance in units of the mean level spacing $1/\bar{\rho}_{\rm sp}(E)$.
Corresponding energy correlation functions are defined, {\em e.g.}, for  Green functions and scattering matrix elements. Instead of energies, also spatial or time-like correlators are of relevance in various branches of physics.

The quantum objects entering such correlators can commonly be semiclassically represented in terms of sums over (periodic) trajectories, similar to that in the trace formula 
(\ref{eq:SP-Gutzwiller}) (see {\em e.g.} the reviews~\cite{Richter00,Jalabert00,Waltner10} for such semiclassical correlation functions in mesoscopic physics). 
Hence, a semiclassical approach to $n$-point correlators naturally leads to $n$-fold coherent summations over amplitudes evolving along, in principle, infinitely many trajectories. Coping with such multiple infinite sums over highly oscillatory objects seems, at first glance, hopeless. However, an intrinsic strength of semiclassical theory lies in the fact that systems with diffusive or ergodic classical dynamics often do not require the computation of specific trajectories (nor is it always desirable). Instead, invoking ergodicity and uniformity of chaotic phase space implies powerful classical sum rules that permit a treatment of the orbits in a statistical manner. As no system-specific information is required, such approaches naturally lead to universal features of quantum-chaotic dynamics and may provide physical laws applicable to whole classes of quantum systems, exclusively characterized by means of their respective symmetry class.

The semiclassical evaluation of multiple sums over paths is illustrated for the prominent case of the two-point correlator, Eq.~(\ref{eq:SP-2-point}); the corresponding treatment of four-point objects is in Sec.~\ref{sec:OTOC} on out-of-time-order correlators.  Replacing $\rhosp$ in Eq.~(\ref{eq:SP-2-point}) by its semiclassical approximation, Eq.~(\ref{eq:SP-Gutzwiller}), one obtains
\begin{eqnarray}
 C(\epsilon)& \simeq &
  \frac{1}{\bar{\rho}_{\rm sp}(E)^2}  \left(\frac{1}{\pi\hbar}\right)^2 \times \nonumber \\
 && \left\langle 
    \sum_{\gamma} \sum_{\gamma'} A_\gamma A_{\gamma'}^\ast \ {\rm e}^{(i/\hbar) 
    [S_\gamma(E) - S_{\gamma'}(E)) + (T_\gamma(E)+T_{\gamma'}(E))\epsilon/ (2\pi \bar{\rho}_{\rm sp})]} 
  \right\rangle_E \ .
    \label{eq:SP-2-point-sc}
\end{eqnarray}
Here, $S(E+E') \simeq S(E) + T(E) E'$ with $T(E) =    \partial S/  \partial E$ (the orbit's period). The contributions of periodic orbit pairs $\gamma, \gamma'$ that exhibit an action difference $\Delta S(E)$ in the phase factor are handled separately from those that do not.  Of course, $\Delta S(E)$ vanishes for the joint contributions of the specific orbit pairs $\gamma = \gamma'$, i.e. {\it the diagonal contributions}. In addition, if a system is invariant with respect to some symmetry, for instance time-reversal symmetry, then that symmetry is reflected in multiplicities of symmetry related orbits with classical action degeneracies as well.  The resultant constructive interference encodes the symmetry's influence on the quantum system.  In effect, this can be considered as part of the diagonal contributions.

In the semiclassical limit, the phases $S_\gamma(E)/\hbar$ oscillate rapidly upon varying $E$, and only terms with sufficiently small action differences 
\begin{equation}
\Delta S(E) = S_\gamma(E)-S_{\gamma'}(E)
\label{eq:delta_S}
\end{equation}
can survive the energy averaging $\langle \ldots \rangle_E$. Using the classical sum rule of
Hannay and Ozorio de Almeida~\cite{Hannay84},
\begin{equation}
    \sum_{\gamma} 
    \frac{1}{{\rm det}({\bf M}_\gamma-{\bf I})}
    f_\gamma(T_\gamma)  \simeq \int_{T_0} dT \frac{f(T)}{T} \, , 
    \label{eq:SP-HOdA}
\end{equation}
that follows from the assumption of uniform phase space exploration by unstable periodic orbits, Berry computed the  diagonal contribution
\begin{equation}
 C_d(\epsilon) \simeq  \frac{1}{\bar{\rho}_{\rm sp}(E)^2}  \left(\frac{1}{\pi\hbar}\right)^2
 \int_{T_0} dT \ T \  e^{iT\epsilon/ (2\pi \hbar \bar{\rho}_{\rm sp})} 
    \label{eq:SP-2-point-diag}
\end{equation}
to the two-point correlator~\cite{Berry85}. He thus derived the spectral rigidity found in RMT semiclassically.  For the spectral form factor $K(\tau)$ (with $\tau \!=\! T/\tH$), the Fourier transform of $C(\epsilon)$, the diagonal approximation leads to the linear "ramp": $K(\tau) = \eta\tau$, 
with $\eta =$ 2 and 1 for systems with and without time reversal symmetry, respectively. Berry's analysis provided the first intimate theoretical link between RMT and the semiclassical theory of chaos.

Apart from the diagonal terms there is an enormous number of off-diagonal orbit pairs in a chaotic system due to the exponential proliferation of the number of periodic orbits with increasing period, $T_\gamma(E)$. Most of the orbit pairs consist of periodic orbits with actions that are uncorrelated.  Summing over them and performing the energy average, they collectively have a vanishing average, including the effects of ``accidental'' nearly equal actions $S(E)$.  However, from RMT it had been known that for  case of time-reversal invariant systems, there had to be further universal spectral correlations beyond those related to the diagonal term~\cite{Bohigas91}:
\begin{equation}
K^{\rm GOE} (\tau) = 
\left\{
\begin{array}{ll}
2\tau -\tau \log(1+2\tau) & {\rm if} \quad \tau < 1 \, , \\
2 - \tau \log \frac{2\tau +1 }{2\tau -1} & {\rm if} \quad \tau >1 \, . \\
\end{array}
\right. 
    \label{eq:SP-FormFac}
\end{equation}
Hence to describe such universal RMT features, one had to find orbit pairs with non-random action differences~\cite{Argaman93}. Although it was expected that such non-vanishing contributions come from a relatively small number of pairs of correlated orbits, for a long time it was unclear how these orbit correlations could emerge from an ergodic phase space structure.

\subsubsection{Braided classical orbit bundles and encounters}
\label{sec:SP-bunches}

 Henri Poincar\'e had already recognized in 1899 that
chaotic motion, which ergodically fills the classical
configuration space or phase space in a uniform manner and is often mistakenly equated with ``stochasticity'', is indeed subject to structural principles. In his {\it Les M\'ethodes Nouvelles de la M\'ecanique  C\'eleste}~\cite{Poincare99} appears the notion that arbitrarily long trajectory segments can be approximated with arbitrary accuracy by pieces of a periodic orbit~\footnote{``\'Etant donn\'ees [$\dots$] une solution particuli\`ere quelconque de ces \'equations, on peut toujours trouver une solution p\'eriodique (dont la p\'eriode peut, il est vrai, \^etre tr\`es longue), telle que la diff\'erence entre les deux solutions soit aussi petite que l'on veut, pendant un temps aussi long qu'on le veut.''}. 
For this reason, sometimes periodic orbits are considered as a ``skeleton'' or backbone of chaotic dynamics~\cite{Cvitanovic91}, along which all the non-periodic orbits must wind; see also~\cite{Li20, Li17a, Li18}. Research since 2000 has brought to light how this ``skeleton'' is constructed and that chaotic dynamics is subject to further principles of order: (periodic) orbits do not appear as independent individual entities but in pairs, as first discovered by~\cite{Sieber01,Richter02,Sieber02}, and more generally in densely packed bundles~\cite{Mueller09}. This hidden classical property of periodic orbits in chaotic systems turned out to play a central role for understanding universal spectral properties.

According to the popular notion of chaos, chaotic classical movement is extremely unpredictable. Two closely adjacent paths diverge exponentially $\sim e^{\lsp t}$, with the SP positive Lyapunov exponent(s) $\lsp$ as the divergence rate.  However, this statement does not include all aspects of symplectic Hamiltonian dynamics: exponentially diverging motion happens locally on or in the neighborhood of local unstable manifolds in phase space, and there exists its complement, motion along stable manifolds where initial phase space distances exponentially decrease. The combination of these two structural elements of Hamiltonian dynamics lies behind the formation of (periodic) orbits in braided bundles. On the other hand, the probability of a solitary chaotic orbit's existence decreases exponentially with its length.


\begin{figure}
    \centering
  \includegraphics[width=0.4\linewidth]{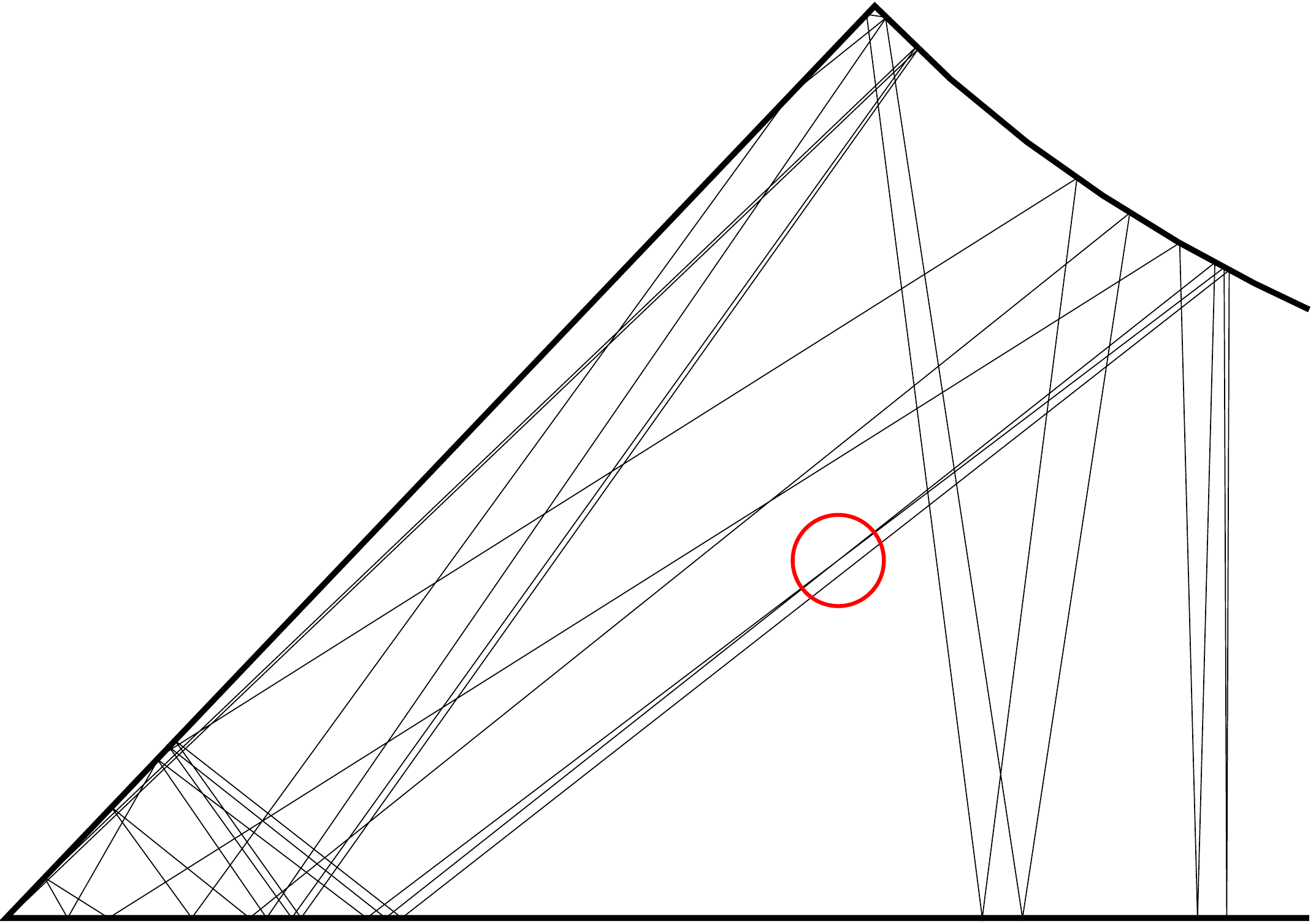}
    \includegraphics[width=0.4\linewidth]{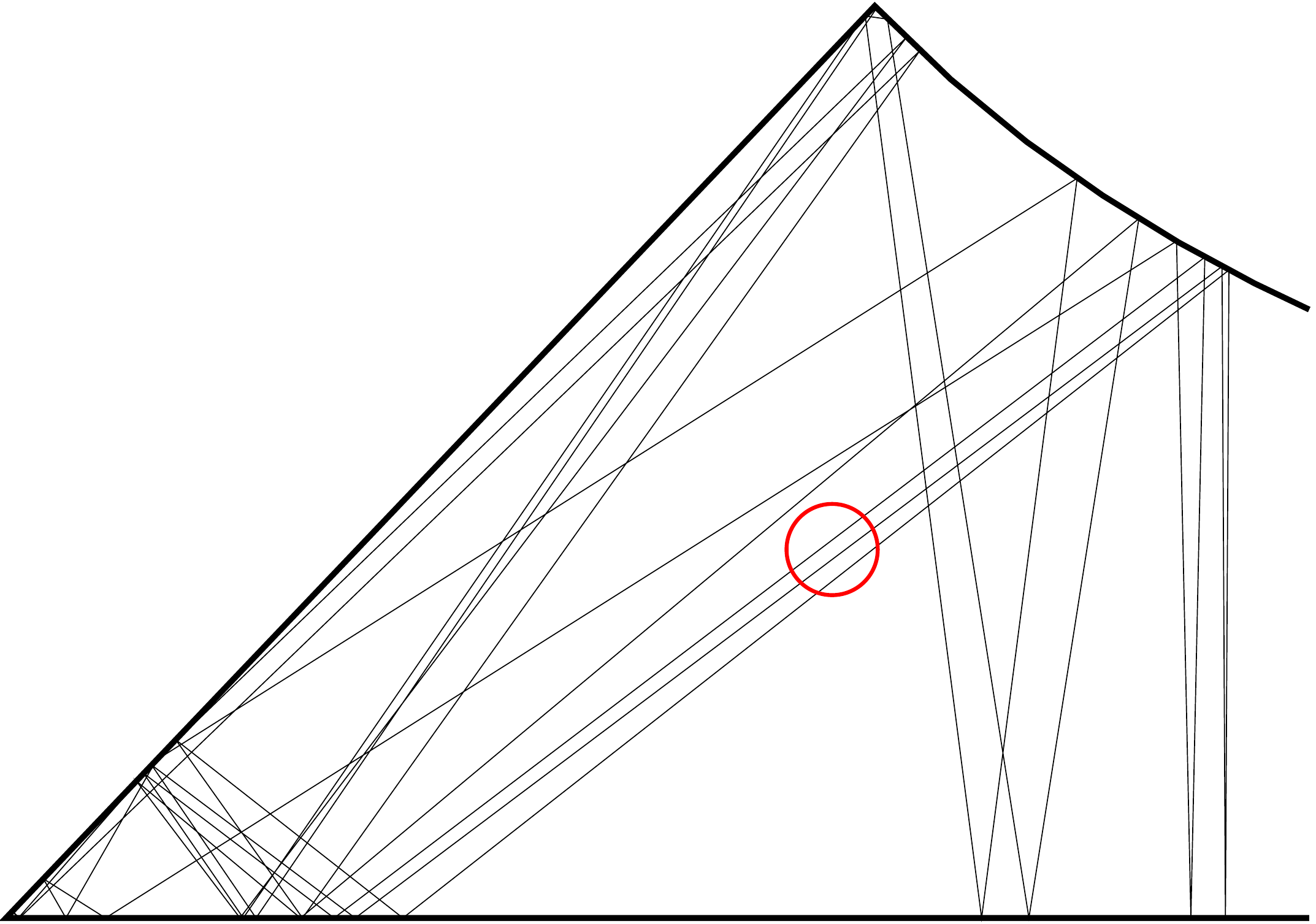}
    \caption{\label{fig:PO-example}
   {\bf "Where's Waldo"?} | 
   Example of a correlated pair of two periodic orbits in the hyperbola billiard, essentially differing from each other in the regions marked by the red circle, where the left orbit exhibits a self-crossing while the right partner orbit does not cross.  This orbit pair illustrates a two-encounter (here in two-dinensional configuration space). The corresponding encounter region, centered around the crossing, extends along the orbits over a scale of $v\tE$  comprising various reflections at the boundaries, depending on $\log \hbar^{-1}$
     (adapted from Ref.~\cite{Haake11} with permission; courtesy of M.~Sieber).
  }
\end{figure}

\begin{figure}
    \centering
  \includegraphics[width=0.55\linewidth]{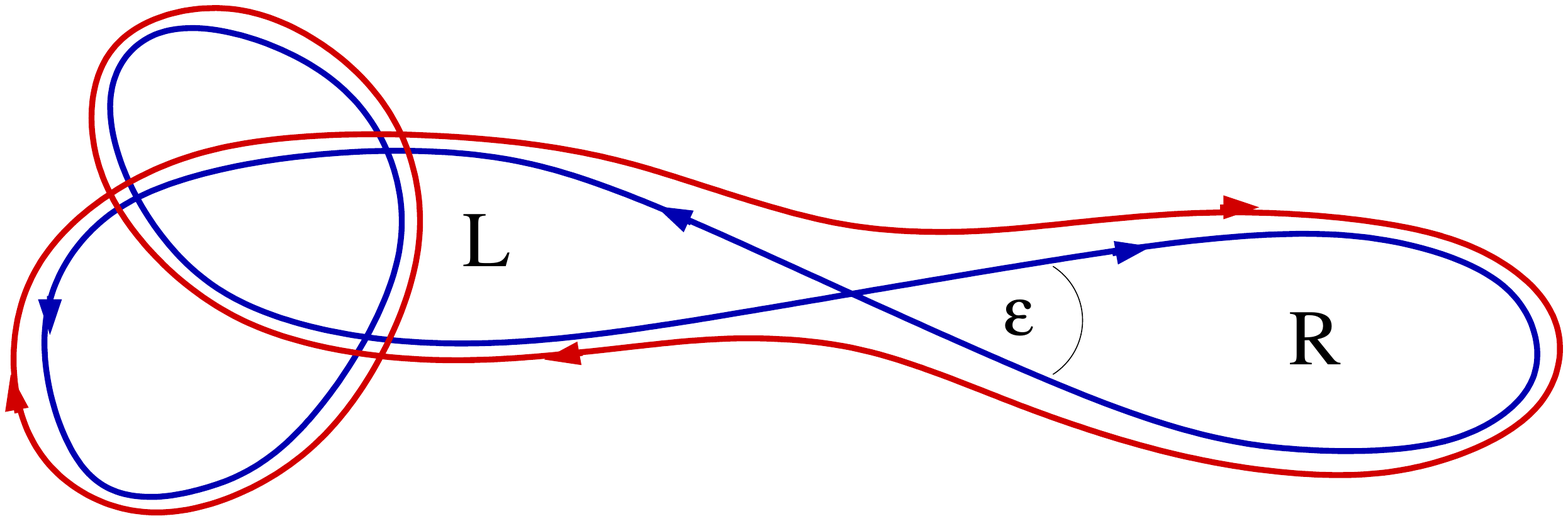}
    \includegraphics[width=0.4\linewidth]{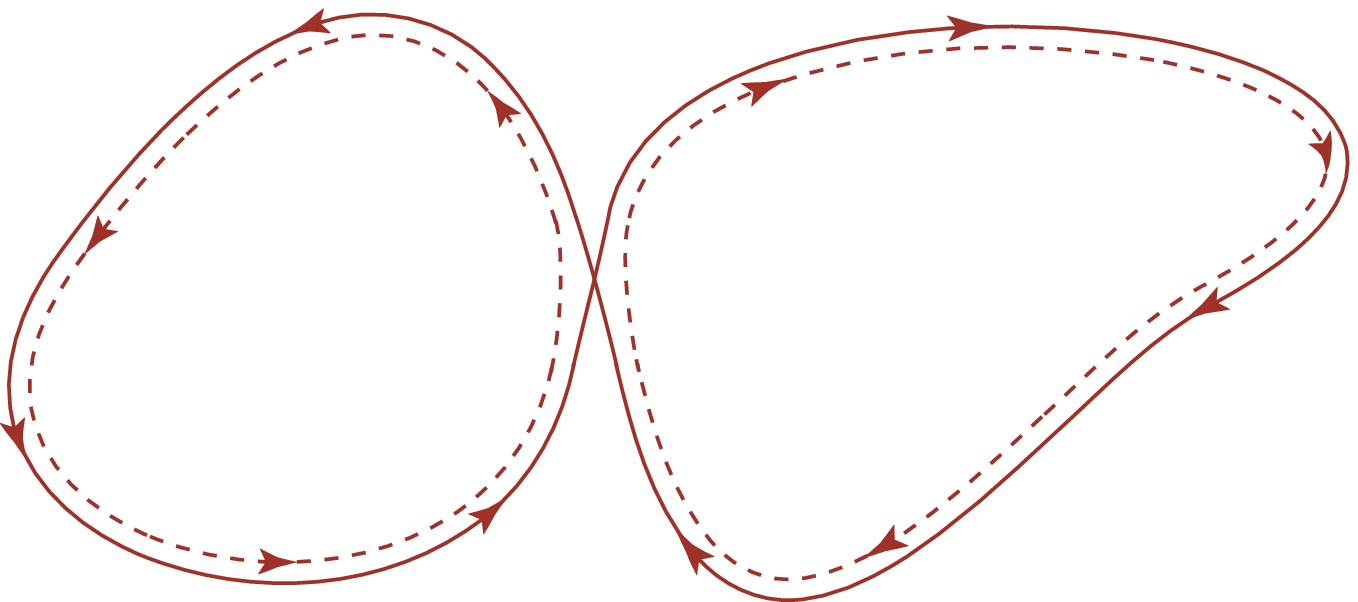}
       \caption{\label{fig:pseudo-orbit}
   {\bf Braided periodic orbit pairs} |
   Left: Scheme of a fundamental pair of classically correlated long periodic orbits linked to each other through a common self-encounter region~\cite{Sieber01}.
   Right:
   Two periodic orbits (dashed lines) forming a pseudo orbit of order 2  contributing to the pseudo-orbit expansion of the spectral determinant in Eq.~(\ref{eq:pseudo1}) according to the rules in Eq.~(\ref{eq:pseudo2}). In this particular case where the composing orbits almost touch and therefore define an encounter region, they are correlated with the longer eight-shaped orbit (solid). (Right panel from Ref.~\cite{Mueller09}.)
   }
\end{figure}

The fact that braided bundles generally tend to arise due to the symplectic phase space structure is best illustrated by considering just two orbits. Figure~\ref{fig:PO-example} shows a representative example of such a pair of periodic
trajectories in the hyperbolic billiard known to exhibit chaotic dynamics.
Since the configuration space in the billiard interior is bounded,
long periodic trajectories necessarily have many self-crossings,
including those with a small angle between the intersecting segments
(see the self-crossing marked by a red circle in the left panel of Fig.~\ref{fig:PO-example}).
The right panel shows an almost identical partner trajectory,
which differs in topology from the reference trajectory only in the area of the self-encounter
in that the partner orbit has no intersection.
Such a trajectory doublet is shown in the left panel of Fig.~\ref{fig:pseudo-orbit} again schematically. 
One of the paths has a crossing in the configuration space under a small angle
$\epsilon$. From this the corresponding partner trajectory can be uniquely constructed
by matching trajectory segments associated with the local stable and unstable manifold of the reference orbit~\cite{Sieber01}. 
Then for each such long orbit, a partner orbit starting and
ending (exponentially) close to the first one exists. 
For the fundamental braided orbit pair shown in the left panel of Fig.~\ref{fig:pseudo-orbit}, each of the two paths around the intersection has a {\em self-encounter}, where its segments are close to each other in configuration space. Outside the encounter region the two loop-like connecting pieces ("L" and "R" in Fig.~\ref{fig:pseudo-orbit}, called ''links'') are almost indistinguishable for the two trajectories, since they are exponentially close. In a rough but helpful simplification, one can consider the links (loops) of both orbits as the same, whereas the two
possibilities of their interconnection in the encounter region allows for constructing and distinguishing the two different orbits. 

The close similarity of periodic orbits forming a pair, such as those depicted in Fig.~\ref{fig:PO-example} and sketched in Fig.~\ref{fig:pseudo-orbit} implies a tiny difference $ \Delta S(E)$, Eq.~(\ref{eq:delta_S}), in their classical action, and accordingly, a small phase difference
$(i/\hbar) (S_\gamma(E)-S_{\gamma'}(E))
$
entering the semiclassical expression, Eq.~(\ref{eq:SP-2-point-sc}), for the spectral correlator.
Placing a transversal
Poincaré surface of section inside the encounter and considering the differences between the two
points where the resepctive encounter stretches pierce through that section, the distances between the piercings can be decomposed into stable and unstable components $s,u$. They determine the approximate action difference of the two partner orbits in a two-encounter as 
$\Delta S \approx su$~\cite{Mueller09} (see also \cite{Sieber01,Sieber02}).  In Ref.~\cite{Li17b} exact geometric relations are given for $\Delta S$ in terms of the properties of Moser invariant curves in the homoclinic structure underlying encounters.  The relative scale of the correction between $\Delta S \approx su$ and the exact result is exponentially small.

In constructing a partner path by switching at a self-encounter, it may happen that not one periodic trajectory, but two or more shorter ones form in such a way that their combination corresponds to the original orbit as a whole.  Such a way of combining orbits is  a particular instance of a pseudo-orbit~\cite{Keating91}, a composite object where a set of orbits combine their stability  and actions in a specific way shown later in Eqs.~(\ref{eq:pseudo1},\ref{eq:pseudo2}). Figure~\ref{fig:po-pair-HR} and the right panel of Fig.~\ref{fig:pseudo-orbit} show simple examples.
Such composites connecting orbits with pseudo-orbits also play a role in the next subsection. While the orbit pair to the left requires time-reversal symmetry, the bundle to the right also exists for the non-time reversal symmetric case.


\begin{figure}
    \centering
    \includegraphics[width=0.70\linewidth]{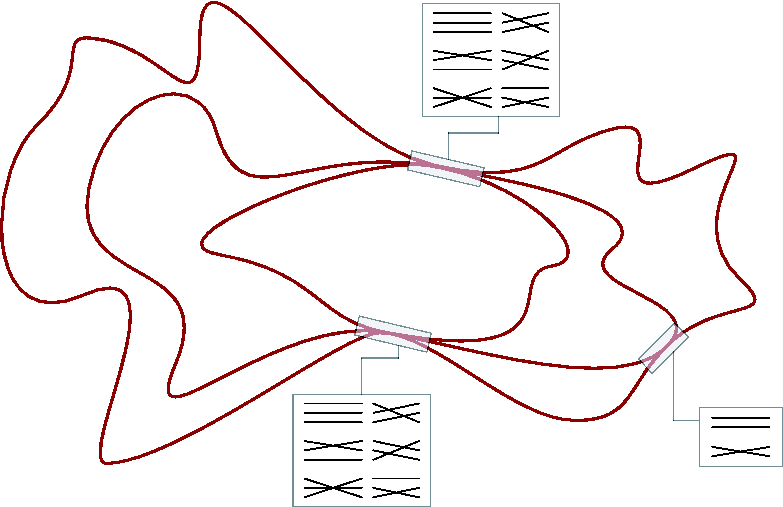}
    \caption{\label{fig:PO-bunches}
   {\bf Braided periodic orbit bundles} | 
  Example of a bundle of 72 periodic-orbit structures (single periodic orbits or pseudo-orbits composed of shorter periodic orbits) with nearly equal lengths and actions differing in two three-encounters and one two-encounter.
The illustration deliberately conveys the impression that it is
only a single orbit. Only in the boxes the different $l!$ interconnections within the encounter regions appear to be resolved. 
(From Ref. \cite{Mueller09}).
   }
\end{figure}

The notation of links connecting self-encounters is helpful for devising the general mechanism for "constructing" (pseudo-) orbits via close self-encounters~\cite{Haake18,Mueller09}.  Every long periodic orbit necessarily has many close self-encounters in configuration space. Not only two, but also three or generally $l$ orbital segments can temporarily approach each other, thus defining an $l$ encounter.  The corresponding $l$ links outside of the encounter can be interconnected in $l!$ different ways through the $l$ encounter, defining a bunch of $l!$ different trajectories. Given one, the Hamiltonian phase space structure assures the existence of all of these orbits. Inasmuch as a long periodic orbit has many close self-encounters $k$,
each of which realizes $l_k!$ possible switchings, 
such a trajectory is a member of a group of trajectories with a total number given by the product of all factors, $N=\Pi_k l_k!$.  Figure \ref{fig:PO-bunches} shows a bundle of $N=(3!)^2 2!=72$ trajectory structures,
generated from two three- and one two-encounter. This bundle comprises individual periodic orbits as well as pseudo-orbits of nearly the same total period~\footnote{In principle, the periodic orbits sketched in Fig.~\ref{fig:PO-bunches} may contain higher repetitions of the entire orbits or parts of them. However, it has been shown~\cite{Waltner19} that trajectories with multiple partial traversals do not contribute (to leading order) to the spectral two-point correlator dsicussed below.}.

The existence of encounters and the construction scheme outlined above is not restricted to periodic orbits but holds in general also for open trajectories~\cite{Richter02, Li20}, with relevance for instance in quantum chaotic transport and scattering; see also Secs.~\ref{sec:SP-Ehrenfest} and \ref{sec:OTOC}. The underlying mechanism of forming orbit bundles is the same in all cases. 
Generally, the longer open or periodic orbits become, the more close encounters they have with other orbits,  leading to the notion that in the long-time-limit all orbits form whole nets weaving the classical phase space with a fine mesh: in the sense of Poincaré's original conception.


\subsubsection{Quantum spectral universality}
\label{sec:SP-universality}


The issue of orbit bundles does not naturally arise in classical SP physics, but their relevance for quantum physics is immediately obvious: due to the close similarity of all members of an orbit bundle, e.g.~as depicted in Fig.~\ref{fig:PO-bunches}, the members exhibit near-degenerate actions and are too highly correlated to ignore when energy averaging.  This was discovered and first worked out by Sieber and one of the authors~\cite{Sieber01} for the case of correlated orbits pairs forming a two-encounter (Fig.~\ref{fig:pseudo-orbit}, left).  Their analysis provided the leading quadratic contribution to the GOE spectral form factor,
Eq.~(\ref{eq:SP-FormFac}), beyond the linear ramp, thereby revealing symplectic chaotic dynamics as the semiclassical origin of RMT behavior.  Based on these insights, Haake and members of his group worked out the encounter calculus that allows one to classify and compute general encounter structures, and used it in a ``tour de force'' approach for systematically calculating the semiclassical theory for the two-point correlator and  spectral form factor, respectively.  In the following, the major steps of their approach are outlined, which can be generalized to the many-body context; see Sec.~\ref{sec:spec-stat}. All details can be found in Haake's textbook~\cite{Haake18}.

Historically the semiclassical calculation of spectral correlators, that starts with the classic 1985 paper by Berry \cite{Berry85}, was based on their representation in terms of the bare trace formula as in Eq.~(\ref{eq:SP-2-point-sc}). In this representation the only structures of relevance are consequently built from pairs of {\it orbits}. The initial success of the semiclassical program for  spectral universality based on the enumeration, classification, and calculation of the pertinent encounter structures contributing to the spectral form factor was, however, restricted to times shorter than the Heisenberg time \cite{Mueller04}. Deriving the behavior of spectral fluctuations beyond this point requires understanding the semiclassical mechanisms that account for quantum unitarity and its non-perturbative effects.  Whereas a version of the trace formula in which unitarity can be studied and/or implemented remains elusive, the so-called spectral determinant
\begin{equation}
    Z(E)=B(E)\, {\rm det}(E-\hat{H})
\end{equation}
(where $B(E)$ is a real function of the energy $E$ without real zeros) provides a powerful periodic orbit expansion.  There unitarity can be explicitly enforced, and it offers a more convenient starting point for a semiclassical calculation based on action correlations aiming to include post-Heisenberg time effects. The price to pay is that the whole enumeration problem now involves pairs of {\it pseudo orbits}.

The starting point of this analysis is the formal identity
\begin{equation}
  \log {\rm det}(E-\hat{H})={\rm Tr~}\log (E-\hat{H} ) 
\end{equation}
that, together with the definition of the spectral resolvent 
\begin{equation}
    R(E+i0^{+})={\rm Tr~}(E+i0^{+}-\hat{H})^{-1} 
\end{equation}
at the complex energy $E^{+}=E+i0^{+}$, allows one to write
\begin{equation}
   Z(E^{+})\sim \exp{\left(\int^{E^{+}}R(E)dE\right)} \, .
\end{equation}
Here the symbol $\sim$ indicates that an arbitrary integration constant producing a multiplicative term absorbed in the function $B(E)$ is omitted. The semiclassical approximation to the spectral determinant is readily obtained by means of the semiclassical representation of the resolvent as a sum over periodic orbits a la Gutzwiller: Integrating Eq.~(\ref{eq:SP-Gutzwiller}) yields, to leading order in $\hbar$,
\begin{equation}
\label{eq:Z}
    R_{\rm sp}(E)=-i\pi \bar{N}_{\rm sp}(E)-i\sum_{\rm po}\frac{A_{\rm po}(E)}{T_{\rm po}}{\rm e}^{i S_{\rm po}(E)/\hbar} \, .
\end{equation}
A careful analysis of this object, beautifully done by Berry and Keating \cite{Berry90} requires the consistent treatment of the sum over repetitions implicit in the trace formula. A simplified version where repetitions are neglected at both the level of the density of states and the spectral determinant is then obtained by simply expanding the exponential in (\ref{eq:Z}). Noticing that for primitive orbits $T_{\rm ppo}=T_{\rm po}$ and therefore $A_{\rm po}/T_{\rm po}=F_{\rm po}$  depends only on the stability of the orbit, one then naturally regroups the terms of the exponentiated sum and orders it by the number of primitive orbits that compose them. The resulting expression is then  a sum 
\begin{equation}
\label{eq:pseudo1}
   Z_{\rm sp}(E^{+})\sim {\rm e}^{-i\pi \bar{N}(E)}\sum_{\rm pso}(-1)^{n_{\rm pso}}F_{\rm pso}{\rm e}^{iS_{\rm pso}/\hbar} 
\end{equation}
over pseudo orbits (pso) of increasingly large order $n$ with
\begin{equation}
\label{eq:pseudo2}
   F_{\rm pso}=\prod_{\rm ppo}^{n}F_{\rm ppo}, {\rm \ \ \ and \ \ }S_{\rm pso}=\sum_{\rm ppo}^{n}S_{\rm ppo},
\end{equation}
including the empty pseudo-orbit $F_{0}=1,S_{0}=0$.


\begin{figure}
    \centering
    \includegraphics[width=0.8\linewidth]{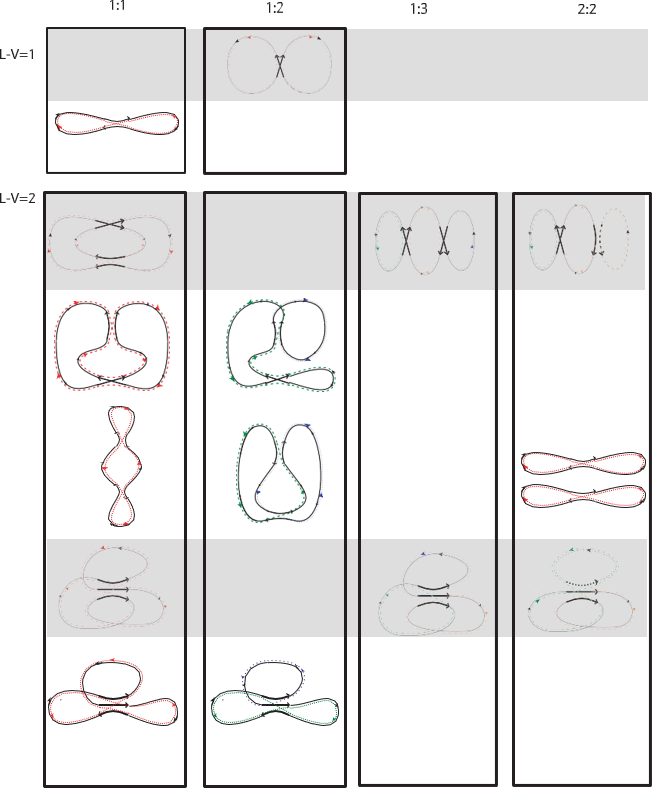}
    \caption{\label{fig:PO-geometries}
   {\bf
   Tableau of periodic-orbit bundles} | Some of the pseudo-orbit pairs whose correlated actions are responsible for the generating function of spectral correlators in Eq.~(\ref{eq:generZ}).
   (From Ref. \cite{Mueller09}.)
   }
\end{figure}
 
 
The implementation of quantum unitarity at the semiclassical level, so far beyond control due to the very nature of the sum over periodic orbits as being formally divergent, follows now in two steps. First, the correct analytical structure of the resolvent as a meromorphic function of the complex energy (correspondingly the density of states being a distribution given as by the usual sum over Dirac-delta peaks) is imposed by imposing the exact relation
\begin{equation}
    R(E^{+})=\frac{d}{dE^{+}} \log Z(E^{+})=\left(\frac{1}{Z(E')}\frac{d}{dE^{+}}Z(E^{+})\right)_{E' \to E^{+}}, \end{equation}
on the corresponding semiclassical approximations $R_{\rm sp}, Z_{\rm sp}$. To finally implement quantum unitarity in full, in a second step one reinforces the reality of the quantum mechanical energies by constructing a spectral determinant that is real for real energies. This condition can be implemented as well at different levels of rigor, the lowest being simply the replacement
\begin{equation}
\label{eq:RS}
  Z_{\rm sp}(E) \to \bar{Z}_{\rm sp}(E)={\cal R}Z_{\rm sp}(E) {\rm \ \ \ for \ real \ } E  
\end{equation}
that, however, makes the definition of how to perform the limit $\bar{Z}_{\rm }(E^{+} \to E)$ ambiguous. 

In a remarkable paper \cite{Heusler07}, the resulting 2-point spectral correlator, Eq.~(\ref{eq:SP-2-point}) based on the improved resolvent
\begin{equation}
\label{eq:RS-res}
    R_{\rm sp}(E^{+})=\left[\frac{d}{dE^{+}}\sum_{A,B}(-1)^{n_{A}}\left({\cal R}F_{A}(E^{+}){\rm e}^{iS_{A}/\hbar}\right)F_{B}(E'){\rm e}^{iS_{B}(E')}\right]_{E'\to E^{+}}
\end{equation}
was computed by incorporating the encounter calculus to include correlated quadruplets of pseudo-orbits of any order appearing in the generating function
\begin{equation}
\label{eq:generZ}
    {\cal Z}(E_{A},E_{B},E_{C},E_{D})=\left\langle\frac{\bar{Z}_{\rm sp}(E_{A})\bar{Z}_{\rm sp}(E_{B})}{Z_{\rm sp}(E_{C})Z_{\rm sp}(E_{D})}\right\rangle
\end{equation}
that then leads to the spectral correlators by differentiation and identification
\begin{equation}
\label{eq:RR}
  \langle R_{\rm sp}(E_{A})R_{\rm sp}(E_{B})\rangle=\left(\frac{\partial^{2}}{\partial E_{A} \partial E_{B}}  {\cal Z}(E_{A},E_{B},E_{C},E_{D})\right)_{(E_{C},E_{D}) \to (E_{A},E_{B})}.
\end{equation}

The calculation of correlated quadruplets of pseudo-orbits requires generalizing the methods initially devised for orbit correlations in \cite{Mueller04} to include now multiple correlated orbits within pseudo-orbits. These correlations are defined through the familiar mechanism where orbits with systematically small action difference $\Delta S$, Eq.~(\ref{eq:delta_S}), are obtained by reshuffling the segments of the orbits inside an encounter, and such differences are consequently characterized by the number and type of the encounters. Two important aspects of the encounter structure of correlated pseudo-orbits are their total number ($V$) and the total number of orbit segments that approach within all encounters ($L$). Terms in the semiclassical expansions are then typically labelled by the function $g=L-V$. 

To gain intuition about the encounter mechanism in the context of pseudo-orbits, consider first some of the lowest orders in the expansion of the generating function ${\cal Z}$. Besides the diagonal approximation, that is accounted for separately, the first correlated pseudo-orbits correspond to the sets $A= \{\gamma \},B=\{\gamma '\}, C=D=\{\}$ with $\gamma,\gamma'$ a pair of correlated orbits. Neglecting highly oscillatory terms proportional to ${\rm e}^{2i\bar{N}_{\rm sp}(E)}$ that vanish under average, this pairing can be obtained in two different ways. The enumeration of all possible pairs of correlated orbits was already achieved in \cite{Mueller04}. It starts with the lowest order of one 2-encounter, the Sieber-Richter pair corresponding to $L=2,V=1$ that, following the diagramatic rules of encounter calculus, contributes to the spectral correlation with a term proportional to $1/(E_{A}-E_{B})^{g}$ with $g=L-V=1$, as depicted in Fig.~(\ref{fig:PO-geometries}) on the leftmost top diagram. As this contribution requires the existence of time-reversal invariance symmetry, it is simply not present in the unitary case.

Generalizing this situation, the contributions from correlated orbits with higher and higher structures and increasingly larger $g$ can be considered. This is nothing but the encounter expansion one obtains from the usual orbit (instead of pseudo-orbit) approach.  There are cancellations between contributions for all $g>1$ in the unitary case. This is shown for the particular case $g=2$ where the contributions from the only possible diagrams allowed by broken time-reversal invariance (grey shadow) have $L=4, V=2$ and $L=3,V=1$ and exactly cancel each other. The other possible diagrams admited in the case of preserved time-reversal invariance end up giving the corresponding non-zero contribution to order $1/(E_{A}-E_{B})^{3}$ in accordance with the perturbative expansion of the universal RMT result. 

Genuine pseudo-orbit correlations, beyond what can be obtained by pairing of two orbits, are shown in the $n:n'$ columns of Fig.~(\ref{fig:PO-geometries}), where $(n,n')\ne(1,1)$ indicates the order of the pseudo-orbits involved. The key observation is that at the perturbative level all such contributions must be cancelled against each other order by order in both orthogonal and unitary symmetry classes, as the result from simple orbit correlations $n=1,n'=1$ was already shown in \cite{Mueller04} to give the correct perturbative part of the universal result. The explicit verification of such cancellation mechanism crucially depends on the $(-1)^{n}$ factors in Eq.~(\ref{eq:RS-res}) and was carried on in \cite{Heusler07}.

The result of the semiclassical evaluation of Eq.~(\ref{eq:RR}) is finally split into two contributions arising from the two possible identification of arguments $(E_{C},E_{D}) \to (E_{A},E_{B})$ or $(E_{C},E_{D}) \to (E_{B},E_{A})$. As shown in \cite{Heusler07}, the first, i.e.~diagonal, identification simply reproduces the result obtained from the representation in Eq.~(\ref{eq:SP-2-point-sc}) with the form of a power expansion in the small parameter $1/(E_{A}-E_{B})$, and it is therefore denoted as perturbative. Interestingly, the cross identification $(E_{C},E_{D}) \to (E_{B},E_{A})$ results in a characteristic oscillatory dependence ${\rm e}^{-2\pi i\bar{\rho} (E_{A}-E_{B})}$ that is obviously non-perturbative, also denoted as the oscillatory contribution. The result of the combined calculation of the semiclassical diagrams for both the perturbative and oscillatory contributions turns out to correctly reproduce term by term the same precise structure in the universal RMT result.

The reduction of the large set of pseudo-orbit correlations back to the perturbative result obtained from orbit pairs relies both on the consistence of pseudo-orbit vs orbit correlations, and on massive cancellations, with the lowest order examples shown in the table in Fig.~(\ref{fig:PO-geometries}). Interestingly, as shown in~\cite{Waltner09}, in the case of ratios instead of spectral determinant products, the origin and interpretation of these cancellations is related instead with so-called curvature effects.

Besides for the spectral two-point correlator and form factor, respectively, during the last 20 years the encounter calculus has been developed for and applied to higher spectral correlation functions~\cite{Mueller18} and many other observables, including scattering, quantum transport, and quantum echoes, to name just a few.
In Secs.~\ref{sec:SP-Ehrenfest} and \ref{sec:OTOC} we will partly review these activities.


\subsection{Ehrenfest phenomena}
\label{sec:SP-Ehrenfest}

The formation of orbit bundles with quasi-degenerate actions is intimately connected with encounters as a structural element of chaotic Hamiltonian dynamics braiding the orbits involved. For trajectory pairs with action difference of order $\hbar$, the encounter time $t_{\rm enc}$ corresponds to the Ehrenfest time
\begin{equation}
\tEsp  
= \frac{1}{\lsp} \log \frac{L}{\lambda_{\rm dB}}
= \frac{1}{\lsp} \log \frac{S}{\hbar}  \, , 
 \, 
\label{eq:sp-Ehrenfest}
\end{equation}
where the label ``sp'' is used to delineate the $\hbar$-dependent Ehrenfest time in the SP context from the $\heff$-dependent many-body Ehrenfest or scrambling time, Eq.~(\ref{eq:scrambling}), discussed in Sec.~\ref{sec:OTOC}. 
Notably, $\tEsp$ links classical and quantum scales, namely the largest positive Lyapunov exponent $\lsp$ with $\hbar$ or the ratio between linear system size $L$ and (de Broglie) wave length $\lambda_{\rm dB}$, respectively.  In Eq.~(\ref{eq:sp-Ehrenfest}), $L/\lambda_{\rm dB}$ can be replaced by $S/\hbar$ where the classical action $S$ can be viewed as scale for a corresponding phase space section.  Rewriting Eq.~(\ref{eq:sp-Ehrenfest}) as 
$L = \lambda_{\rm dB}\exp{(\lsp \tEsp)}$ provides a simple interpretation: $\tEsp$ corresponds to the time it takes an initial minimum uncertainty Gaussian density of phase points of width $\lambda_{\rm dB}$ to spread to a scale of the system size $L$
in a possibly higher dimensional chaotic system governed by $\lsp$. Beyond the logarithmically short time scale $\tEsp$ interference necessarily sets in~\cite{Chirikov81} and the Ehrenfest theorem~\cite{Ehrenfest27} soon fails (for a recent review of early work of Chirikov and coworkers on $\tEsp$-effects for the standard map, see~\cite{Shepelyansky20}).
 
 In Sec.~\ref{sec:SP-correlations}, RMT-type universality was  deduced by formally taking the semiclassical limit of large $\tH$ for fixed $\tau\!=\! t /\tH$, {\em i.e.},  involving increasing times $t$. In turn, this implies that encounter times, $\tEsp$, collapse to zero, since they scale logarithmically with $\hbar$.  However, measurements and numerical calculations commonly show fascinating quantum chaotic phenomena in the regime of small but non-vanishing $\hbar$, {\em i.e.}~non-vanishing $\tEsp$, implying deviations from RMT-type universality~\cite{Berry85}. Here a brief non-technical overview is given over this regime that is perfectly amenable to semiclassical methods, but beyond RMT approaches; for a detailed review-type account of the underlying semiclassical theory of Ehrenfest effects with an exhaustive account of the literature, see {\em e.g.} Ref.~\cite{Waltner12}.

 After it had been demonstrated in the early $90$'s that -- contrary to the prevailing perspective --  it was possible to develop advanced methods to adequately treat post-Ehrenfest quantum dynamics purely semiclassically~\cite{Tomsovic91b, Oconnor92, Sepulveda92, Tomsovic93}, Ehrenfest phenomena were investigated, particularly for observables relevant in mesoscopic quantum systems. There the $\tEsp$-dependence was considered for a large variety of scattering, transport, spectral, and quantum decay properties of chaotic conductors for which representative examples are given in the following.

\subsubsection{Quantum transport}

\begin{figure}
    \centering
  \includegraphics[width=0.45\linewidth]{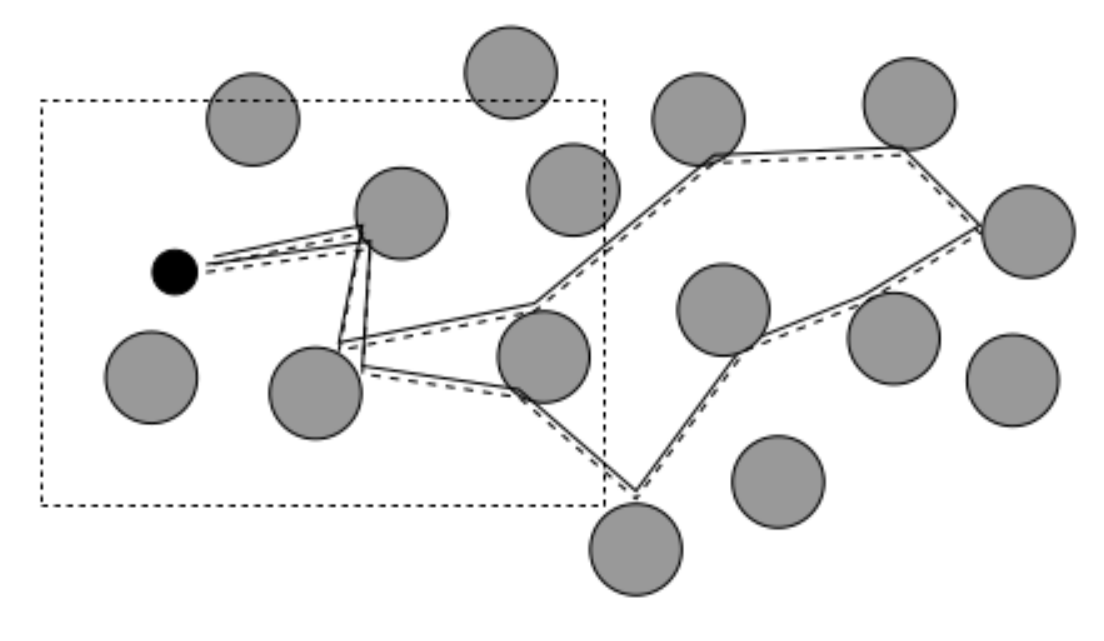}
  \hspace{2cm}
    \includegraphics[width=0.2\linewidth]{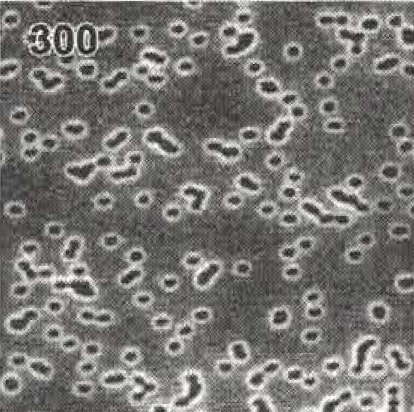}
    \caption{\label{fig:Lorentz}
   {\bf Ehrenfest effect on weak localization in a Lorentz gas} | 
   Left: Sketch of a pair of paths contributing to coherent backscattering in a ballistic system composed of randomly placed disks. Trajectories of a mimimal wave packet of size $\lambda_{\rm dB}$, marked as black dot, separate on scales of $\tEsp$, Eq.~(\ref{eq:sp-Ehrenfest}), up to a distance of the size $L$ of the classical scatterers, providing a mechanism for splitting the initial wave packet and leading to coherent backschattering due to constructive interference of time-reversed back-reflected paths.
   Right: Experimental realization of a Lorentz gas built from an irregular array of "antidots" in a  high-mobility two-dimensional electron system 
(Left and right  panel from Refs.~\cite{Richter00} and \cite{Yevtushenko00}, respectively).
   }
\end{figure}

For the Lorentz gas, a prototypical model of randomly placed disks acting as chaotic classical scatterers in an otherwise ballistic two-dimensional system~\cite{Gaspard98},
$\tEsp$-signatures in weak localization were first theoretically discussed in Ref.~\cite{Aleiner96prb}. Based on a ballistic $\sigma$-model, and invoking averaging over weak disorder, this approach accounted for correlations in the initial chaotic dynamics (dashed box in Fig.~\ref{fig:Lorentz}) up to $\tEsp$. For later times dynamics merges into uncorrelated diffusive behavior in the Lorentz gas setting. The combined mechanism of initial chaotic spreading, followed by diffusive backscattering lead to the prediction \cite{Aleiner96prb}
\begin{equation}
\Delta \sigma \simeq -\frac{e^2}{\pi\hbar} \exp{\left(-\frac{
\tEsp }{t_\phi}\right)} \, \ln \left(
\frac{t_\phi}{t_{\rm e}}\right)
\label{eq:delta-sigma}
\end{equation}
for the weak localization correction to the two-dimensional conductivity.
In Eq.~(\ref{eq:delta-sigma}), $t_{\rm e}$ and $t_\phi(T)$ denote the elastic scattering time and the temperature-dependent phase breaking time, respectively.
The subsequently observed unusual exponential temperature dependence of $\Delta \sigma(T)$  for a ballistic electron gas in between randomly placed antidots (right panel of Fig.~\ref{fig:Lorentz}) allowed for experimentally detecting and extracting the Ehrenfest time~\cite{Yevtushenko00} using Eq.~(\ref{eq:delta-sigma}). In view of Eq.~(\ref{eq:sp-Ehrenfest}), it is then possible to estimate the {\em classical} Lyapunov exponent $\lsp$ of such an electron antidot billiard from the {\em quantum} weak localization correction $\Delta \sigma$.
See Ref.~\cite{Schneider13} for later work on how the Ehrenfest time effectively poses a short-time threshold for the trajectories contributing to the interaction correction  in antidot lattices.


After these earlier studies, and parallel to the development of the encounter calculus for spectral correlators introduced above, the particular relevance of action correlations and braided orbit bundles for chaotic quantum transport has quickly become evident. In Ref.~\cite{Richter02} the leading-order weak localization correction to the classical magneto-transmission of ballistic mesoscopic conductors was computed and related to off-diagonal contributions and interference of encounter-correlated lead-connecting trajectories. This finding is in agreement with RMT and solved a decade long issue concerning missing current conservation in semiclassical transport theory (based on the diagonal approximation~\cite{Blumel88,Baranger91}) for chaotic disorder-free conductors. Various subsequent theoretical works have extended semiclassical quantum transport theory on the basis of encounter calculus or related approaches. These include the semiclassical calculation of higher-order corrections to the classical conductance of ballistic cavities~\cite{Heusler06}, shot noise in chaotic conductors~\cite{Lassl03,Schanz03,Braun06,Whitney06,Mueller07,Brouwer07}, weak anti-localization in spin-orbit coupled semiconductor cavities~\cite{Zaitsev05,Zaitsev05a,Bolte07}, higher transport moments and correlators~\cite{Berkolaiko12,Mueller07,Novaes13,Berkolaiko11,Bereczuk21,Novaes22}, Wigner-Smith time delay~\cite{Berkolaiko11,Kuipers07,Kuipers08,Kuipers14,Novaes15}, role of tunnel barriers~\cite{Waltner12a,Kuipers13,Bento21,Oliveria22}, and extensions to the ac-conductance~\cite{Petitjean09}.

Moreover, for transport through chaotic conductors Ehrenfest phenomena were again the focus of interest. It turned out that the weak localization correction to the classical conductance indeed shows a characteristic exponential suppression, reading (for large number of scattering channels)~\cite{Adagideli03}
\begin{equation}
\Delta G = -2\frac{e^2}{h} \, e^{\tEsp/\tD } 
\label{eq:delta-sigma-Ehrenfest}
\end{equation}
in terms of the dwell time, $t_{\rm D}$, characterizing the typical time for a particle that enters the scattering region to leave again. Equation~(\ref{eq:delta-sigma-Ehrenfest}) has a straight-forward interpretation: if $\tEsp >\tD$, a charge carrier will leave the chaotic conductor with dwell time $\tD$ before an encounter could be formed, thereby suppressing the conductance correction on scales smaller than the encounter time. Reference~\cite{Adagideli03}, focussing on the transmission, has been complemented and generalized by further trajectory-based approaches~\cite{Jacquod06,Rahav06,Brouwer06,Altland07,Waltner10}  that also considered $\tEsp$-effects on the complementary weak localization mechanism for the quantum reflection. In addition, shot noise measured in terms of the Fano factor exhibits such a $\tEsp$-behavior implying that it vanishes in the strict limit of large Ehrenfest times~\cite{Whitney06,Brouwer07}; see Ref.~\cite{Waltner11} for the generalization to full counting statistics.  Corresponding techniques allow for computing $\tEsp$-modifications of the RMT form factor~(\ref{eq:SP-FormFac}) giving rise to deviations from the linear "ramp" for the GUE case~\cite{Brouwer06b,Waltner10}.

Contrary to the suppression of weak localization in the classical limit, in a chaotic quantum dot the variance of the conductance comprises $\tEsp$-independent terms, as was first numerically observed in~\cite{Tworzydlo04, Jacquod04}. The fact that the variance measuring the size of mesoscopic conductance fluctuations prevails in the classical limit is also captured by semiclassical theory~\cite{Brouwer06}. It arises from trajectories spending a long time in the vicinity of periodic orbits inside the cavity.  Since the associated dwell times are arbitrarily long, these trajectories overcome the mechanism that $\tEsp$ provides a short-time suppression for quantum effects. Moreover, conductance fluctuations in ballistic conductors differ from those of disordered conductors at large $\tEsp$; see~\cite{Brouwer07} for a unified semiclassical treatment of chaotic quantum dots and extended systems exhibiting diffusive dynamics at long time scales, such as depicted in Fig.~\ref{fig:Lorentz}. 

 \subsubsection{Quantum decay}
 
Naturally, the imprints of the Ehrenfest time should appear most directly in the time domain, i.e.~in explicitly time-dependent quantities. Quantum mechanical decay of an open chaotic quantum system, such as a cavity with a hole in the boundary, is predominantly governed by classical exponential decay with dwell time $\tD$. However, a semiclassical interference mechanism similar to that of coherent backscattering leads to an enhancement of the quantum probability compared to the classical survival probability~\cite{Waltner08}, confirming RMT predictions~\cite{Frahm97}. Going beyond RMT one can semiclassically compute explicit $\tEsp$-effects on the time-dependent quantum decay. The spatially integrated probability density inside the open quantum system decreases as
\begin{equation}
\rho(t) \simeq e^{-t/\tD} + e^{-(t-\tEsp)/\tD} \
\frac{(t-2\tEsp)^2}{2\tD\tH} \ \Theta(t-2\tEsp) \, 
\label{eq:decay}
\end{equation}
where the second contribution is the leading term in a series in $1/\tH$ of quantum corrections arsing from 
special pairs of interfering, correlated open trajectories with encounters located at their ends~\cite{Waltner08}.
\footnote{
These trajectories also provide the key mechanism establishing an appropriate semiclassical version of the continuity equation~\cite{Kuipers09}.}
From times $t>2\tEsp$ on, the quantum decay is delayed compared to the classical decay $e^{-t/\tD}$ because it takes a minimum time $2\tEsp$ to form encounters thereby generating quantum interference effects.  This unique $\tEsp$-behaviour has been confirmed by numerical wave packet simulations~\cite{Waltner08}.

\subsubsection{Proximity effect in Andreev billiards}

Finally, post-Ehrenfest interference mechanisms are at the heart of the formation of an induced gap in the density of states of a chaotic quantum dot in proximity to a superconductor.
Such an ``Andreev billiard''~\cite{Kosztin95, Adagideli02} possesses the interesting peculiarity that the suppression of its density of states at the Fermi energy crucially depends on whether the dynamics of its classical counterpart is integrable or chaotic~\cite{Melsen96}.
The spectrum of a chaotic Andreev billiard was expected, according to RMT, to exhibit a hard gap to scale with the energy $\sim \hbar/\tD$, where $\tD$ is the average dwell time a particle moves in the billiard between successive Andreev reflections~\cite{Melsen96}. On the contrary, semiclassical theory based on the 
diagonal approximation yielded only
an exponential suppression of the density of states~\cite{Lodder98,Schomerus99,Ihra01} pointing to an obvious discrepancy that attracted much theoretical interest; see {\em e.g.}~\cite{Adagideli02a,Micklitz09} for first attempts to account for $\tEsp$-aspects, and~\cite{Beenakker05} for a review.

Later it was shown~\cite{Kuipers10,Kuipers11,Engl11}, using encounter calculus, how classical orbit correlations lead to the formation of the hard gap, as predicted by RMT in the limit of negligible Ehrenfest time, and how the influence of a finite $\tEsp$ causes the gap to shrink until the classical regime of exponential suppression is reached. Notably, this crossover is not smooth; instead, for intermediate $\tEsp$ a second pronounced gap was predicted to appear around $E\sim \hbar/\tEsp$ that would be a particularly striking feature of $\tEsp$-effects.

The Ehrenfest time has reappeared  as `scrambling time' in a new guise in the many-body context. In Sec.~\ref{sec:OTOC} the saturation of out-of-time-order correlators at Ehrenfest time scales are discussed as another clear-cut $\tE$-singature, marking the change in interference governed by pre- and post-Ehrenfest semiclassical dynamics.

\section{Semiclassical theory of bosonic many-body quantum chaos}
\label{sec:SC-MB}


\subsection{van Vleck propagator for bosonic systems}


It might seem that the semiclassical approximation, based on the unaltered kinematic structure of quantum mechanics (Hilbert spaces, linear time evolution, observables as Hermitian operators, entanglement, etc.) supplemented by the asymptotic analysis of the propagator, requires only minimal modifications to be ushered into the realm of interacting many-body (MB) systems. Due to the product form of the total Hilbert space in such systems, the corresponding modification of the MB propagator would be simply accounted for by adapting the classical limit into the high dimensional phase space characteristic of MB classical systems.  However, this picture does not account for a kinematic aspect of purely quantum origin, namely quantum indistinguishability, that imposes severe restrictions on the allowed MB identical particle system states by demanding that they have a well defined transformation rule under particle label permutations~\cite{Sakurai94}. This restriction on the allowed states is completely alien to the world of classical mechanics where identical particles can be made indistinguishable only in a statistical sense, whereas the fundamental microscopic dynamics enables, in principle, tracking of each particle's identity simply by following its path in phase space. This feature is even essential at the non-interacting level where it has macroscopic effects, such as the stability of fermionic matter~\cite{Dyson67a, Dyson67b, Lieb76}, the phenomena of Bose-Einstein condensation~\cite{Bose24, Einstein24}, the Hong-Ou-Mandel effect~\cite{Hong87}, and related coherence effects of much recent interest, boson sampling~\cite{Aaronson10}. It is therefore critical to incorporate it in any semiclassical program aimed at MB quantum systems.

In the spirit of the semiclassical theory of particle systems, quantum indistinguishability can be implemented by the application of (anti-) symmetrization operators~\cite{Sakurai94} directly on the arguments of the van Vleck-Gutzwiller propagator for distinguishable particles. In this way, the fermionic (F), bosonic (B) propagators are explicitly given by
\begin{equation}
    K^{\rm F, B}_{\rm sp}({\bf r}_{f},{\bf r}_{i},t):=\frac{1}{N!}\sum_{{\cal P}}\epsilon^{{\cal P}}K_{\rm sp}({\cal P}{\bf r}_{f},{\bf r}_{i},t) 
\end{equation}
involving a sum over the $N!$ elements ${\cal P}$ of the permutation group acting on the particle labels of the $Nd$-dimensional configuration eigenstate $|{\bf r}\rangle$, weighted by their parity $\epsilon^{{\cal P}}$ where $\epsilon=-1\ (1)$ for fermions (bosons). 
The advantage of the semiclassical approximation in this first-quantized representation is that it is solely based on the semiclassical approximation for distinguishable particles, which does not know anything about the F/B nature of the degrees of freedom. 

This first-quantized formalism has been very successful, especially in the framework of  non-interacting MB systems (or weakly interacting ones using perturbation theory), from the mesoscopic theory of quantum transport~\cite{Richter00,Jalabert00} and quantum dots~\cite{Ullmo95} to the description of the scattering of bosonic states through chaotic cavities~\cite{Urbina16}. The problem is substantially more involved if interactions are taken into account due to the very different nature of the sum over classical paths inherent in the semiclassical propagator and the sum over permutations accounting for indistinguishability. This lack of compatibility is clearly seen in the limit $N \gg 1$  that becomes an essentially impossible task due to the (factorial) proliferation of terms in the (anti-) symmetrization process, even if one could efficiently account for the classical distinguishable limit.

The path towards extending semiclassical methods into the realm of quantum interacting systems of identical particles naturally profits from the change of perspective given by the description of such systems in terms of {\it quantum fields}~\cite{Negele98}. On the conceptual level, understanding particle states as excitations of a quantum field means that the individual identity of the distinguishable degrees of freedom, now an extra conceptual baggage without any physical relevance, is immediately absent from the description: instead of building MB states out of (anti-) symmetrized states of distinguishable particles, one specifies states by saying how many particles occupy any given set of single-particle (SP) orbitals irrespective of their individual identities. 

The space of quantum states labelled by such configurations is the familiar Fock space, and the goal of this section is to review the conceptual and technical steps that enable the adaptation of the semiclassical program into this new landscape, as well as its new regime of validity, advantages, and limitations. 


\subsubsection{Semiclassical derivation}

Following standard references~\cite{Negele98}, the construction of the Fock space begins with selecting an arbitrary but fixed set of SP states, denoted from now on as ``sites'' or ``orbitals''
\begin{equation}
  \phi_{i} {\rm \ \ with \ \ \ }i=1,\ldots,d  
\end{equation}
where $d$ is the (maybe infinite) dimension of the SP (or "local") Hilbert space. The quantum mechanical state $|\Psi\rangle$ of a bosonic MB system is then an element of the Fock space $|\Psi\rangle \in {\cal F}$ of the form
\begin{equation}
    |\Psi\rangle=\sum_{\bf n}\Psi_{{\bf n}}|{\bf n}\rangle
\end{equation}
where the basis states, called Fock states, 
\begin{equation}
 |{\bf n}\rangle=|n_{1},\ldots,n_{d}\rangle {\rm \ \ \ \ with \ \ } n_{i}=0,1,2,\ldots 
\end{equation}
are labelled by occupation numbers $n_{i}$, the eigenvalues of the observables $\hat{n}_{i}$
\begin{equation}
    \hat{n}_{i}|{\bf n}\rangle=n_{i}|{\bf n}\rangle
\end{equation}
that count how many particles occupy the SP states $\phi_{i}$. Observables in ${\cal F}$ are written in terms of the corresponding creation/annihilation operators $\hat{b}^{\dagger},\hat{b}$ defined through their action in the basis Fock states,
\begin{eqnarray}
    \hat{b}_{i}|n_{1},\ldots,n_{i},\ldots,n_{d}\rangle&=&\sqrt{n_{i}-1}|n_{1},\ldots,n_{i}-1,\ldots,n_{d}\rangle \nonumber \\
    \hat{b}_{i}^{\dagger}|n_{1},\ldots,n_{i},\ldots,n_{d}\rangle&=&\sqrt{n_{i}}|n_{1},\ldots,n_{i}+1,\ldots,n_{d}\rangle
\end{eqnarray}
satisfying the important relations
\begin{equation}
   \left[\hat{b}_{i},\hat{b}_{j}\right]=0 {\rm \ \ , \ \ }\left[\hat{b}_{i},\hat{b}_{j}^{\dagger}\right]=\hat{1}\delta_{i,j} {\rm \ \ and \ \ } \hat{n}_{i}=\hat{b}_{i}^{\dagger}\hat{b}_{i}.
\end{equation}

As a rule, Hermitian operators that are quadratic forms in $\hat{b}^{\dagger},\hat{b}$ represent single-particle observables, whereas two-body interactions are represented by combinations of fourth order. The general form of the Hamiltonian describing a system of bosons evolving under the influence of external potentials and pairwise interactions is then
\begin{equation}
\label{eq:Ham}
    \hat{H}=\sum_{i,j}h_{i,j}\hat{b}^{\dagger}_{i}\hat{b}_{j}+\sum_{i,j,i',j'}v_{i,j,i',j'}\hat{b}^{\dagger}_{i}\hat{b}_{j}\hat{b}^{\dagger}_{i'}\hat{b}_{j'}\ .
\end{equation}
Correspondingly, the Fock-space propagator (assuming for simplicity time-independent external and interaction potentials) is defined as usual by
\begin{equation}
    K({\bf n}^{(f)},{\bf n}^{(i)},t)=\langle {\bf n}^{(f)}|{\rm e}^{-\frac{i}{\hbar}\hat{H}t}|{\bf n}^{(i)}\rangle
\end{equation}
and our goal is to first identify the MB version of the semiclassical regime $\hbar_{\rm eff}=1/N\to 0$, and then, starting from a suitable path integral representation of $K$, perform the steps to derive the Fock space version of the van Vleck-Gutzwiller semiclassical sum over classical paths.

The first difficulty in attempting this program is the very fact that there is not a path integral between Fock states, at least not in the usual sense of time slicing and inserting representations of the unit operator in terms of Fock states. The issue here is clear: the Fock states define a discrete basis. Historically, this problem was addressed by shifting to the coherent state basis on ${\cal F}$~\cite{Negele98} defined as the common eigenstates of the (non-Hermitian) annihilation operators, and labelled by a set of continuous complex numbers
\begin{equation}
    \hat{b}_{i}|{\bf z}\rangle=z_{i}|{\bf z}\rangle \ .
\end{equation}
They admit a realization of the unit operator in ${\cal F}$
\begin{equation}
  \frac{1}{(2\pi i)^{d}}\int \prod_{i}dz_{i}dz^{*}_{i} |{\bf z}\rangle \langle {\bf z}| = \hat{1} 
\end{equation}
in a form suitable for the usual time-slicing path integral construction. 

The resulting form of the coherent state path integral for bosonic quantum fields has been extensively used as a basis for semiclassical approximations~\cite{Negele98, Klauder78, Baranger01}, so its use to derive a van Vleck-Gutzwiller type of approach is quite appealing. Although, one conceivable drawback of coherent states is that the resulting saddle point equations do not generally admit real solutions, thus requiring the complexification of the classical limit of the theory~\cite{Baranger01}. This approach has recently been implemented and successfully applied to describe quantum dynamics in cold atomic systems in optical lattices~\cite{Tomsovic18}, but its conceptual and technical foundation differs in some essential ways from the van Vleck-Gutzwiller approach taken into Fock space. A main feature of the usual (coordinate) path integral in single-particle systems, which differs from  the complexified phase space inherent in the coherent state approach, is its characteristic Hamiltonian classical limit in terms of real phase space coordinates.  This happens to be naturally consistent with the boundary value problem of ordinary time-dependent WKB theory~\cite{Maslov81}. If one is interested in maintaining the reality of the dynamical variables, the widely used coherent state path integral turns out not to be the ideal starting point for the van Vleck-Gutzwiller derivation.

Interestingly, a more direct approach follows the actual historical development of semiclassical methods for SP systems, where the wave packet propagator (constructed from the translations of the harmonic oscillator ground state) was a later development that came only after the construction of semiclassical approximations in a configuration representation culminating with ordinary time-dependent WKB theory~\cite{Keller58, Maslov81}.  

A path integral in Fock space that provides a semiclassical approximation with real paths can be based on MB states of operators that have three key properties of the familiar configuration (or momentum) operators in SP systems. First, they must have a real, continuous, and unbounded spectrum. Second, they must have, in some precise sense, a classical limit where they play the role of {\it half} of a canonical pair. Third, when the corresponding realization of the unit operator is inserted in the time-sliced propagator, they must produce a real action functional admitting real solutions when extremized with fixed boundary conditions on the paths. A very natural choice is then the manifestly Hermitian combinations
\begin{equation}
\label{eq:quad1}
    \hat{q}_{i}=\frac{\hat{b}^{\dagger}_{i}+\hat{b}_{i}}{\sqrt{2}} {\rm \ \ , \ \ } \hat{p}_{i}=\frac{\hat{b}^{\dagger}_{i}-\hat{b}_{i}}{\sqrt{2}i}
\end{equation}
again reminding us of the relation between the creation/annihilation and position/momentum operators for the SP harmonic oscillator \cite{Sakurai94}. In fact, these pairs can be easily shown to fulfill the relations
\begin{equation}
    \left[\hat{q}_{i},\hat{q}_{j}\right]=\left[\hat{p}_{i},\hat{p}_{j}\right]=0 {\rm \ \ and \ \ } \left[\hat{q}_{i},\hat{p}_{j}\right]=i\delta_{i,j}
\end{equation}
of canonically conjugate operators.  Together with their eigenbases that satisfy all the required conditions as can be seen by direct computation, they are at the center of the construction of the semiclassical approximation for bosonic matter fields. Following the terminology of quantum optics where these canonical pairs are of common use as Hermitian versions of the standard field operators, we refer to them as {\it quadrature operators} or simply quadratures~\cite{Scully97}. Armed with the quadratures and their eigenbases
\begin{equation}
\label{eq:quad2}
    \hat{q}_{i}|{\bf q}\rangle=q_{i}|{\bf q}\rangle {\rm \ \ and \ \ } \hat{p}_{i}|{\bf p}\rangle=p_{i}|{\bf p}\rangle
\end{equation}
nicely satisfying \cite{Engl16}
\begin{equation}
    \langle {\bf q}|{\bf p}\rangle=\frac{{\rm e}^{i{\bf q}\cdot {\bf p}}}{(2\pi)^{d/2}}
\end{equation}
it is possible to express the exact path integral by the usual method of time-slicing and inserting unity.  However, note that inserting the quadrature definitions, Eq.~(\ref{eq:quad1}), into the generic form of the Hamiltonian, Eq.~(\ref{eq:Ham}), leads to Hamiltonians of a very different character than the ``kinetic plus potential energy'' (mechanical type) often found in non-relativistic SP systems. In fact, despite their identical formal properties, the configuration and momentum quadratures do not represent anything like position and momentum, and they can be considered as just a technical tool used to develop a path integral with the desired properties. For systems of massive bosons, they are not observable in the strict sense (a property they share with coherent states~\cite{Bartlett07}), and therefore the construction of the propagator between physical (Fock) states must be also eventually addressed.

Given the above considerations, the construction of the path integral form of the propagator between configuration quadratures is implemented similarly to the standard methodology~\cite{Schulman81} with a few key modifications.  In a nutshell, one first time slices the evolution into a large number of factors of the form ${\rm e}^{-i\delta t \hat{H}/\hbar}$ with $\delta t \to 0$, and inserts the representation of the unit operator in Fock space in terms of the $q$-quadratures. The form of the Hamiltonian in Eq.~(\ref{eq:Ham}), very different from the usual kinetic-plus-potential-energy with the only dependence on the momentum being at most quadratic, demands a careful treatment of the resulting matrix elements. This is conveniently achieved by 
\begin{equation*}
 \langle {\bf q}|{\rm e}^{-\frac{i}{\hbar}\delta t \hat{H}}|{\bf q}'\rangle = \int d{\bf p}\langle {\bf q}|{\rm e}^{-\frac{i}{\hbar}\delta t \hat{H}}|{\bf p}\rangle \langle{\bf p}|{\bf q}'\rangle = \int d{\bf p}\langle {\bf q}|{\bf p}\rangle{\rm e}^{-\frac{i}{\hbar}\delta t \frac{\langle {\bf q}|\hat{H}|{\bf p}\rangle}{\langle {\bf q}|{\bf p}\rangle}}\langle{\bf p}|{\bf q}'\rangle+{\cal O}(\delta t)
\end{equation*}
therefore introducing extra integrations over momentum quadratures, to get the so-called Hamiltonian (or phase space) form of the propagator \cite{Negele98}
\begin{equation}
    K({\bf q}^{(f)},{\bf q}^{(i)},t):=\langle{\bf q}^{(f)}|{\rm e}^{-\frac{i}{\hbar}\hat{H}t}|{\bf q}^{(i)}\rangle=\int {\cal D}[{\bf q}(s),{\bf p}(s)]{\rm e}^{iR[{\bf q}(s),{\bf p}(s)]}
\end{equation}
where the integral is now defined over the space of paths $({\bf q}(s),{\bf p}(s))$. In this representation, only the paths in configuration quadrature endpoints are constrained, 
\begin{equation}
{\bf q}(s=0)={\bf q}^{(i)}, {\bf q}(s=t)={\bf q}^{(f)}.
\end{equation}
whereas the momentum quadrature endpoints are completely unconstrained. Finally, the real-defined action functional is given in its Hamiltonian form by
\begin{equation}
  R[{\bf q}(s),{\bf p}(s)]=\int ds \left[{\bf p}(s) \cdot \dot{{\bf q}}(s)-H_{\rm cl}({\bf q}(s),{\bf p}(s))/\hbar\right]  
\end{equation}
where the classical symbol is obtained from the Hamiltonian operator  expressed with all $\hat{p}$ operators moved to the right of the $\hat{q}$ ones as 
\begin{equation}
H_{\rm cl}({\bf q},{\bf p})=\frac{\langle {\bf q}|\hat{H}|{\bf p}\rangle}{\langle {\bf q}|{\bf p}\rangle}.
\end{equation}
Note that the action functional in Fock space is dimensionless, an aspect that reflects once again how quadrature operators are not related with any physical coordinate/momentum in any sense beyond the formal analogies with their SP counterparts.  A nice feature is the very natural way that the Hamiltonian symbol appearing in the exact path integral is obtained from the quantum operator. Preparing the road for the semiclassical approximation where ordering effects lead to subdominant contributions, and denoting 
\begin{equation}
\label{eq:Tom1}
\hat{H}=H(\hat{{\bf b}}^{\dagger},\hat{{\bf b}}),
\end{equation}
the phase space function appearing in the path integral is given by  
\begin{equation}
\label{eq:Tom2}
H_{\rm cl}({\bf q},{\bf p})=H\left(\frac{{\bf q}+i{\bf p}}{\sqrt{2}},\frac{{\bf q}-i{\bf p}}{\sqrt{2}}\right).
\end{equation}

Before commencing with the Fock space version of Gutzwiller's celebrated analysis of the exact path integral that culminates in the derivation of the semiclassical propagators, two deeply intertwined aspects must be addressed, namely the identification of the semiclassical regime where an asymptotic analysis of the path integral can be meaningfully applied, and the connection of the quadrature propagator to physical Fock states.  First, the semiclassical regime in Fock space is {\it not} defined by $\hbar \to 0$.  Here, as in many other important situations, the fundamental nature of the quadrature operators as formally, but not physically, the MB version of the canonical position and momentum operators in particle systems plays a key role. In contrast to the first quantized approach, inspection of the action functional reveals that in Fock space, $\hbar$ is simply a constant that transforms energies like $H$ into frequencies, but does not play the fundamental role of defining the small parameter upon which the asymptotic analysis is built on. 

In order to identify a suitable asymptotic parameter, focus on the action of the quadrature propagator on the physical Fock states.  Consider first the exact relation
\begin{equation}
\label{eq:transfo}
    K({\bf n}^{(f)},{\bf n}^{(i)},t)=\int d{\bf q}^{(f)}d{\bf q}^{(i)}\langle {\bf n}^{(f)}|{\bf q}^{(f)}\rangle K({\bf q}^{(f)},{\bf q}^{(i)},t)\langle {\bf q}^{(i)}|{\bf n}^{(i)}\rangle
\end{equation}
where the transformation overlaps $\langle q|n\rangle$ are formally derived from the algebraic properties of quadrature states as in~\cite{Engl16}.  This gives 
\begin{equation}
\label{eq:CT}
   \langle q|n\rangle=\frac{{\rm e}^{-q^{2}/2}}{\pi^{1/4}\sqrt{2^{n}n!}} H_{n}(q)\, .
\end{equation}
The transformation kernel in Eq.~(\ref{eq:CT}) above is essentially the same as the corresponding results for the familiar harmonic oscillator in terms of the Hermite polynomials $H_{n}(x)$. It is worth mentioning, however, that in the present second-quantized framework all operators involved in the change of basis, namely quadratures and number operators, are strictly dimensionless. This is in contrast to a first-quantized framework where a typical classical action,  $S_{\rm typ}$, has dimensions and enters the definition of a unitless $\hbar_{\rm eff}=\hbar/S_{\rm typ}$.  Thus, the identification of an $\hbar_{\rm eff}$ for the MB case does not involve a classical action scale. 

Armed with Eq.~(\ref{eq:CT}), one can obtain the propagator between Fock states based on the path integral in quadrature representation.  The semiclassical analysis, and the identification of a proper asymptotic regime, begins with the analysis of this kernel in the limit of large occupations $n \gg 1$. In this case, a well known asymptotic form \cite{Gradshteyn00}
\begin{equation}
   \langle q|n\rangle \simeq A(q,n)\cos{(F(q,n)+\pi/4)} 
\end{equation}
holds, where $A(q,n)$ is a smooth prefactor and following \cite{Engl16}
\begin{equation}
   F(q,n)=\int dq\sqrt{2n+1 - q^{2}}
\end{equation}
is naturally identified as the generating function of the classical canonical transformation between the canonical pairs $(q,p)$ and $(n,\theta)$ with
\begin{equation}
    q+ip=\sqrt{2n}{\rm e}^{i\theta}.
\end{equation}
Using this generating function, the phase space variables $(q,p)$ labelling  quadrature eigenstates on each orbital are maximal for $q^{2}+p^{2}=2n$ if the quadrature propagator is applied to Fock states with large occupation numbers. Thus,  acting on a Fock state with large occupations $n_{i} \gg 1$, the overlap between quadrature and Fock states suggests the scaling $q_{i}\propto \sqrt{n_{i}},p_{i} \propto \sqrt{n_{i}}$. 

For many systems of interest, the Hamiltonian has an additional conservation property already easily seen in the general Hamiltonian of Eq.~(\ref{eq:Ham}), namely the operator representing the total number of particles in the system
\begin{equation}
    \hat{N}=\sum_{i}\hat{n}_{i}
\end{equation}
is conserved, a constraint that is fundamental in the case of massive bosons~\cite{Bartlett07}.  A consequence of this symmetry is that the Fock state propagator is different from zero if and only if
\begin{equation}
    N^{(f)}=N^{(i)}=N
\end{equation}
and therefore all possible dynamical processes and amplitudes are restricted to the subspace of ${\cal F}$ fixed by the particular eigenvalue $N$ of $\hat{N}$. This observable appears as a real numerical constant parameterizing the propagator.  Furthermore, simple combinatorial arguments imply that for asymptotically large $N$ and fixed number of orbitals $d$, the vast majority  of Fock states satisfy $n_{i} \simeq N/d$. In essence, as long as Fock states with occupations bounded away from zero are considered, the regime of validity of the scaling with the large occupations is simply given by setting a large enough total number $N$ of particles with the quadrature variables scaling as
\begin{equation}
    (q_{i},p_{i})=\sqrt{N}(Q_{i},P_{i}).
\end{equation}

The effect of this scaling with the total number of particles on the quadrature propagator is reduced, except for considerations about the functional measure that can be accounted for by a convenient regularization, to its effect on the action functional as
\begin{equation}
  R[\sqrt{N}{\bf Q}(s),\sqrt{N}{\bf P}(s)]=\int ds \left[N{\bf P}(s) \cdot \dot{{\bf Q}}(s)-H_{\rm cl}(\sqrt{N}{\bf Q}(s),\sqrt{N}{\bf P}(s))\hbar\right].  
\end{equation}
Given the specific homogeneity properties of the second-quantized Hamiltonian of Eq.~(\ref{eq:Ham}), 
\begin{equation}
  H_{\rm cl}(\sqrt{N}{\bf Q}(s),\sqrt{N}{\bf P}(s))=NH_{\rm cl}({\bf Q}(s),{\bf P}(s))  
\end{equation}
as long as the interaction matrix elements are rescaled by 
\begin{equation}
   v=\tilde{v}/N\ ,
\end{equation}
a requirement arising from a very intuitive observation: for large occupations, $N \gg 1$, the interaction term in Eq.~(\ref{eq:Ham}) trivially dominates the dynamics given its natural scaling with $N^{2}$. It would otherwise overwhelm the $N$-dependence of the single particle part. This observation suggests that a meaningful limit is only achieved by 
\begin{equation}
    N \to \infty,\quad v \to 0,\quad vN=\tilde{v}=const\ .
\end{equation}
Thus, 
\begin{equation}
\label{eq:action}
  R[\sqrt{N}{\bf Q}(s),\sqrt{N}{\bf P}(s)]=N\int ds \left[{\bf P}(s) \cdot \dot{{\bf Q}}(s)-{\cal H}({\bf Q}(s),{\bf P}(s))/\hbar\right]  
\end{equation}
where ${\cal H}$ denotes the classical Hamiltonian with rescaled interaction strength and therefore dynamics that are {\it fully independent of $N$}. The form of Eq.~(\ref{eq:action}) gives the formal identification $\hbar_{\rm eff}=1/N$ and $N \to \infty$ as the semiclassical regime of systems with a large number of interacting bosons.   

There is one remaining ingredient in Eq.~(\ref{eq:transfo}) to be written in terms of scaled variables, the part corresponding to the transformation kernels. One readily sees that
\begin{equation}
    F(\sqrt{N}Q,n)=NF(Q,\rho),\quad \rho=n/N\ ,
\end{equation}
thus bringing the quadrature propagator, when projected on initial and final Fock states with large total number of particles, into a linear combination of integrals of the form
\begin{eqnarray}
\label{eq:full}
   K({\bf n}^{(f)},{\bf n}^{(i)},t)&=&\int d{\bf Q}^{(f)}d{\bf Q}^{(i)} \prod_{i}A(Q_{i}^{(f)},\rho_{i}^{(f)}){\rm e}^{iNF(Q_{i}^{(f)},\rho_{i}^{(f)})}A(Q_{i}^{(i)},\rho_{i}^{(i)})  \nonumber \\
   &\times& {\rm e}^{iNF(Q_{i}^{(i)},\rho_{i}^{(i)})}\int_{{\bf Q}(0)={\bf Q}^{(i)}}^ {{\bf Q}(t)={\bf Q}^{(f)}}{\cal D}[{\bf Q}(s),{\bf P}(s)]{\rm e}^{iN{\cal R}[{\bf Q}(s),{\bf P}(s)]}\ .
\end{eqnarray}
The asymptotic limit for $N\to \infty$ naturally emerges since the corresponding action functional ${\cal R}[{\bf Q}(s),{\bf P}(s)]$ and phase functions $F(Q,\rho)$ are {\it independent of $N$}.  Therefore, both the stationary phase condition $\delta {\cal R}=0$ that defines a consistent classical limit when supplemented with the boundary conditions ${\bf Q}(0)={\bf Q}^{(i)},{\bf Q}(t)={\bf Q}^{(f)}$, and the canonical transformation performing the change of phase space coordinates $(Q,P) \to (\rho,\theta)$ can be performed using stationary phase analysis.

Performing explicitly the variations over the $Q,P$ paths we get the corresponding Hamilton's equations
\begin{eqnarray}
    \frac{\delta {\cal R}}{\delta {\bf Q}}=0 &\rightarrow& \hbar\frac{d}{ds}{\bf P}=-\frac{\partial {\cal H}}{\partial {\bf Q}} \nonumber \\ 
    \frac{\delta {\cal R}}{\delta {\bf P}}=0 &\rightarrow& \hbar \frac{d}{ds}{\bf Q}=\frac{\partial {\cal H}}{\partial {\bf P}}
\end{eqnarray}
which, using Eqs.~(\ref{eq:Tom1}) and (\ref{eq:Tom2}), are  recognized as the real and imaginary parts of the mean field equations 
\begin{equation}
\label{eq:MFE}
    i \hbar\frac{d}{ds}\psi_{i}(s)=\frac{\partial}{\partial \psi^{*}_{i}}{\cal H}_{\rm MF}({\bf \psi},{\bf \psi}^{*})
\end{equation}
that now emerge neatly as a classical limit of the quantum field theory, where the mean field Hamiltonian is defined in terms of the classical symbol $H_{\rm cl}$ as 
\begin{equation}
 {\cal H}_{\rm MF}({\bf \psi},{\bf \psi}^{*})= H_{\rm cl}({\bf Q},{\bf P}).
\end{equation}
Finally, the mean field Hamiltonian is a function of the complex classical fields
\begin{equation}
    \psi_{i}=\frac{Q_{i}+iP_{i}}{\sqrt{2}}
\end{equation}
that together with $\psi^{*}$ parameterize the manifold in the phase space of the classical limit with the constraint
\begin{equation}
    \sum_{i}\rho_{i}=1.
\end{equation}

We see then that in our construction, the classical limit is identical to the celebrated mean-field equations well known from the theory of interacting bosonic systems in their discrete (lattice) version. In fact, for a Hamiltonian with the form
\begin{equation}
    \hat{H}=\sum_{i}e_{i}\hat{n}_{i}-J\sum_{i}\left(\hat{b}^{\dagger}_{i}\hat{b}_{i+1}+\hat{b}^{\dagger}_{i+1}\hat{b}_{i}\right)+\frac{U}{2}\sum_{i}\hat{n}_{i}(\hat{n}_{i}-1)
\end{equation}
a continuous limit with suitable redefinitions reads
\begin{equation}
\label{eq:classlim}
   i \hbar \frac{d}{ds}\psi(x,s)=\left(-\frac{\hbar^{2}}{2m}\frac{\partial^{2}}{\partial x^{2}}+V(x)\right)\psi(x,s)+U|\psi(x,s)|^{2}\psi(x,s) 
\end{equation}
which is the familiar Gross-Pitaevskii equation \cite{Negele98} widely used to describe interacting bosonic systems. Now that the  the mean field equations are identified as the underlying theory playing the role of the classical limit of the MB quantum theory in the asymptotic regime $N \gg 1$, the relationship between quantum, semiclassical, and mean-field descriptions becomes transparent. In the same vein as Hamilton's classical equations for particles not being able to describe quantum interference simply because the (phase) space of classical states does not allow for superpositions, the mean-field solutions cannot by themselves contain genuine MB quantum interference. Here, as with the SP case, to understand the connection between multiple mean-field solutions and their interference requires the full machinery of the semiclassical approximation.

The first point where a semiclassical approximation drastically departs from the mean field limit is in its way of treating the information provided by the mean-field equations. Invariably, mean-field methods are based on the propagation of a {\it single} solution of the mean-field equations, fully and uniquely determined by its initial condition $\psi(s=0)$. Quite to the contrary, in this approach the classical limit of the interacting MB quantum theory is {\it not} an initial value problem, but actually a two-point boundary value problem consistently determined by the mean field equations supplemented with the boundary conditions
\begin{equation}
   {\bf Q}(s=0)={\bf Q}^{(i)},{\bf Q}(s=t)={\bf Q}^{(f)} 
\end{equation}
and thus generally admitting multiple solutions, except for the non-interacting case.  Transformations can be made to an initial value representation method, but the multiplicity of solutions remains a key feature. Following the conceptual framework of the semiclassical approximation, the role of these multiple mean-field solutions is clear: they describe genuine MB quantum interference. 

The way MB quantum interference is made explicit naturally comes from the application of the stationary phase approximation to the MB propagator, fully justified now by the emergence of the large parameter $N$. This renders all integrals involved  to be highly oscillatory and allows for following formally Gutzwiller's classic calculation \cite{Gutzwiller07}. In a first stage, the (scaled) quadrature path integral is calculated to obtain
\begin{eqnarray}
\label{eq:semquad}
   \int_{{\bf Q}(0)={\bf Q}^{(i)}}^ {{\bf Q}(t)={\bf Q}^{(f)}}{\cal D}[{\bf Q}(s),{\bf P}(s)]&&{\rm e}^{iNR^{(r)}[{\bf Q}(s),{\bf P}(s)]} \\ &&\simeq \sum_{\gamma({\bf Q}^{(i)},{\bf Q}^{(f)},t)}D_{\gamma}({\bf Q}^{(i)},{\bf Q}^{(f)},t){\rm e}^{iN{\cal R}_{\gamma}({\bf Q}^{(i)},{\bf Q}^{(f)},t)} \nonumber
\end{eqnarray}
as a sum over the {\it mean field} solutions $\gamma$ with given boundary conditions, involving semiclassical amplitudes $D_{\gamma}$ that account for the stability of the solutions and other topological features of the Hamiltonian mean-field flow around them and, most importantly, the actions ${\cal R}_{\gamma}$ along each of them.

The semiclassical propagator in quadrature representation, Eq.~(\ref{eq:semquad}), is strictly speaking just an intermediate step in the construction of its Fock state version. Nevertheless, it is a very useful object with very desirable features. The first is the perfect analogy between the quadrature and coordinate representations in the semiclassics of  MB and SP  systems respectively, and the suggestive possibility of directly importing into the MB context several key ideas and results. Specifically, the derivation of the MB trace formula follows identical steps as in Gutzwiller's derivation as discussed ahead. The second concerns the friendly way coherent states are treated in quadrature representation, allowing for a very tractable way to connect these two useful and natural sets of MB states. Last, but not least, it is the very natural way in which systems with negligible interactions are described by a mean-field that defines a simple {\it linear} problem, a very appealing feature that is lost if physical Fock states are used instead~\cite{Engl18}.

For this last change of representation the integrals are performed over quadratures, Eq.~(\ref{eq:full}), in the stationary phase approximation. Leaving aside lengthy technical details that can be found in~\cite{Engl16,Engl15a}, the final result is~\cite{Engl14b}
\begin{equation}
   K({\bf n}^{(f)},{\bf n}^{(i)},t)\simeq \sum_{\gamma({\bf n}^{(i)},{\bf n}^{(f)},t)}A_{\gamma}({\bf n}^{(i)},{\bf n}^{(f)},t)){\rm e}^{iN{\cal R}_{\gamma}({\bf n}^{(i)},{\bf n}^{(f)},t)} 
   \label{eq:MB-propagator}
\end{equation}
where the sum extends over the solutions $({\bf n}_{\gamma}(s),{\boldsymbol \theta}_{\gamma}(s))$ of the mean-field equations with boundary conditions
\begin{equation}
    |\psi_{i}(s=0)|^{2}=n_{i}^{(i)} \quad , \quad |\psi_{i}(s=t)|^{2}=n_{i}^{(f)}
\end{equation}
and actions ${\cal R}_{\gamma}$. The semiclassical amplitudes are explicitly given by
\begin{equation}
   A_{\gamma}({\bf n}^{(i)},{\bf n}^{(f)},t)=\left[{\rm det}\left| \frac{N}{2\pi}\frac{\partial^{2} {\cal R}_{\gamma}({\bf n}^{(i)},{\bf n}^{(f)};t)}{\partial {\bf n}^{(i)} \partial {\bf n}^{(f)}}\right|\right]^{1/2}{\rm e}^{-i\frac{\pi}{2}\mu_{\gamma}}
\end{equation}
where the index $\mu_\gamma$ is the Maslov index of the trajectory $\gamma$~\cite{OzorioBook}.

The semiclassical approximation of the Fock state propagator is a starting point for the semiclassical analysis of both dynamical and stationary properties of MB quantum systems of interacting bosons.  The propagator~(\ref{eq:MB-propagator}) is not restricted to chaotic dynamics, but also allows, in principle, for investigating the imprint of more complex, {\it e.g.} mixed regular-chaotic, phase space dynamics and the consideration of system-dependent properties unique to individual bosonic MB systems.

Having at hand both the semiclassical propagator and a well defined classical (mean field) limit,  a fundamental conceptual aspect can be addressed, namely, the meaning of MB quantum chaos.  Since the asymptotic analysis automatically provides as a limit a theory with a well defined Hamiltonian structure~\cite{OzorioBook} given by Eq.~(\ref{eq:classlim}), the quantum ramifications of  mean field chaotic dynamics can be rather precisely investigated and interpreted. Therefore, for systems of interacting bosons with large occupations the MB quantum manifestations of MB mean field chaos can be placed on a firm theoretical foundation.

The following passage summarizes the directions that, starting with the semiclassical propagator in Fock space or its variants, have been pursued by several groups during the last years in an attempt to understand the quantum signatures of mean field integrability and chaos.

\subsubsection{Relationship to the truncated Wigner approximation} 
\label{sec:twa}

It is possible to view the way the classical limit plays a role in the quantum mechanical description of a many-body system in three stages. At the most primitive level, expectation values of time-dependent (Heisenberg) observables $\hat{A}(t)=A(\hat{{\bf b}}^{\dagger}(t),\hat{\bf b}(t))$ defined with respect to an initial coherent state $|{\bf z}\rangle=|{\boldsymbol \psi}\rangle$ are obtained from the classical limit simply by transporting the classical symbol along the solution ${\boldsymbol \psi}(t), {\boldsymbol \psi}(0)={\boldsymbol \psi}$ of the mean field equations (\ref{eq:MFE}),
\begin{equation}
    \langle {\boldsymbol \psi}|\hat{A}(t)|{\boldsymbol \psi}\rangle \simeq A({\boldsymbol \psi}(t),{\boldsymbol \psi}^{*}(t)) \ ,
\end{equation}
which defines the strict mean field approximation.

The second stage adds a little more sophistication in which the classical solutions are still used directly to guide the quantum evolution, but zero-point motion underlying the quantum state is incorporated that evolves under the mean field flow,
\begin{eqnarray}
  \langle {\boldsymbol \psi}|\hat{A}(t)|{\boldsymbol \psi}\rangle \simeq \int d{\boldsymbol \Psi}d{\boldsymbol \Psi}^{*} {\rm e}^{-|{\boldsymbol \Psi}-{\boldsymbol \psi}|^{2}} A({\boldsymbol \Psi}(t),{\boldsymbol \Psi}^{*}(t)),
\end{eqnarray}
giving the celebrated truncated Wigner approximation (TWA) \cite{Polkovnikov2011}. The pure mean field approximation is then obtained as a particular case when the classical symbol $A_{\rm cl}$ is smooth and the integral is well approximated by taking ${\boldsymbol \Psi}\simeq {\boldsymbol \psi}$.  Both the mean field and TWA fail to account for coherent effects due to path interference. The former because it is based on a single unique classical solution, and the latter because it is based on adding probabilities instead of amplitudes. In essence, both approximations are {\it classical}. 

The third stage is to incorporate fully the semiclassical approximation.  It accounts for interference effects explicitly and completely by the use of the sum over amplitudes.  In the exact expression
\begin{equation}
    \langle {\boldsymbol \psi}|\hat{A}(t)|{\boldsymbol \psi}\rangle=\sum_{{\bf n},{\bf n}',{\bf m},{\bf m}'}{\boldsymbol \psi}_{{\bf n}}^{*}{\boldsymbol \psi}_{{\bf n}'}A_{{\bf m},{\bf m}'}K({\bf n},{\bf m},t)K^{*}({\bf n}',{\bf m}',t)
\end{equation}
where 
\begin{equation}
   {\boldsymbol \psi}_{{\bf n}}= \langle {\boldsymbol \psi}|{{\bf n}}\rangle {\rm \ \ , and \ \ }A_{{\bf m},{\bf m}'}=\langle {\bf m}|\hat{A}|{\bf m}'\rangle,
\end{equation}
the substitution of $K$ by its semiclassical approximation given in Eq.~(\ref{eq:MB-propagator}) does the trick. The key object to analyze is the product
\begin{equation}
\label{eq:doublesum}
   K({\bf n},{\bf m},t)K^{*}({\bf n}',{\bf m}',t)\simeq \sum_{\gamma,\gamma'}A_{\gamma}A_{\gamma'}^{*}{\rm e}^{iN({\cal R}_{\gamma}-{\cal R}_{\gamma'})} 
\end{equation}
where $\gamma$ labels mean field paths joining ${\bf n}$ with ${\bf m}$ in time $t$, and similarly for $\gamma'$. The TWA is readily obtained (in its polar form where $\psi=\sqrt{n}{\rm e}^{i\theta}$) from the diagonal approximation where the action  is linearized for the $\gamma'=\gamma$ terms~\cite{Engl14}.  Genuine many-body interference arises from the off-diagonal contributions $\gamma' \ne \gamma$. It is then a question to be addressed in a case-to-case basis of how much off-diagonal information, which demands a great effort to evaluate, is necessary to describe the physical phenomena of interest. This may range from the explicit and precise description of every quantum fluctuation around the classical background as done in \cite{Tomsovic18} well beyond the Ehrenfest time, to the selective use of restricted off-diagonal contributions to capture robust interference effects as in \cite{Schlagheck19}.

It is worth noting that the derivation of the TWA here relies on the more fundamental semiclassical approximation for MB amplitudes. As such, it is expected that the foundations of the TWA may suffer from ambiguities in systems where either the classical limit and/or semiclassical regime cannot be precisely defined. Extremely important systems such as spin half chains and Fermi-Hubbard models indeed represent such cases. Progress towards a formal construction of the TWA in these cases has been an active field recently (see \cite{DAVIDSON17} and references therein), with successful applications to SYK models \cite{Schmitt19} and spin chains \cite{Schachenmayer15}.


\subsubsection{A first application: coherent backscattering in Fock space}
\label{sec:cbs}

 The capacity of semiclassical propagators to describe quantum interference comes from the natural way such effects are described within the path integral formalism where the coherent superposition of a multitude of paths carrying different phases produce the interference patterns. In the semiclassical regime the largely uncontrolled proliferation of all possible quantum paths is replaced by a sum over specific mean field paths, making the mechanism of interference much more explicit.  Applied to MB Fock space the discrete sum over paths is now a coherent sum of amplitudes associated with the discrete set of mean-field solutions. Following the success of the single-particle semiclassical approach in describing the leading-order interference term in disordered conductors~\cite{Akkermans07}, the so-called weak localization correction due to interference between pairs of time-reversed paths, the corresponding MB effect arising from the superposition of amplitudes associated with two such corresponding mean-field solutions is given in \cite{Engl14b}.  There it is shown that such coherent backscattering produces a characteristic enhancement of the return probability in MB Fock space. 
 
 \begin{figure}
    \centering
  \includegraphics[width=0.8\linewidth]{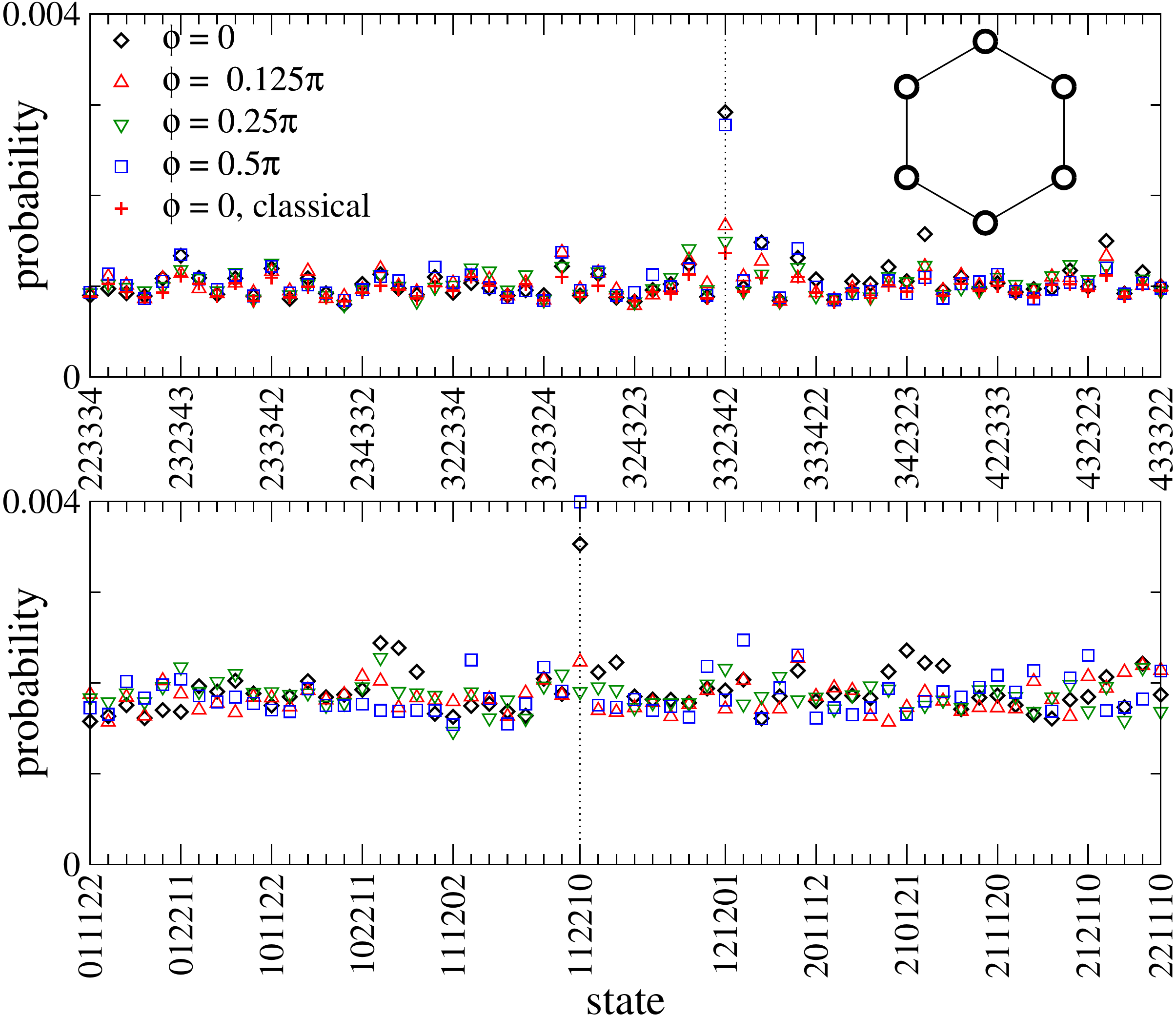}
    \caption{ \label{Fig:CBSI}
   {\bf
   Many-body coherent backscattering in Fock space.
  } Numerical simulation of the transition probability in Fock space for a ring-shaped 6-site Bose-Hubbard ring (upper right inset). The initial state ${\bf n}^{(i)}$ (indicated by the vertical line) is propagated for a time larger than the equilibration time and the probability of finding the system in the final state ${\bf n}^{(f)}$ indicated in the horizontal axis is calculated by exact numerical solution of the quantum dynamics.  The observed enhancement of the transition probability when ${\bf n}^{(f)}={\bf n}^{(i)}$ over the classical uniform background is a purely coherent effect that is suppressed by the application of a gauge field $\phi$ that destroys the time-reversal invariance of the system. (From Ref.~\cite{Engl14b}.)}
\end{figure}

 The numerical confirmation of this coherent MB effect is shown in Fig.~\ref{Fig:CBSI}. Here, the probability $P({\bf n}^{(i)},{\bf n}^{(f)},t)=|K({\bf n}^{(i)},{\bf n}^{(f)},t)|^{2}$ of finding the system in the final Fock state ${\bf n}^{(f)}$ at time $t$ initially prepared in the state ${\bf n}^{(i)}$ is obtained by solving numerically the quantum dynamics of a 6-site Bose-Hubbard ring in the regime of chaotic mean field dynamics. After a relatively short relaxation time scale, the tendency toward  uniform equilibration is clearly visible where all transition probabilities roughly reach the same classical value (also well described by the TWA). The only exception happens for ${\bf n}^{(f)}={\bf n}^{(i)}$ in which a non-classical enhancement is clearly observed. Furthermore, if time-reversal invariance of the system is broken by means of a gauge field parametrized by $\phi$, this enhancement, a hallmark of coherent backscattering due to quantum interference among classical paths related by time reversal, disappears.  
\begin{figure}
    \centering
  \includegraphics[width=0.8\linewidth]{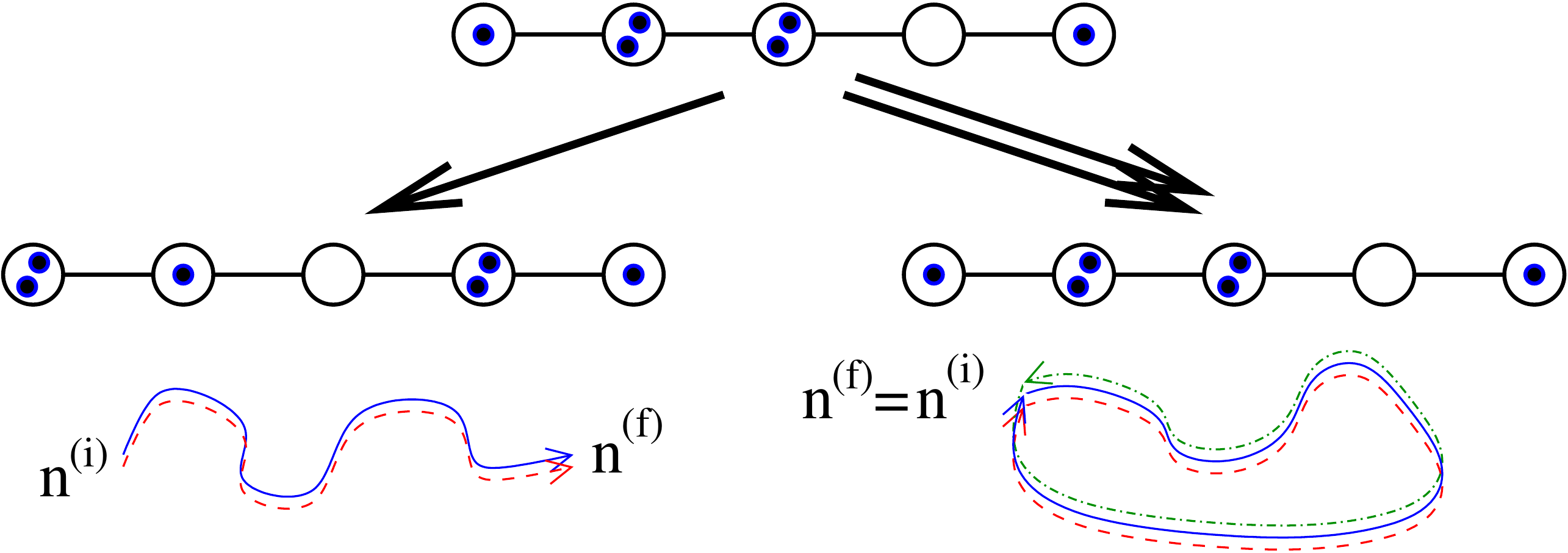}
    \caption{ \label{Fig:CBSII}
   {\bf
   Many-body coherent backscattering in Fock space.
  } The semiclassical propagator used to calculate the transition probability between Fock states $|K({\bf n}^{(i)},{\bf n}^{(f)},t)|^{2}$ naturally leads to the consideration of double sums over mean field equation solutions. Under local averaging only robust, smooth contributions survive. Generically, these smooth contributions simply correspond to interference from pairs of identical amplitudes corresponding to the same mean field solutions. For ${\bf n}^{(f)}={\bf n}^{(i)}$, however, {\it two} different solutions related by time-reversal constructively interfere, a purely quantum effect due to coherent superposition of amplitudes.
  (From Ref.~\cite{Engl14b}.)}
\end{figure}
 The semiclassical explanation of this effect starts with the double-sum over mean field solutions in Eq.~(\ref{eq:doublesum}), and the realization that only for ${\bf n}^{(f)}={\bf n}^{(i)}$, there is an off-diagonal contribution from orbits $\gamma,\gamma'$ related by time reversal as depicted in Fig.~\ref{Fig:CBSII}.


\subsection{Spectral properties}
\label{sec:spec-prop}

Beyond its use for calculating dynamical properties of observables, such as coherent backscattering,
the MB van Vleck-Gutzwiller propagator (\ref{eq:MB-propagator}) represents the fundamental building block for a semiclassical theory of spectral properties of MB bosonic systems.
In this vein a calculation of the MB density of states is summarized, which leads to a MB version of Gutzwiller's trace formula following Ref.~\cite{Engl15}. It turns out that
periodic mean field solutions of the nonlinear wave equations play the role of the
classical single-particle periodic orbits in Gutzwiller’s original trace formula.
Based on the MB trace formula RMT-type universal spectral correlations can be addressed through the lens of semiclassical theory as proposed in \cite{Engl15,Dubertrand16}. There post-Ehrenfest phenomena and the encounter calculus in the MB context naturally arises again.


\subsubsection{Many-body Gutzwiller trace formula}
\label{sec:MB-Gutzwiller}

Given the success of Gutzwiller's trace formula, Eq.~(\ref{eq:SP-Gutzwiller}), over the past 50 years, it is quite natural to investigate the corresponding MB extensions. A straightforward generalization consists in  increasing the particle number $N$ for the usual semiclassical limit $\hbar \rightarrow 0$, {\em i.e.} the horizontal crossover in Fig~\ref{fig:sc-limits}. However, for large $N$ this approach would require periodic orbits in a vast $6N$-dimensional phase space, and in addition dealing with the (anti-)symmetrization. Hence in deriving a semiclassical approximation for the MB density it is prefereable to resort to the complementary limit $\heff = 1/N \rightarrow 0$ following \cite{Engl15}.

 Either starting from the quadrature or coherent state representation of the semiclassical propagator, under the assumption of chaotic mean field dynamics, a series of further stationary phase calculations leads eventually to the semiclassical approximation for the MB density of states in the form~\cite{Engl15}
\begin{equation}
    \rho(E,N) \simeq \bar{\rho}(E,N) \ + \ \rho^{\rm osc}(E,N) =
    \bar{\rho}(E,N) \ + \ 
    \sum_{\rm pm}A_{\rm pm}{\rm e}^{iNS_{\rm pm}(E)} \, .
    \label{eq:MB-Gutzwiller}
\end{equation}
The first (Weyl) term is given, to leading order in $\heff$, by the phase space volume of the corresponding mean field energy shell,
\begin{equation}
    \bar{\rho}(E,N)=\left(\frac{N}{2\pi}\right)^{d}\int \ d{\bf q} \ d{\bf p}\ \delta(E-H_{{\rm cl}} ({\bf q},{\bf p}))\ \delta(N-N({\bf q},{\bf p})) \, ,
\end{equation}
where $H_{{\rm cl}}$ is defined in Eq.~(\ref{eq:Tom2}).
The sum in Eq.~(\ref{eq:MB-Gutzwiller}) extends over all {\it periodic} solutions 
\begin{equation}
    ({\bf q},{\bf p})(t=0)=({\bf q},{\bf p})(t=T) 
\end{equation}
(for some period $T$)
of the classical mean field  equations 
\begin{equation}
\label{eq:MBequations}
    \hbar \frac{d{\bf q}}{dt}=\frac{\partial H_{\rm cl}({\bf q},{\bf p})}{\partial {\bf p}}, {\rm \ \ \ } \hbar \frac{d{\bf p}}{dt}=-\frac{\partial H_{\rm cl}({\bf q},{\bf p})}{\partial {\bf q}}
\end{equation}
with fixed energy $E$ and particle number 
\begin{equation}
 N({\bf q},{\bf p})=\frac{1}{2}\sum_{i}(q_{i}^{2}+p_{i}^{2}) \, .
\end{equation}
This implies that unstable periodic mean field modes (pm) in Eq.~(\ref{eq:MB-Gutzwiller}) play the role of the classical periodic orbits in the single-particle context. In close analogy, the modes's actions take the form
\begin{equation}
    S_{\rm pm}(E) = \oint_{\rm pm} {\bf p}({\bf q},E)\cdot d{\bf q}
    \label{eq:MB-action}
\end{equation}
and the amplitudes read, as in Eq.~(\ref{eq:SP-stab}),
\begin{equation}
    A_{\rm pm}(E) = \frac{T_{\rm ppm}(E)}{|{\rm d
    et}({\bf M}_{\rm pm}(E) - {\bf I} |^{1/2}} {\rm e}^{-i\mu_{\rm pm} \frac{\pi}{2}}\ ,
\end{equation}
in terms of the period $T_{\rm ppm}$ of the primitive mean field mode, the monodromy matrix ${\bf M_{\rm pm}}$ depending on the stability of the mode, and its Maslov index $\mu_{\rm pm}$. The resulting semiclassical MB trace formula for discrete quantum fields incorporates genuine MB interference including that required to build up the discreteness of the MB spectrum, which arises from the coherent sum over phases associated with periodic mean field solutions.
This is in close analogy to Gutzwiller's expression for the single-particle case.

Here, a few further remarks are in order (for details see also \cite{Engl15}):
(i) notice that the range of validity in energy extends down to lowest energies because $\heff$ and not $\hbar$ controls the semiclassical limit, and thus Eq.~(\ref{eq:MB-Gutzwiller}) holds true even for MB ground states; (ii) due to the existence of the continuous symmetry related to particle number conservation, symmetry-projected semiclassical densities of states were considered to get an expression for the MB spectral density
within each sector with fixed total particle number;
(iii) by using quadrature states of the field the above derivation does not employ the coherent state representation that requires a complexification of the theory's classical limit \cite{Dubertrand16}; (iv) because in the non-interacting case the quantum problem reduces to a harmonic system, the trace formula is still applicable since the corresponding periodic mean field solutions (of the linear Schrödinger equations) turn out to be isolated in the generic case where the single-particle energies defining their frequencies are incommensurate; (v) the trace formula, Eq.~(\ref{eq:MB-Gutzwiller}), may also shed light on the the fact that MB systems often exhibit incoherent, possibly chaotic, single-particle dynamics, while at the same time show collective motion~\cite{Guhr98,Hammerling10}; and (vi) there is a certain conceptual analogy between the semiclassical MB propagator and the corresponding MB density of states as sums over mean field solutions on the one hand and configuration interaction methods, constructing MB wave functions as linear combinations of Slater determinants, {\em i.e.} fermionic mean field solutions on the other.

The MB trace formula allows, in principle, for computing an approximate MB density of states\footnote{In Ref.~\cite{Tomsovic18} a spectrum of a Bose-Hubbard system (with $N=$40 atoms) was computed with high accuracy, using corresponding MB semiclassical techniques.} for MB energies and particle numbers that are out of reach of usual numerical MB techniques.  Moreover, the close formal relation between the semiclassical single-particle, Eq.~(\ref{eq:SP-Gutzwiller}), and MB, Eq.~(\ref{eq:MB-Gutzwiller}), trace formulas implies that many insights and results known for quantum chaotic single-particle dynamics can be taken over into the MB context as summarized in the following for spectral fluctuations.


\subsubsection{Many-body encounters and universal spectral correlations}
\label{sec:spec-stat}

In Sec.~\ref{sec:SP-universality} the semiclassical foundations of RMT-type spectral universality are outlined for chaotic single-particle dynamics, reflected in the BGS conjecture~\cite{Bohigas84}.  
Although it might be evident to consider that this reasoning simply carries over to the quantum chaotic MB case, a formal justification has been missing. 
Also it was not straightforward how the encounter calculus would be generalized to the MB case.

For the usual semiclassical limit $\hbar \rightarrow 0$, the encounter formalism has been shown to be applicable and to lead to RMT results for any phase space dimensionality~\cite{Turek05}, and hence also to the $6N$ dimensions of an $N$-particle system in 3 spatial dimensions. However, MB generalizations require some care. For instance, for a non-interacting MB system with chaotic single-particle dynamics, {\em e.g.} $N$ non-interacting particles in a billiard, the MB density of states is composed of independent single-particle spectra -- with conserved single-particle energies as associated constants of motion -- and thus do not obey RMT-statistics. The spectral statistics are Poissonian in the infinite dimensional limit; for recent work showing rich spectral features due to finite dimensionalities see \cite{Liao20}.  Correspondingly, in the complementary limit $\heff \rightarrow 0$ the non-interacting case also features non-generic spectral fluctuations that do not correspond to the expected Poissonian spectra of integrable systems. This is a
consequence of the field theoretical context where
the free field corresponds to a peculiar linear system that is non-generic since it is not merely integrable.  There, treating the quasi-integrable case due to the effect of a small interaction within a semiclassical perturbation theory may provide a useful approach.

Consider strongly interacting MB systems with an associated chaotic mean field dynamics characterized by a largest MB Lyapunov exponent $\lambda$.
For single-particle dynamics universal spectral correlations arise from interference between periodic orbits with quasi-degenerate actions and periods beyond the Ehrenfest time $\tE^{\rm (sp)} = (1/\lsp) \log (S/\hbar)$, see Eq.~(\ref{eq:sp-Ehrenfest}).
For quantum chaotic large-$N$ MB systems in the limit $\heff \rightarrow 0$, correspondingly genuine MB quantum interference is governed by another log-time scale, the Ehrenfest time 
\begin{equation}
    \tE = \frac{1}{\lambda} \ \log N \, , 
    \label{eq:scrambling}
\end{equation}
also referred to as the scrambling time in the MB context~\cite{Swingle16}.

This very close formal analogy between the single-particle $\hbar\rightarrow 0$ and the MB $\heff \rightarrow 0$ regimes -- based on corresponding trace formulas and hierachy of times scales -- allows for the straightforward generalization of the bosonic MB spectral form factor semiclassical calculation~\cite{Engl15, Dubertrand16} by applying the encounter calculus summarized in Sec.~\ref{sec:SP-universality}. This amounts to replacing $\hbar$ by $\heff$, the Lyapunov exponent $\lsp$ by $\lambda$, the Ehrenfest time 
$ \tE^{(sp)}$ by $\tE$, Eq.~(\ref{eq:scrambling}), single-particle phase space by the $2L$-dimensional phase space of the lattice, and the single-particle density of states $\rho_{\rm sp}^{\rm osc}(E)$ by
$\rho^{\rm osc}(E;N)$, Eq.~(\ref{eq:MB-Gutzwiller}).
Encounters between different (periodic) mean field modes take on the role of encounters between classical (periodic) orbits. This implies the interpretation that interfering periodic mean-field solutions of Eqs.~(\ref{eq:MBequations}) with quasi-degenerate actions $S_{\rm pm} (E)$, Eq.~(\ref{eq:MB-action}),
lead to the emergence of universal MB spectral fluctuations, in close correspondence with the reasoning in the single-particle case. This includes in particular the same RMT-type expressions for the spectral MB two-point correlator $R(\Delta E;N) \sim \langle \rho^{\rm osc}(E;N) \rho^{\rm osc}(E+\Delta E;N)\rangle$ and its associated spectral form factor\footnote{Note that $R(\Delta E;N)$ contains interesting new
parametric correlations with regard to (changes in)
particle number and or interaction strength.}.

These conclusions drawn from the semiclassical MB encounter formalism coincide both with long known results from nuclear shell models and data~\cite{Brody81, Bohigas83}, embedded Gaussian ensembles~\cite{KotaBook}, including those restricted to two-body interactions, as well as with recent results showing random-matrix behaviour of the spectral form factor for a periodically kicked Ising spin-1/2 model without a semiclassical limit (to leading orders in $t/t_H$ \cite{Bertini18,Kos18}) and for a Floquet spin model~\cite{Chan18}. Moreover, Wigner-Dyson-type level statistics have recently been numerically shown with appropriate parameters for a discrete Bose-Hubbard system~\cite{Kolovsky04, Dubertrand16} and the SYK-model~\cite{Garcia17} in the large $N$-limit.  They help close an important missing theoretical link in the sense that the k-body restricted interaction ensembles, i.e.~embedded random matrix ensembles, have resisted analytic proofs of their asymptotic behaviors~\cite{Benet01, Asaga01, Srednicki02} unlike the original classical random matrix ensembles.  Semiclassical results for spectral determinants~\cite{Keating91, Heusler07, Keating07, Mueller09, Waltner09, Waltner19} in terms of pseudo-orbits carry over to the MB case, i.e.~the semiclassical finding \cite{Waltner19} that pseudo- mean field paths with durations $t > t_H$ necessarily must involve multiple partial traversals that do not contribute to the MB spectrum.

It is worth, highlighting again the relevance of the scrambling time scale $\tE$, Eq.~(\ref{eq:scrambling}), and the role of encounters for entangling mean field modes: The semiclassical MB propagator and the trace formula, as sums over mean field paths, contain genuine MB interference thereby giving rise to MB correlations. The encounter calculus is involving ergodic averages distilling out of all these paths, otherwise mostly random interference terms, that prevail for an observable after energy or spatial averages. Each encounter diagram, such as those shown in Fig.~\ref{fig:PO-geometries}, represents all interferences resulting from certain types of coupled mean field trajectory configurations with quasi-degenerate actions. If we think of entanglement as coupling between different product states, as mean field solutions are, then encounters generate entanglement that resists (energy) averaging. After starting to propagate a wave packet along a separable initial (periodic) mean field path, it will split at an encounter, acting like a rail switch and entangling the mean field paths that come close to each other in the encounter region in Fock phase space. The time scale of this entanglement process is given by $\tE$.  Whereas the relevance of encounters for entanglement arises naturally, developing tools to measure the degree of entanglement created through encounter structures remains open. 
This would also distingish encounter-mediated entanglememt growth from some sort of entanglement captured through TWA approaches introduced in Sec.~\ref{sec:twa}.  Quantum unitaries acting as interconnects in quantum networks can be viewed as mimicking certain encounter structures of a quantum chaotic dynamical system. For such random unitary dynamics entanglement growth has been measured~\cite{Nahum17}.

To conclude, MB semiclassical methods developed for systems with underlying chaotic dynamics can provide a direct theoretical derivation of universality in the spectral statistics of large-$N$ MB systems, and the applicability of RMT more generally.  In the regime of exponentially unstable mean field chaos in the classical limit, the local fluctuations of MB spectra comply with RMT predictions.  Though beyond the objective of this contribution, keep in mind that MB semiclassical methods accomplish more than this.  They apply to individual systems and system specific non-statistical quantities as well. In particular, if some system displayed unexpected properties from an RMT perspective, it would still be expected that the semiclassical theory would capture that behavior.  This is especially true of system specific MB dynamics before, up to, and just beyond the Ehrenfest time scale, a time scale that has collapsed to zero in RMT.


\subsection{Out-of-time-order-correlators and commutators}
\label{sec:OTOC}


\subsubsection{Concept and heuristic semiclassical reasoning}
\label{sec:concept}

\begin{figure}
    \centering
  \includegraphics[width=0.8\linewidth]{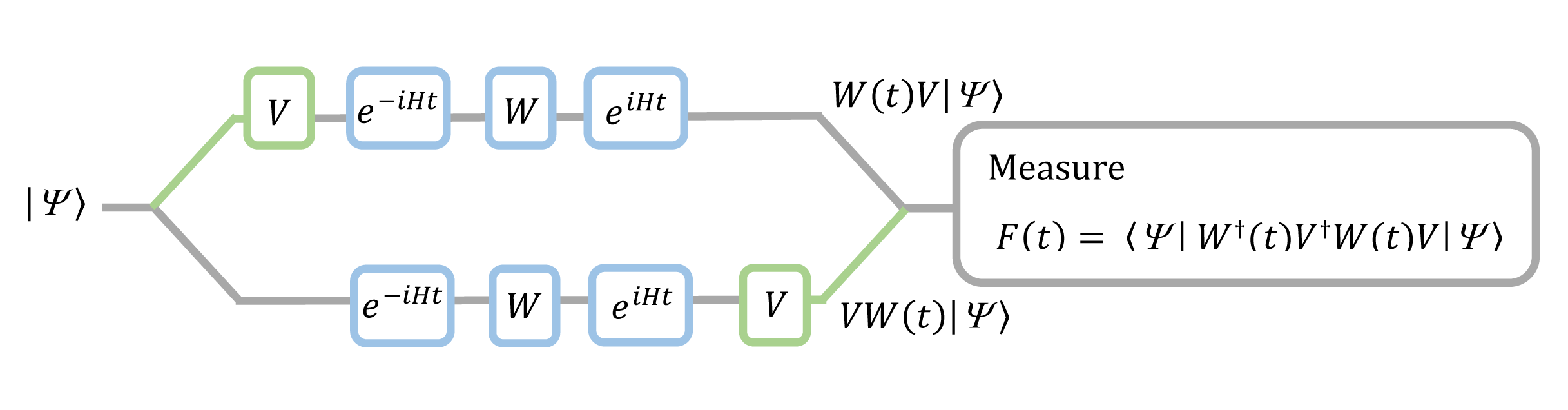}
    \caption{\label{fig:OTOCorr}
   {\bf
   Scheme of an out-of-time-order correlator} | $F(t)$ can be viewed as the overlap between two MB states arising from the operation of $V(0)$ and $W(t)$ on $|\Psi\rangle$ in different time order, {\em i.e.},  $F(t)$ contains four non-time ordered operators.
  }
\end{figure}

Based on correlations between periodic mean field modes and by invoking ergodicity and classical hyperbolicity, it was just discussed how spectral RMT-type universality can be semiclassically explained in the MB context. For the GOE spectral form factor beyond Ehrenfest timescales, terms based on MB interference and organized through a hierarchy of encounters of mean field modes, provide higher-order $\heff$-corrections to Berry's diagonal contribution.
This section presents another prime example for ergodic MB interference, so-called out-of-time-order correlators~\cite{Larkin69,Shenker14,Maldacena16} 
\begin{equation}
  F(t)= \langle \Psi | 
   \hat{W}^\dagger(t) \ \hat{V}^\dagger(0)
   \hat{W}(t) \ \hat{V}(0)  | \Psi \rangle \, ;
  \label{eq:OTOCorr_definition}
\end{equation}
see Fig.~\ref{fig:OTOCorr}, and their closely related relatives, out-of-time-order commutators (OTOCs)~\cite{Maldacena16}
\begin{equation}
  C(t) = 2 - {\rm Im}(F(t)) = 
\langle \Psi | 
  \left[ \hat{W}(t), \hat{V}(0)  \right]^\dagger
  \left[ \hat{W}(t),\hat{V}(0)  \right] | \Psi \rangle \, .
  \label{eq:OTOComm_definition}
\end{equation}
Both, $F(t)$ and $C(t)$ comprise two forward and two backward propagations by means of the Heisenberg operator $\hat{W}(t) = \exp{(-i\hat{H}t/\hbar)} \hat{W}(0)  \exp{(i\hat{H}t/\hbar)}$.

Consider a Bose-Hubbard system in which the local measurement of an atom at a given site perturbs (locally) the MB system. The squared commutator $C(t)$ of such a suitable (local) operator $\hat{W}(t)$
with another (local) operator $\hat{V}(0)$ measures the temporal growth of $\hat{W}$, including its increasing complexity at another site at a certain distance. Hence the initial local (quantum) information is spread and scrambled across the huge Hilbert space of the interacting MB system with its vast number of degrees of freedom~\cite{Sekino08}, making OTOCs the measures of choice for quantifying growing complexity and instability of MB systems, thereby with relevance for quantum computing \cite{Mi21}.  Although OTOCs require rewinding time and their implementation is experimentally challenging, already several measurement protocols exist~\cite{Zhu16,Swingle16,Li16,Garttner16,Dominguez21}. For a recent comprehensive tutorial on OTOCs; see \cite{Xu22}).


OTOCs represent one of the most direct quantum analogues
of classical measures for hyperbolic instability of chaotic dynamics.  For the single particle case invoking a heuristic classical-to-quantum correspondence for small $\hbar$ and replacing the commutator (\ref{eq:OTOComm_definition})
for pre-Ehrenfest times by Poisson brackets $\{\cdot,\cdot\}$ generates a leading-order Moyal expansion, e.g.~$\hat{W}=\hat{p}_i$ and $\hat{V}=\hat{q}_j$
\cite{Larkin69,Maldacena16},
\begin{equation}
 C(t) \longrightarrow
  |\rmi \hbar|^2 \left\{p_i,q_j(t) \right\}^{\!2}
  \! \simeq \! \hbar^2 \! 
  \left(\frac{\partial q_j(t)}{\partial q_i}\!
  \right)^{\!2} \propto
   \hbar^2 \rme^{2\lambda t} \, .
  \label{eq:OTOC_Moyal}
\end{equation}
The leading off-diagonal monodromy matrix element $ \partial q_j(t) / \partial q_i$
is replaced by an exponential growth determined by the classical single-particle Lyapunov exponent $\lambda_{SP}$.  This close quantum-to-classical correspondence for quantum chaotic single-particle dynamics is intriguing, and thus there is also the quest for establishing a corresponding MB quantum-to-classical correspondence, {\em i.e.}, a MB version of such a quantum butterfly effect. This, in particular, includes an interpretation of the quantum mechanical growth rate of OTOCs for MB systems with a classical limit.  Most notably their growth is bounded in time and OTOCs saturate due to a MB interference mechanism setting in at the Ehrenfest time, i.e.~scrambling time. It gives rise to an interference term that is of the same order as the corresponding diagonal contribution \cite{Rammensee18}.  Such distinct features at $\tE$ render OTOC evolution a hallmark of Ehrenfest phenomena.
%


\subsubsection{Pre- and post-Ehrenfest times: exponential growth and saturation}
\label{sec:prepost}

To derive the OTOC pre-Ehrenfest growth and to illustrate these genuine MB interferences consider again Bose-Hubbard systems describing $N$ interacting bosons
with creation (annihilation) operators $\cop{}$ ($\anop{}$) at sites $i\!=\!1,\ldots,L$.
Evaluate the OTOC (\ref{eq:OTOComm_definition}) for the position and momentum
quadrature operators, Eq.~(\ref{eq:quad1}),
\begin{equation}
    \hat{q}_i =
    (\anop{i}\!+\! \cop{i})/\sqrt{2N} \quad , \quad
   \hat{p}_i = (\anop{i}\!-\!  \cop{i})/ (\sqrt{2N}\rmi ) \, , 
   \end{equation} 
which are
related to the occupation operators $\hat{n}_i$ through
$(\hat{q}_i^2 \!+\!\hat{p}^2_i)/2 = \heff(\hat{n}_i \!+\! 1/2)$.
The OTOC, Eq.~(\ref{eq:OTOComm_definition}), reads
\begin{equation}
  C(t)\!=\!\Braket{\Psi| \!
  \left[ \hat{p}_i,\hat{U}^\dagger(t)\hat{q}_j\hat{U}(t)\right] \!
  \left[ \hat{U}^\dagger(t)\hat{q}_j\hat{U}(t),\hat{p}_i\right] \!
  |\Psi }
  \label{eq:pq_OTOC_w_Ut}
\end{equation}
in terms of the MB time evolution operator
$\hat{U}(t)=\exp(-\rmi \hat{H} t / \hbar)$.
In Eq.~(\ref{eq:pq_OTOC_w_Ut}) consider an initial coherent state $\Ket{\Psi}$ localized in both quadratures.

The semiclassical derivation is based on approximating $\hat{U}(t)$ by its asymptotic form for
small $\heff$, the MB version~\cite{Engl14b,Engl15}, Eq.~(\ref {eq:MB-propagator}), of the van Vleck-Gutzwiller propagator.  The corresponding sum runs over all mean-field solutions $\gamma$ of the
equations of motion
$\rmi \hbar \partial \Phi/\partial t = \partial H_{\mathrm{cl}}/\partial \Phi^\ast$
of the classical Hamilton function~(\ref{eq:Tom2})
that denotes the mean-field limit of $\hat{H}$ 
for $\heff=1/N\rightarrow 0$:
 \begin{equation}
 \label{eq:BHham}
        H_{\mathrm{cl}}
        \left(\vec{q},\vec{p}\right)
        = \sum_{i, j}h_{ij}  \Phi_i^*\Phi_j
        +\sum_{i, j, i', j'}V_{i j i' j'}
        \Phi_i^*\Phi_{i'} \Phi_j^*\Phi_{j'} \, .
       \end{equation}
In the coherent sum over mean-field paths in Eq.~(\ref{eq:semquad}) the phases are given by classical actions
$R_\gamma(\vecfin{q},\vecinit{q};t) \! = \!
  \int_0^t \rmd t
    [
       \vec{p}_\gamma(t)\cdot\vec{q}_\gamma(t)
      -H^{\mathrm{cl}}
      \left(\vec{q}_\gamma(t),\vec{p}_\gamma(t)\right)/\hbar
    ]
$
along $\gamma$, and the weights $A_\gamma$ reflect their classical (in)stability.

In order to make a connection to RMT-type universality assume that the mean-field limit exhibits uniformly hyperbolic, chaotic dynamics with the same Lyapunov exponent $\lambda$ at any phase space point.
Evaluating Eq.~(\ref{eq:pq_OTOC_w_Ut}) in position quadrature representation, inserting unit operators, 
and using Eq.~(\ref{eq:semquad}) for the propagator $K$ gives a general semiclassical MB representation of the OTOC.  To leading order in $\hbar_{\rm eff}$, the derivatives $\hat{p}_i = -\rmi \heff \partial /  \partial q_i$ only act on the phases $R_\gamma$ in $K$. Employing the relations
$
 \vecinit{p}_\gamma = -
  \partial R_{\gamma}/\partial \vecinit{q}
$
generates for the OTOC~\cite{Rammensee18}, Eq.~(\ref{eq:pq_OTOC_w_Ut}), 
\begin{eqnarray}
    C(t) \simeq & 
    \!\! \int\! \rmd^n q_1'\! \int \! \rmd^n q_2\!
    \int \! \rmd^n q_3'\! \int \! \rmd^n q_4\!
    \int \rmd^n q_5'
    \Psi^{*}\!\left(\vec{q}_1'\right)\Psi\!\left(\vec{q}_5'\right)
     \nonumber \\
    & \quad  \times \!\!\!\!\!
    \sum_{
    \alpha': \vec{q}_1' {\rightarrow}
    \vec{q}_2
    }
    \quad
\sum_{ 
    \alpha : \vec{q}_3' {\rightarrow}\vec{q}_2
    }\!\!\!\!
    A_{\alpha'}^*  A_{\alpha}
    \rme^{(\rmi/\heff)\left(\!-R_{\alpha'}+R_{\alpha}\right)}
    \left(\init{p}_{\alpha',i}\!-\!\init{p}_{\alpha,i} \right)\fin{q}_{2,j}
     \nonumber
     \\
  & \quad \times \!\!\!\!\!
    \sum_{
    \beta': \vec{q}_3' {\rightarrow}\vec{q}_4}
   \quad  \sum_{
    \beta : \vec{q}_5' {\rightarrow}\vec{q}_4}
 \!\!\!\!
    A_{\beta'}^*  A_{\beta}
    \rme^{(\rmi/\heff) \left(\!-R_{\beta'}+R_{\beta}\right)} \!
    \left(\init{p}_{\beta,i}\!-\!\init{p}_{\beta',i} \right)\fin{q}_{4,j} \, .
  \label{eq:OTOC_sc_integral_representation}
\end{eqnarray}
The four time evolution operators
in Eq.~(\ref{eq:pq_OTOC_w_Ut}) have been converted into
fourfold sums over mean-field trajectories of time $t$ linking the various initial and final position quadratures.  In the semiclassical expression, Eq.~(\ref{eq:OTOC_sc_integral_representation}), the operators $\hat{p}_i$ and $\hat{q}_j$ are replaced
by their classical counterparts $\init{p}_{\gamma,i}$ and $\fin{q}_{\gamma,j}$.
The commutators translate into  differences of initial momenta of trajectories
not restricted to start at nearby positions.

\begin{figure}
  \begin{center} 
  \begin{tabular}{ccc}
  \includegraphics[width=0.5\linewidth]{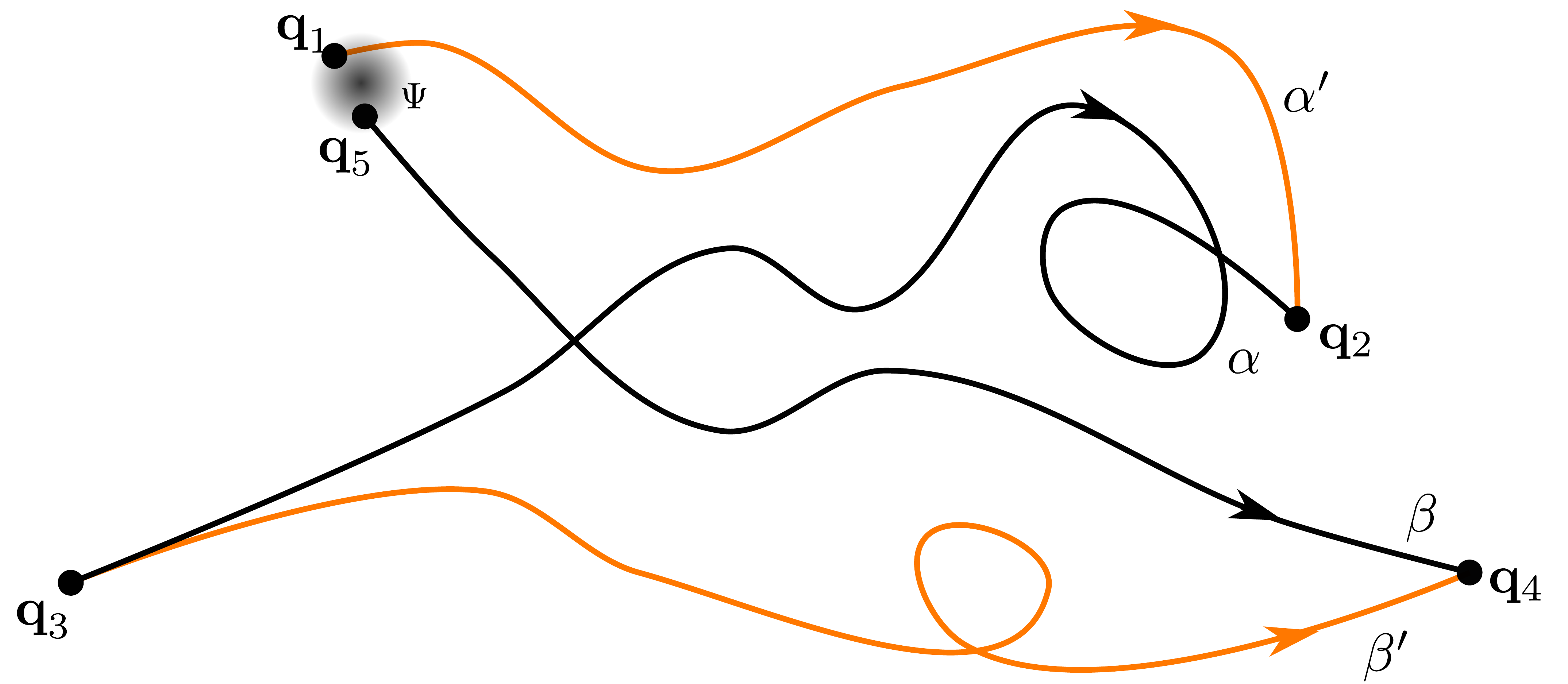}
      & &
  \includegraphics[width=0.4\linewidth]{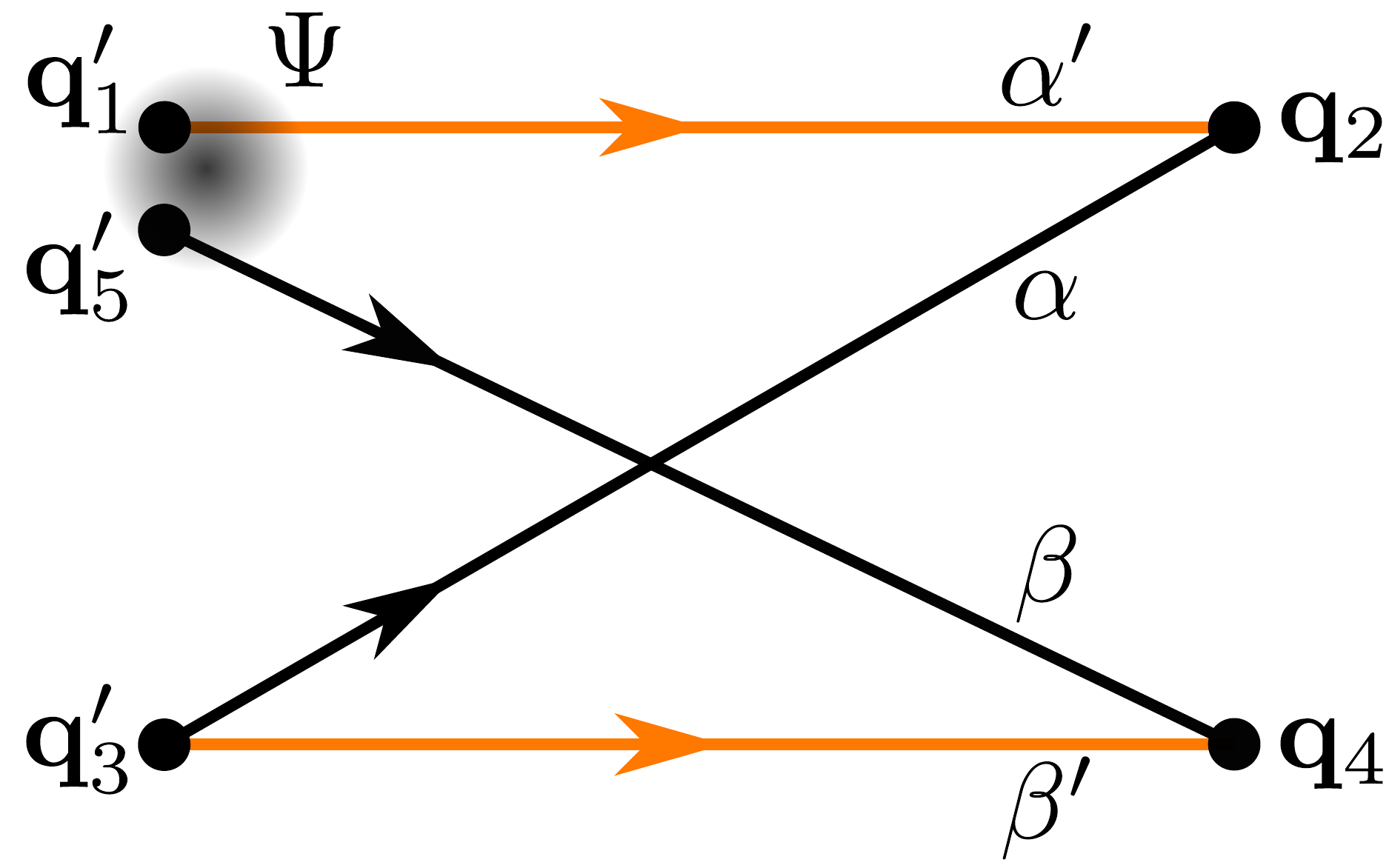} 
     \\ (a)&&(b) \\
 \includegraphics[width=0.4\linewidth]{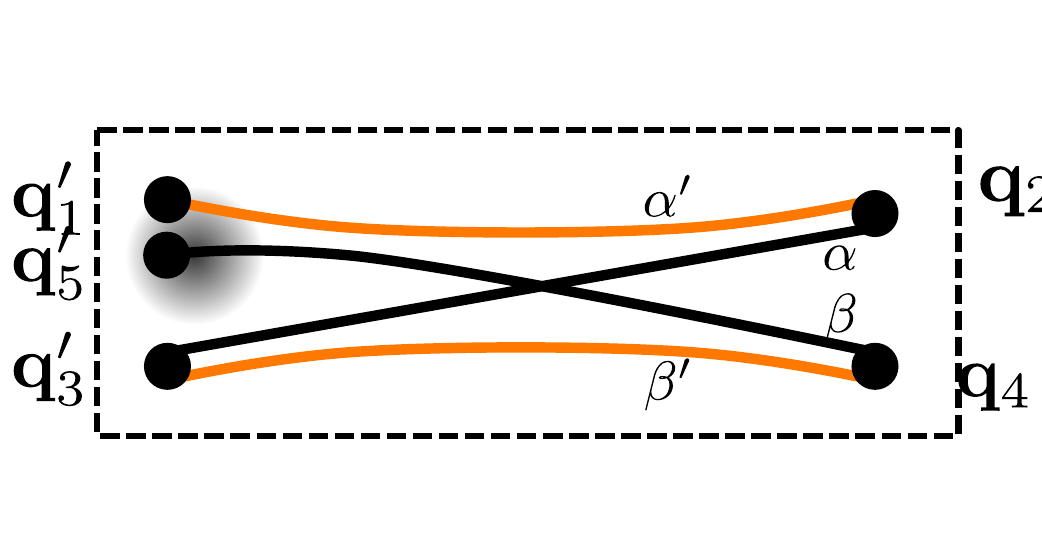}
  & & 
\includegraphics[width=0.4\linewidth]{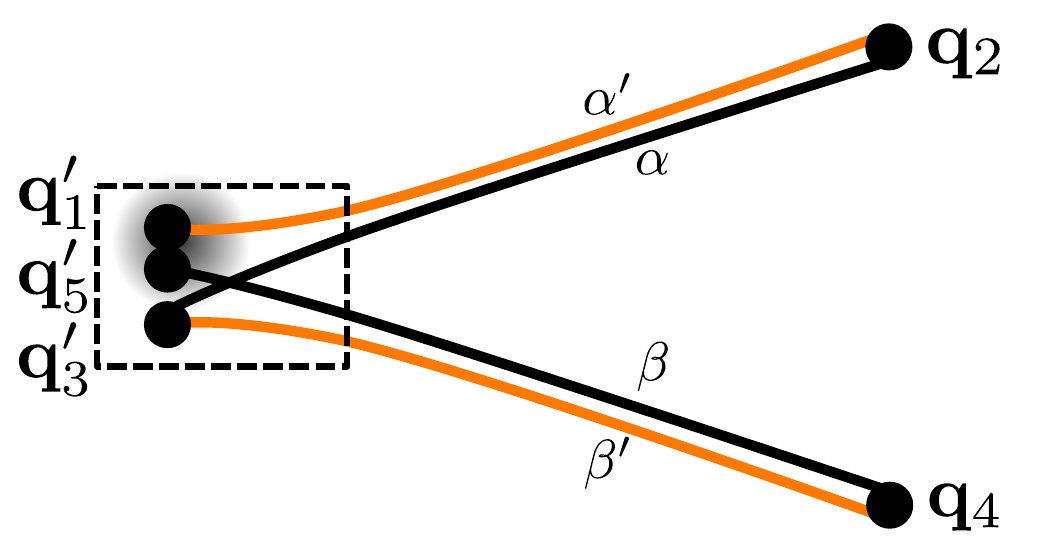}
\\   (c)&&(d) \\
\includegraphics[width=0.4\linewidth]{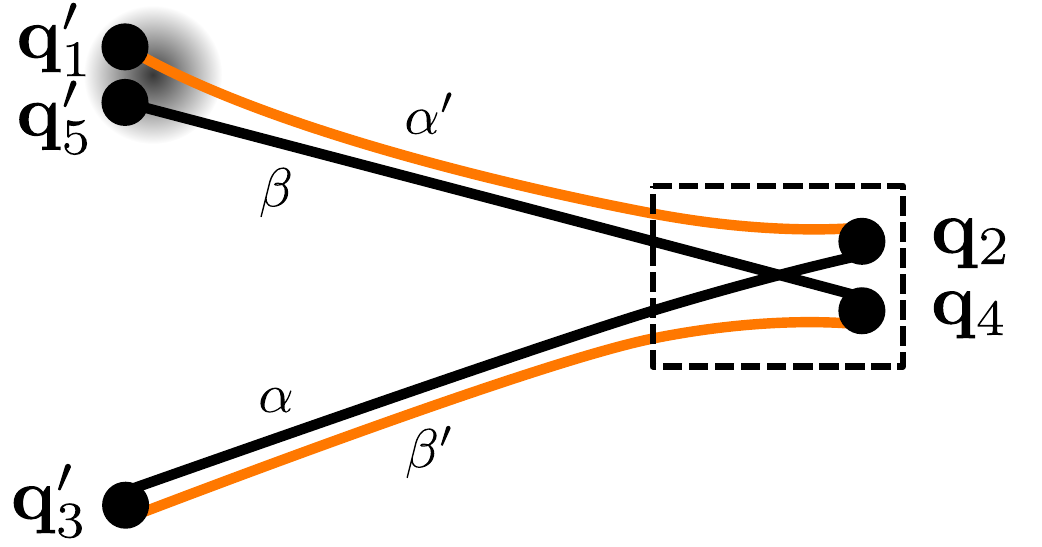}
   &  &
 \includegraphics[width=0.4\linewidth]{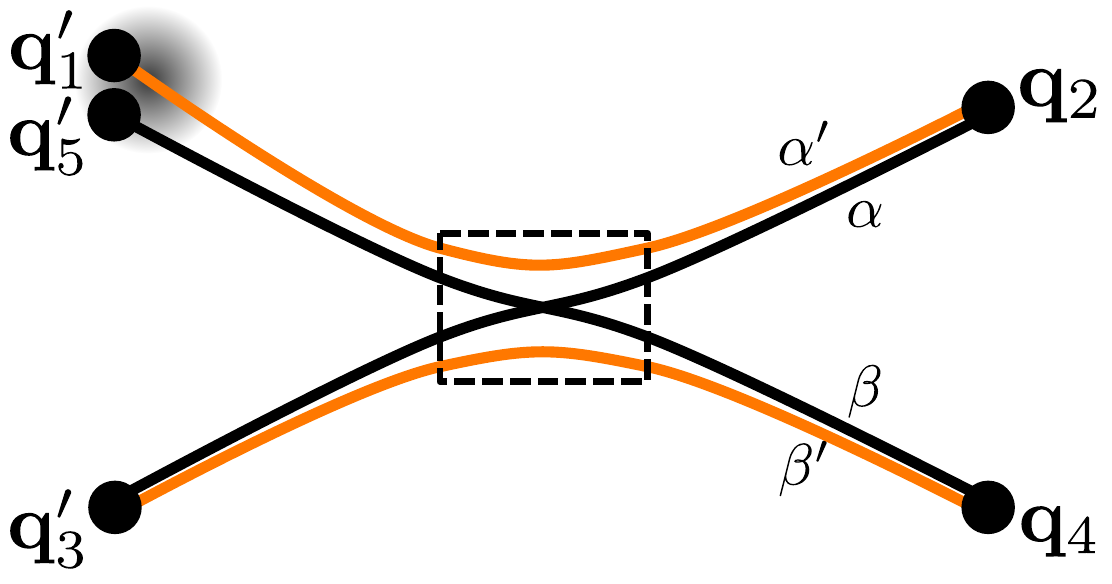}
      \\    (e)&&(f)
     \end{tabular}
  \end{center}
      \caption{
      {\bf Configurations of interfering mean-field paths that contribute to the OTOC $C(t)$,} Eq.~(\ref{eq:OTOC_sc_integral_representation}).
      (a) arbitrary trajectory quadruple and (b) corresponding general diagram denoting
       forward and backward propagation along black and orange mean-field paths. (c)-(f): Relevant configurations contributing predominantly to $C(t)$:
     The trajectory quadruples reside (c) inside an encounter (marked by dashed box), form a "two-leg"-diagram
     with an encounter (d) at the beginning or (e) at the end, or (f) build a "four-leg"-diagram with the encounter
     in between. (From Ref.~\cite{Rammensee18}.)}
     \label{fig:OTOC_diagrams}
\end{figure}

The geometric connections amongst the trajectories quadruples involved are sketched in Fig.~\ref{fig:OTOC_diagrams}.
Panel (a) shows an arbitrary orbit quadruple and (b) the corresponding diagram. Black and orange arrows refer to contributions to $K$ and $K^\ast$, respectively,  i.e.~forward and backward propagation in time. The grey shaded spots mimic the initial state $|\Psi\rangle$.

In the semiclassical limit $R_\gamma(\vecfin{q},\vecinit{q};t) \! \gg \! \heff$. Hence, the corresponding phase factors in Eq.~(\ref{eq:OTOC_sc_integral_representation})
are highly oscillatory with respect to initial and final positions.  Thus, contributions from arbitrary trajectory quadruples are usually suppressed whereas
correlated trajectory quadruples with action differences such that
$R_\alpha\!-\!R_{\alpha'}\!+\!R_{\beta}\!-\!R_{\beta'} \simeq O(\heff)$ are not averaged out and contribute dominantly to $C(t)$.  For post-Ehrenfest times these are quadruples where for most of the propagation the four paths are pairwise nearly identical except for 
encounter regions where trajectory pairs approach each other, evolve in a 
correlated manner, and exchange their partners. The encounter calculus applies in the high-dimensional phase space associated with the MB Fock space.

To leading order in $\heff$, the relevant quadruples for OTOCs involve a single encounter.  These can be subdivided
into four classes depicted in Fig.~\ref{fig:OTOC_diagrams}(c)-(f).
Diagram (c) represents a bundle of four trajectories staying in close vicinity to each other throughout time $t$, i.e.~forming an encounter marked by the dashed box. This diagram turns out to be dominant for times $t<\tE$, Eq.~(\ref{eq:scrambling}), the time scale for traversing an encounter region, if the associated action differences are of order $\heff$.  Due to exponential divergences in chaotic phase space the dynamics merges beyond the encounter boundary into uncorrelated time evolution of the individual trajectories.  However, the symplectic Hamiltonian structure implies that the exponential separation along unstable phase space manifolds is complemented by motion near stable manifolds.  This enables the formation of pairs of exponentially close trajectories~\cite{Sieber01}, e.g.~paths $\alpha'$ and $\alpha$ or $\beta$ and $\beta'$ in Figs.~\ref{fig:OTOC_diagrams}(d,f).
This mechanism becomes quantum mechanically
relevant for times beyond $\tE$; see the discussion in Sec.~\ref{sec:SP-Ehrenfest}. 
Here it is crucial for understanding post-Ehrenfest OTOC saturation.
Panels (c, d) display diagrams with an encounter at the beginning or end of two such trajectory pairs.
The diagrams in (f) are characterized by uncorrelated motion of four trajectory pairs
before and after the encounter.

The evaluation of Eq.~(\ref{eq:OTOC_sc_integral_representation}) requires a thorough consideration of the
dynamics in and outside the encounter regions.
Inside an encounter, Fig.~\ref{fig:OTOC_diagrams}(c),
the hyperbolic dynamics essentially follows a common mean-field solution: linearization in the vicinity of one reference path allows for expressing contributions from the remaining three trajectories.
The detailed evaluation of the diagrams (d-f) in Fig.~\ref{fig:OTOC_diagrams} is given in Ref.~\cite{Rammensee18}.
It involves the calculation of corresponding encounter integrals based on phase space averages invoking ergodicity.
Diagrams similar to class (f) 
have been earlier considered in the context of shot noise ~\cite{Lassl03,Schanz03,Braun06}
and observables based on quantum chaotic single-particle~\cite{Kuipers10} and MB \cite{Urbina16} scattering.
However, the evaluation of such encounter integrals for OTOCs requires a generalization to high-dimensional MB phase spaces. The occurence of operators (positions and momenta) in the case of OTOCs demand a generalization of the encounter calculus and special treatment,
 depending on whether the initial or final position quadratures are inside an encounter.

Using the amplitudes $A_\gamma$ in
 Eq.~(\ref{eq:OTOC_sc_integral_representation})
to convert integrals over final positions into initial momenta, the OTOC contribution from each diagram is conveniently represented as an ergodic phase-space average
\begin{equation}
  C(t) \simeq
  \int \rmd^n q' \int \rmd^n p' W(\vec{q}',\vec{p'})
 I(\vec{q}',\vec{p}';t) \, .
  \label{eq:PS_average}
\end{equation}
Here,
\begin{equation}
W(\vec{q}',\vec{p'}) \!=\! \int \rmd^n y /  (2\pi)^n
  \Psi^*\!\left(\vec{q}'\!+\! \vec{y}/2 \right)
  \Psi\left(\vec{q}'\!-\!\vec{y}/2 \right)
  \exp[(\rmi)\vec{y}\vec{p}']
\end{equation} 
is the Wigner function~\cite{OzorioBook} and $I(\vec{q}',\vec{p}',t)$ comprises all encounter intgerals.  The detailed evaluation of the encounter integrals $I$ represented by the different diagrams in Fig.~\ref{fig:OTOC_diagrams}, and
thereby $C(t)$, yields the following results for pre- and post-Ehrenfest time evolution~\cite{Rammensee18}:
From diagram (c) it follows for $\lambda^{-1} < t < \tE$
upon ergodic averaging in the semiclassical limit
\begin{equation}
    I(\vec{q}',\vec{p'};t) \simeq F_<(t)  \quad ; \quad 
    F_<(t) \approx e^{2\lambda(t - \tE)} \Theta(\tE-t)
    = \heff^2 e^{2\lambda t} \Theta(\tE-t) \, .
\label{eq:F<}
\end{equation}
Diagram (d) turns out to be negligible, diagrams (e, f) together yield for $t > \tE$
\begin{equation}
    I (\vec{q}',\vec{p'};t) \simeq  F_>(t) \Braket{
      \left(p_i'-p_i\right)^2 } ( \Braket{q_j^2} - \Braket{q_j}^2)  \quad ; \quad F_>(t) = \Theta(t-\tE) \, .
\label{eq:F>}
\end{equation}
Here,
\begin{equation}
    \Braket{f} =  \frac{1}{\Sigma(E)} \int\rmd^n q \int \rmd^n p f(\vec{q},\vec{p}) \delta\!\left(E \!-\!
      \mathcal{H}^{\rm cl}\left(\vec{p},\vec{q}\right)
    \right)
\end{equation}
is the ergodic average with $\Sigma(E)$ the phase space volume of the energy shell at energy $E$.


\begin{figure}
\centering{
  \includegraphics[width=0.8\linewidth]{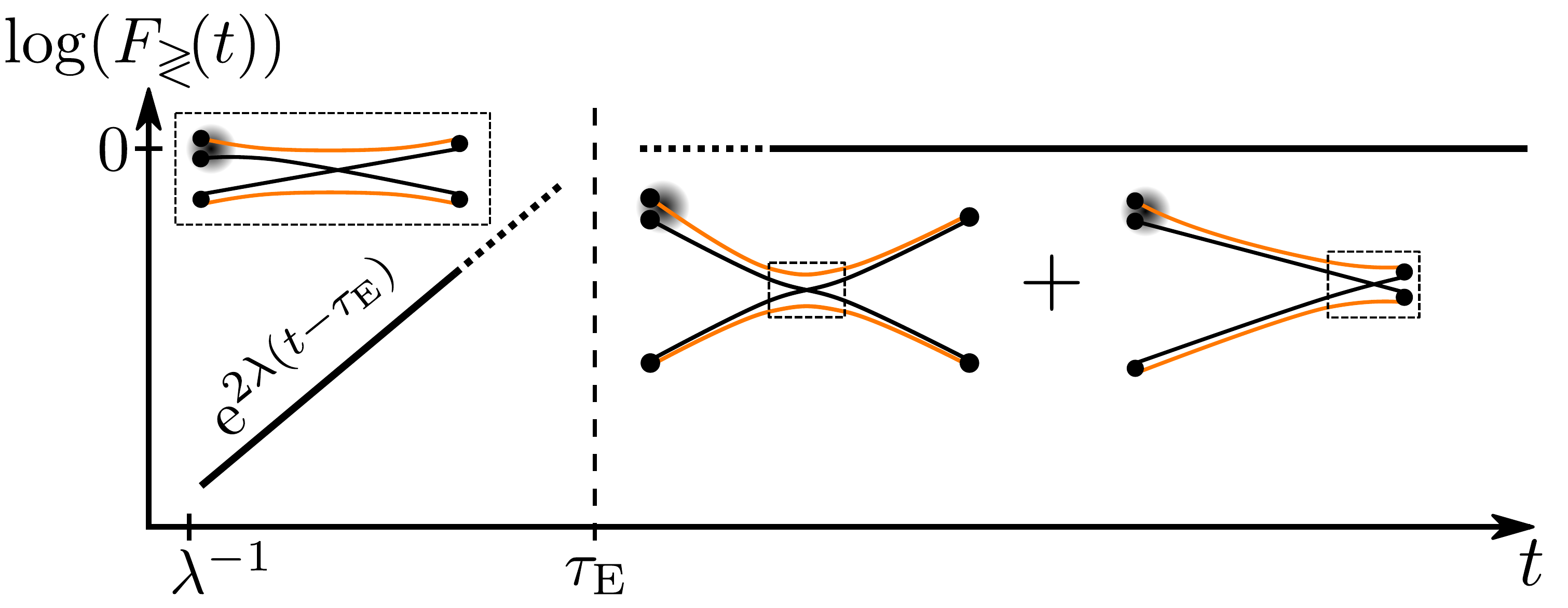}
  }
    \caption{
    \label{fig:OTOCscheme}
 {\bf  Universal contribution to the time evolution of out-of-time-order commutators} | 
Exponential increase according to $F_<(t)$,
Eq.~(\ref{eq:F<}), before and according to $F_>(t)$, Eq.~(\ref{eq:F>}), after the 
 Ehrenfest time $\tE = (1/\lambda) \log N$ marked by the vertical dashed line.
Insets depict diagrams (c), (f) and (e)
from Fig.~\ref{fig:OTOC_diagrams} representing interfering mean-field trajectories.
  (From Ref.~\cite{Rammensee18}.)
  }
\end{figure}

The time-dependences of the universal functions $F_<$ and $F_>$ are sketched in Fig.~\ref{fig:OTOCscheme}.
For $t<\tE$ the  semiclassical evaluation for MB systems confirms the heuristic result, Eq.~(\ref{eq:OTOC_Moyal}). The careful treatment of the encounter dynamics, diagram (c), provides a natural  cut-off (exponential suppression) at $\tE$, absent in Eq.~(\ref{eq:OTOC_Moyal}). It results from the mechanism that the initial phase space area enabling four trajectories to stay close to each other is exponentially shrinking for $t > \tE$. 
The fact that for $t<\tE$ all four mean-field solutions essentially follow in the linearizable vicinity of a common one, see diagramm (c),  indicates that the initial exponential increase of an OTOC of a chaotic MB system can be considered as a property of unstable mean-field dynamics that would also be captured by a truncated Wigner approach. 


\begin{figure}
\centering{
  \includegraphics[width=0.8\linewidth]{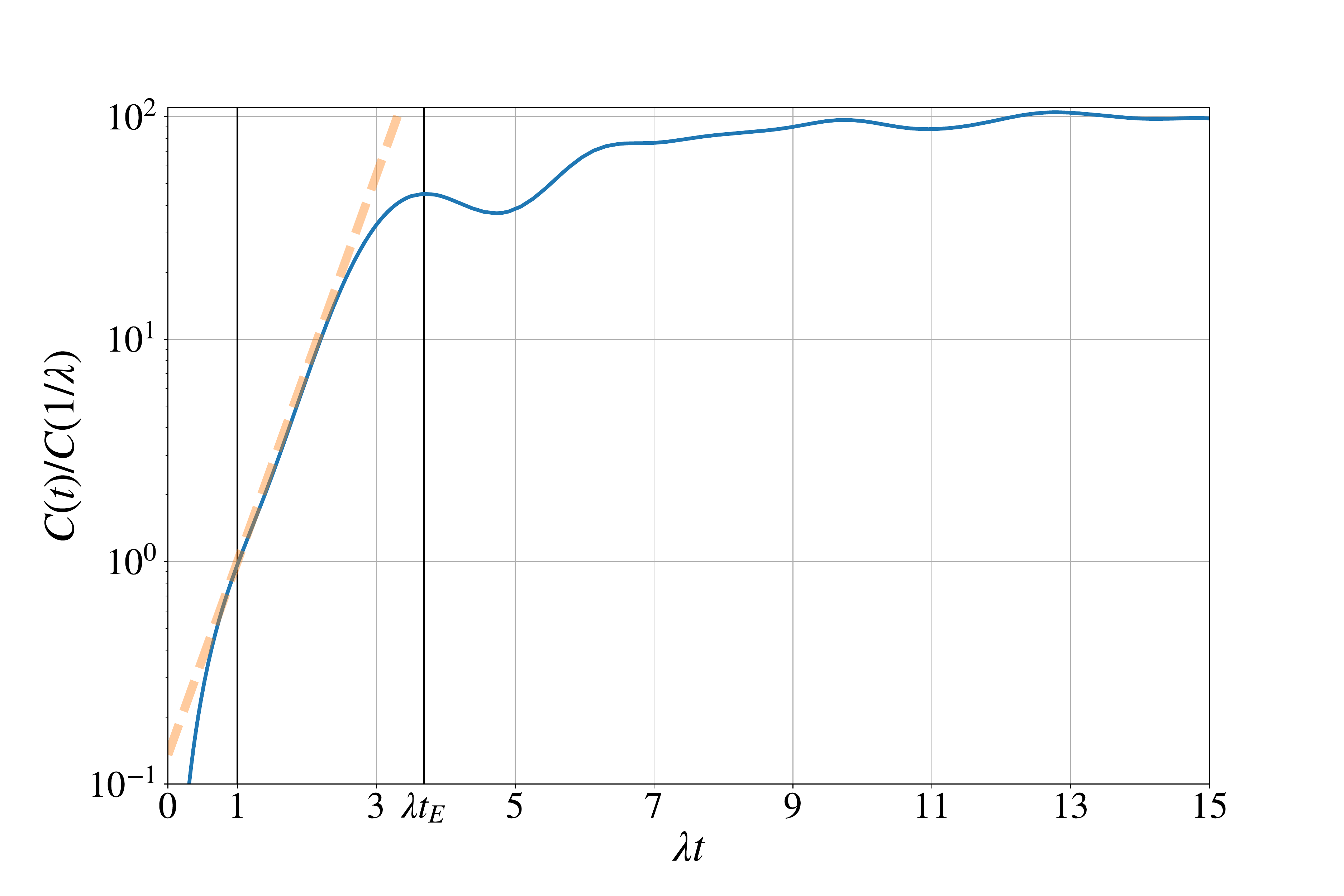}
  }
    \caption{
    \label{fig:OTOCBH}
 {\bf Out-of-time-order commutator of a Bose-Hubbard (BH) system} | Numerically exact calculation of Eq.~(\ref{eq:OTOComm_definition}) for a BH system with four sites and $N=40$ particles. The system is initially described by a coherent state localized near a hyperbolic fixed point of the classical mean-field dynamics. For the choice of parameters $J/NU\simeq \pi/2$ in Eq.~(\ref{eq:BHham}) with hopping $h_{i,j}=J(\delta_{i,j+1}+\delta_{i+1,j})$ and local interactions $V_{i,j,i',j'}=U\delta_{i,j}\delta_{i',j'}\delta_{i,j}$, the corresponding stability exponent is given by $(J/\hbar)\lambda$. For times $1<\lambda t < \log(N)$ (time in units of the typical hopping time between sites $\sim \hbar/J$) a clear exponential growth due to this local hyperbolicity can be observed. 
 (courtesy of Mathias~Steinhuber). }
\end{figure}

On the contrary, the term $F_>(t)$ in Eq,~(\ref{eq:F>}) is suppressed for $t\!<\!\tE$, but is indeed responsible for OTOC saturation.  After the scrambling time $t>\tE$ genuine MB interference sets in captured by encounter diagrams such as in panel (f). This diagram represents successive forward and backward dynamics swapping back and  forth along different encounter-coupled mean-field trajectories. This involves correlated
quantum MB dynamics and the temporal build up of entanglement between mean-field modes.
This mechanism is evidently in a regime where mean-field approaches fail~\cite{Han16}. 
Thus, genuine MB interference
is the quantum mechanism behind the commonly observed saturation of OTOCs at the scrambling or Ehrenfest time.

Note that the expression, Eq.~(\ref{eq:F>}), for the OTOC 
 contains variances of classical quantities, 
 e.g.~the variance of the $j$-th final position quadrature that determine the OTOC saturation level.
Here different types of classical MB dynamics at post-scrambling times, e.g.~diffusive versus chaotic evolution, may lead to a different time-evolution of these classical variances.
As shown in \cite{Rammensee18}, diffusive dynamics implies a linear increase with time, whereas a calculation assuming ergodic dynamics yields $C(t) \approx 2/L^2$ for $t \gg \tE$ with $L$ the number of sites of a Bose-Hubbard system (for $|\Psi\rangle$ being either a localized wave packet or extended chaotic MB state) corresponding to the flat plateau in Fig.~\ref{fig:OTOCscheme}.

Figure~\ref{fig:OTOCBH} shows the OTOC Eq.~(\ref{eq:OTOComm_definition}) with $\hat{V}= \hat{n}_0$ and $\hat{W}=\hat{n}_1$ denoting occupation operators for adjacent sites
obtained quantum mechanically for a 4-site BH system with $N=40$ particles. These numerics confirm the semiclassical predictions.  Up to $\tE$ the OTOC increases exponentially with slope $2\lambda$ where $\lambda$ agrees with the Lyapunov exponent of the (here locally) unstable MB mean-field dynamics of the specfifc BH system. At $t \simeq \tE$ saturation sets in.

The present semiclassical analysis of MB OTOCs in the large-$N$ limit, the vertical limit in Fig.~\ref{fig:sc-limits}, can be readily generalized to systems of $N$ particles in $d$ spatial dimensions in the complementary limit of small $\hbar$, in particular to
the quantum chaotic single-particle case.
Invoking the corresponding Gutzwiller propagator, Eq.~(\ref{eq:vVG}), in $n=d\times N$ dimensions the exponential increase of the OTOC $C_N(t)$ in Eq.~(\ref{eq:F<}) is then governed by the leading Lyapunov exponent $\lambda_N$ of the corresponding classical $N$-particle system.  Saturation sets in at the corresponding Ehrenfest time $(1/\lambda_N) \log S/\hbar$ with $S$ a typical classical action.  For a chaotic phase space average the usual semiclassical limit the saturation value  $C(t) \approx \hbar^2 N/d$ results~\cite{Rammensee18}.  For $N=1$ this short-time growth with $\lambda_1$ has also been independently semiclassically derived in \cite{Kurchan18,Jalabert18}.
The exponential increase and saturation of such single-particle OTOCs was considered in detail in numerical case studies of the kicked rotor~\cite{Rozenbaum17} and quantum maps~\cite{Garciamata18}.


\subsubsection{Extensions}
\label{sec:extensions}

\begin{figure}
    \centering
    \begin{tabular}{ccc}
   \includegraphics[width=0.45\linewidth]{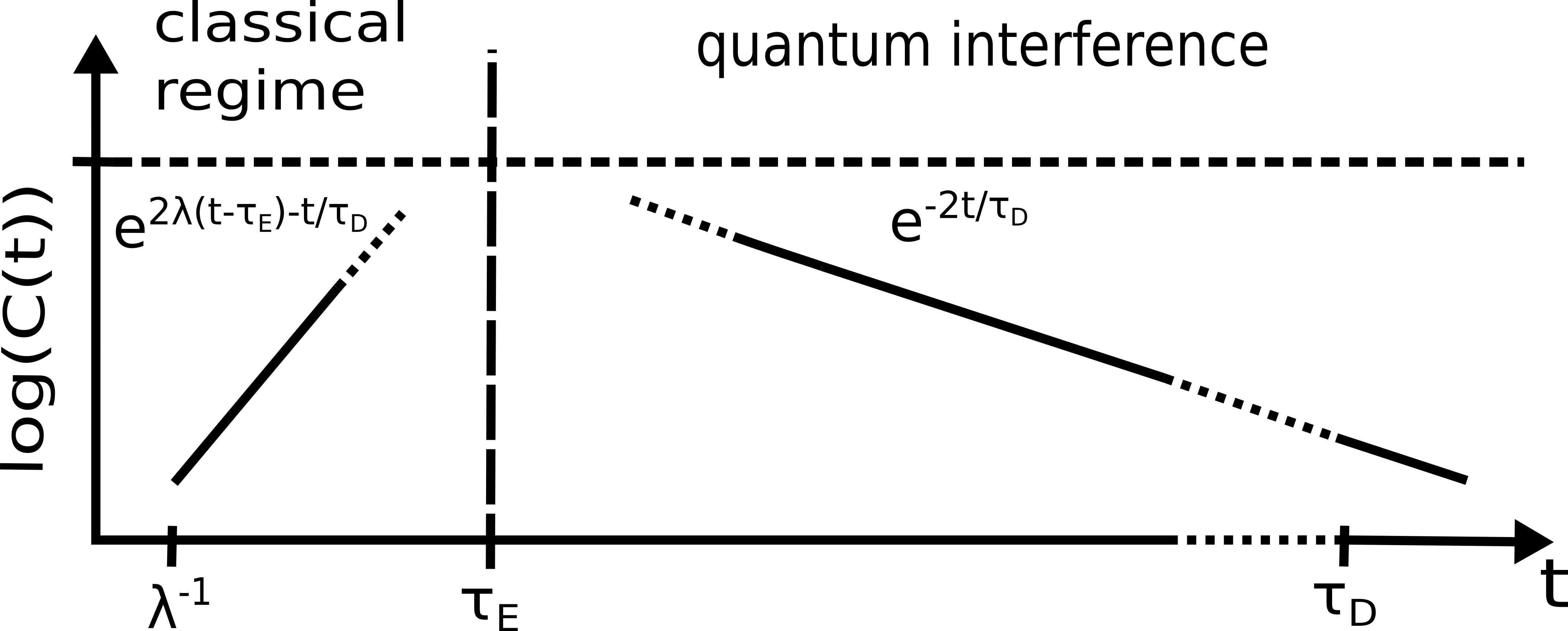}
      & &
   \includegraphics[width=0.45\linewidth]{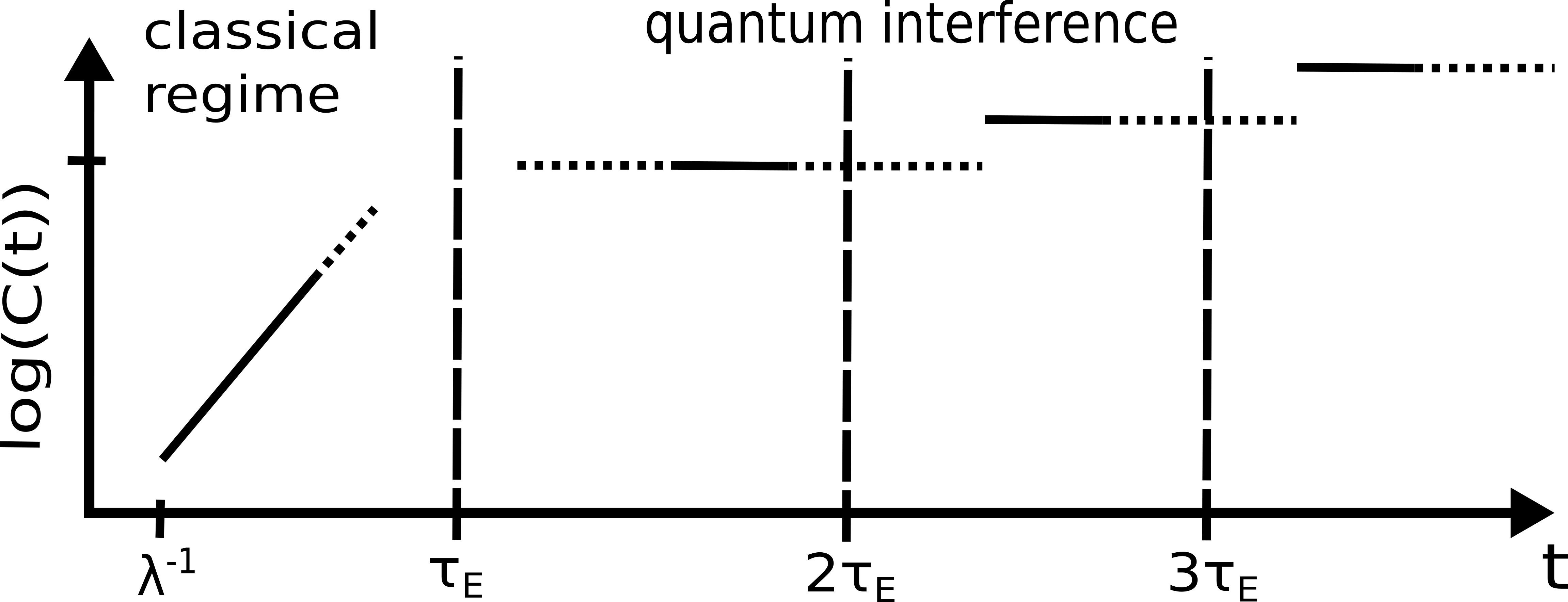}
      \\    (a)&&(b)
     \end{tabular}
    \caption{
    \label{fig:otoc-k}
   {\bf Generalized OTOCs} |
 (a) Semiclassical prediction for the OTOC $C(t)$ of  an open chaotic MB quantum system with decay time $\tD > \tE$. 
For $t < \tE$, the classical regime governed by a dominant mean-field path, the exponent is diminished by the rate $1/\tD$, while in the regime of genuine MB quantum interference ($t > \tE$) the OTOC exhibits an exponential decay with twice that rate, $-2/\tD$.
(b) Sketch of the semiclassical prediction for the $k$-OTOC $C_k(t)$, Eq.~(\ref{eq:k-OTOC}).  
Characteristic steps are expected at multiples of the scrambling time $\tE$. (From Ref.~\cite{Vierl19} with permission.)
  }
\end{figure}

There are two further interesting extensions of semiclassical results to summarize for MB OTOCs.
First it is worth considering how the time dependence of the OTOC changes for open MB quantum systems, specifically for $N$-particle systems where each particle has a large but finite average dwell time $\tD > \tE$ to stay in the system. The corresponding classical decay  can be incorporated into the encounter integrals $I(t)$ in Eq.~(\ref{eq:PS_average}) by means of terms $\sim \exp{(-t/\tD)}$. Most notably, their individual consideration in the encounter diagrams 
\ref{fig:OTOC_diagrams}(c) and (d-f) for $t < \tE$ and $t > \tE$, respectively, yields, to leading order, different decay rates~\cite{Vierl19}
\begin{eqnarray}
    F_<(t) & \sim
 e^{(2\lambda t -t/\tD)} \Theta(\tE-t) \ , \\
 F_>(t) & \sim \ e^{(-2t/\tD)} \ \Theta(t-\tE) \, ,
\label{eq:F-open}
\end{eqnarray}
as depicted in Fig.~\ref{fig:otoc-k}(a). The non-trivial, doubled decay rate $2/\tD$ in the regime of MB interference arises from the structure of the corresponding encounter diagrams (d-f) containing two independent ``legs'' (trajectory pairs) of length $\sim t$ that can lead to particle decay, compared to correlated dynamics centered around one path of length $t$ inside the encounter, diagram (c). Its experimental observation would clearly indicate this subtle and possibly unexpected aspect of MB interfernce.

Second, it is of interest to consider $k$-th order generalizations of the usual OTOC~\cite{Cotler17}, i.e.~$k$-OTOCs
\begin{equation}
  C_k(t) =
\langle \Psi | 
  \left[ \hat{p}_i(0), \hat{q}_j(t) \right]^k | \Psi \rangle \, .
  \label{eq:k-OTOC}
\end{equation}
Note that this definition does not contain absolute values as in the definition, Eq.~(\ref{eq:OTOComm_definition}), of the usual OTOC, i.e.~$C(t) \neq C_2(t)$.
Generalizing the semiclassical encounter calculus to the case of a $k$-OTOC, $k-1$ encounters can be placed into a trajectory structure comprising $2k$ paths. Careful evaluation~\cite{Vierl19} of the leading-order encounter diagrams suggests a stepwise structure for $C_k(t)$ with stepsize $\tE$ as visualized in Fig.~\ref{fig:otoc-k}(b). 


Returning to the usual OTOCs, ergodic quantum MB systems with a chaotic classical limit $\hbar \rightarrow 0$ or 
$\heff \rightarrow 0$ show an OTOC growth with exponent $2\lambda_N$ or $2\lambda$, respectively. However, an exponentially increasing OTOC does not necessarily imply chaotic dynamics, i.e.~OTOCs are not necessarily indicative of chaos.  Important exceptions are large-$N$ MB systems (near criticality) where their quantum dynamics are accompanied by unstable fixed points (separatrices) in the associated MB mean-field dynamics that can even be integrable.  The local fixed point instability $\lambda_s$ also leads to an exponential increase $C(t)\sim e^{2\lambda_s t}$ up to times $\tE$~\cite{Geiger19,Xu20}. Thus scrambling does not necessarily imply chaos.
In such systems the quantum critical states may be viewed as residing close to separatrices that have much in common with encounters. However, due to the integrable classical limit, the fast initial scrambling is followed subsequently by oscillatory behavior between reentrant localization and delocalization of information in Hilbert space~\cite{Geiger19}.

For such quantum critical systems the semiclassical leading-order $1/N$-expansion 
has been refined providing $e^{2\lambda t}/N$ (and not $1/N$) as a renormalized parameter that non-perturbatively rules the quantum-classical transition~\cite{Geiger21}.
For scrambling in many-body hyperbolic systems this provides formal grounds for a conjectured and numerically observed multi-exponential form of OTOCs for the SYK-model~\cite{Kobrin21}.

Apart from quantum chaotic and critical large-$N$ systems, the imprints on OTOCs of non-ergodic dynamics, for example, a mixed (regular-chaotic) phase space in the classical limit, have also been recently considered~\cite{Fortes19}, but are not in the focus of this review.

\section{Perspectives}
\label{sec:persp}

During the past twenty years considerable progress has been made and various breakthroughs have been achieved in laying the foundations of a semiclassical theory that successfully provides the understanding of quantum chaotic universality, or more precisely, of universal (spectral) features of SP and MB quantum systems exhibiting chaotic classical limits. However, as is generally accepted, this RMT-type universality applies to much broader classes of quantum systems, including those without a strict classical limit, 
such as quantum graphs and networks, MB quantum circuits
and spin chains 
(see Sec.~\ref{sec:spec-stat} and~\cite{Fisher22} for a recent review) that often may be represented by ensemble theories, i.e.~theory of disordered systems and RMT.  Why such quantum chaotic systems in a wider sense exhibit the same universal features as quantum chaotic systems in a narrow sense, i.e.~those with a classical limit, partially remains a mystery. 

Leaving aside such conceptual questions, the now existing theoretical framework, reviewed here, opens various interesting perspectives and challenges.  Although semiclassical theory applies to the dynamics of individual systems, enabling a complete understanding of system specific properties and deviations from universal behaviors (at least theoretically), we conclude this review with perspectives of particular relevance to quantum universality:
\begin{itemize}
    \item[(i)] {\em Variety of classical limits --}
in both complementary asymptotic limits considered, the quantum-classical transition is singular.  The limiting, but non-vanishing, values of $\hbar/S \ll 1$ and $\heff = 1/N \ll 1$ imply ever shorter wave lengths and extensive quantum interference.  Classical and mean field physics arise for $\hbar/S \equiv 0$ and $\heff \equiv 0$, respectively.
Such singular asymptotic behavior is generally indicative of fascinating physical phenomena.  The semiclassical theory presented provides the leading-orders (in $\hbar$ and $\heff$) of quantum mechanical contributions: the former dealing with quantum wave interference and the latter with genuine MB quantum interference.  These two quite distinct limits, sketched in Fig.~\ref{fig:sc-limits}, represent two avenues of asymptotic analysis. It may be interesting to consider other limiting procedures where both $\hbar$ and $\heff$ come into play in concert. Indeed, short wavelength approximations (semiclassical methods) are applied universally across classical field theories, e.g.~optics, acoustics, gravity waves, seismic waves, etc... in which the limit leads to ray equations underlying the motion of classical waves.  In turn these rays themselves can be chaotic.  Beginning with quantum fields, it is possible to imagine a kind of ``asymptotics of asymptotics'' in which there are chaotic rays underlying the classical field solutions underlying semiclassically the quantum field solutions.  This is as opposed to just the single limit $\heff\ll 1$ considered here generating some kind of nonlinear, unstable classical field.  Does the asymptotics of asymptotics chaotic limit lead to a distinct kind of MB quantum chaos from that mainly considered in this text?  On another front, following the ``diagonal'' in Fig.~\ref{fig:sc-limits}, {\em i.e.}, sending $\hbar$ and $\heff$ to zero simultaneously appears particularly appealing and challenging, since there is no reason to expect the two limits to commute and could lead in another unique direction. There is a great deal yet to uncover.

\item[(ii)] {\em The limit of dilute local Hilbert spaces --} The construction of the semiclassical propagator relies on two key facts: the existence of a classical limit (defined through extremizing an action identified in turn from the exact path integral) and a semiclassical regime (identified as the small parameter $\hbar_{\rm eff} \to 0$ scaling the action). In situations where either of these two ingredients is not obvious or explicit, the  semiclassical program relies on further assumptions. Three very important situations where such problems appear and await for further progress are semiclassical analysis of fermionic Fock space, the related case of spin-$1/2$ chains, and the extension to fields in the continuum. In systems described by (continuous or discrete) fermionic degrees of freedom, the natural path integral based on Grassman-valued fields leads to Grassman-valued actions \cite{Negele98} where the stationary phase approximation cannot be defined in any sensible manner. This problem reflects in turn the difficulty in identifying an $\hbar_{\rm eff}$ due to the fundamentally quantum character of the Pauli principle, a problem shared by spin systems with low spin with their fundamentally discrete natural basis states (in fact, fermionic and spin-$1/2$ degrees of freedom are rigorously mapped by means of the Jordan-Wigner transformation~\cite{Coleman15}). Progress in this direction can be achieved by forcing a description in terms of bosonic (commuting) classical fields as in \cite{Engl18,Engl15a,Engl14} for both fermionic and spin systems. The subsequent semiclassical program can be formally defined, and in some situations provides extremely accurate results for delicate MB interference effects as shown in \cite{Engl16,Engl18}, but so far it lacks rigorous support. A similar violation of the large local Hilbert space assumption also occurs in bosonic systems if one considers the continuum limit.  Since the number of sites tends to infinity, any finite number of particles will get effectively diluted thus breaking the fundamental assumption of large occupations. Identifying a proper semiclassical regime for the propagator of bosonic fields in the continuum holds the key for a very promising program, as several important results are known for such non-linear classical field equations.  There are the existence of chaos \cite{Brezinova12,Wanzenbock21}, a precise definition of classical mean-field integrability by means of the inverse scattering method, and the corresponding semiclassical quantization based on solitons as building classical blocks \cite{Korepin1993}. Extending this approach into the chaotic regime remains a fundamental and fascinating open problem. 

\item[(iii)] {\em Many-body scarring and deviations from equilibration --} ergodicity is commonly related to the equidistribution of eigenfunctions linked to a chaotic phase space energy shell leading to generic equilibration as predicted by the eigenstate thermalization hypothesis~\cite{Deutsch91, Srednicki94}.  There are various properties and mechanisms that lead to different degrees of ergodicity breaking and, as a result, to the possibility of hindered relaxation.  Corresponding settings possibly include: (i) MB systems without a classical limit, which are subject to additional constraints~\cite{Bernien17,Karpov21}, leading to disconnected sub-Hilbert spaces with reduced dimensions; (ii) MB systems with a classical limit that exhibits a non-ergodic limit, e.g.~mixed phase space structures of co-existing chaotic and regular regions. As an obvious case, quantum states residing on tori associated with locally integrable phase space regions are long-lived and typically decay or MB equilibrate on (exponentially) long time scales. 
 
 However, there is a hierarchy of weak ergodicity breaking mechanisms reflected in deviations from equilibration. In this context, two examples from SP physics are {\em dynamical localization} due to partial transport barriers leading to additional time scales whose effects are to localize eigenfunctions~\cite{Bohigas93} and the concept of {\em scars}, discovered and introduced by Heller~\cite{Heller84} for low-dimensional quantum-chaotic systems. Both represent prime examples of weak ergodicity breaking. For the latter, a quantum eigenstate that is semiclassically anchored on or close to an {\em unstable} periodic orbit \cite{Heller84} is scarred, if the period of the orbit is short and its Lyapunov exponent weak enough, such that a single-particle wave packet launched along this orbit shows distinct recurrences after one period. This can be cast into a rough criterion for scar formation~\cite{Heller84}. This concept naturally requires a classical limit, and is intriguing because it indicates deviations from ergodicity for a fully chaotic system that globally shows eigenstate thermalization, and must be differentiated from the aforementioned quantum states localized in regular regions.  Very recently, scarring in Heller's original sense could be demonstrated for a MB Bose-Hubbard system with a high-dimensional associated classical phase space, including the corresponding generalization of the scar criterion~\cite{Hummel22}. 

Earlier, MB "scars", reflected in persistent oscillations of local observables, were observed in Rydberg-based quantum simulators \cite{Bernien17}, as well as in corresponding numerical simulations \cite{Turner18,Serbyn21}. They were found in spin-chain type MB Hamiltonians that do not possess a natural classical limit.  It remains to be understood how such a ``MB scarring'' of this type can be related to the semiclassical scar mechanism associated with periodic orbits. Furthermore, for systems with a semiclassical limit, it has to be explored, whether true MB scars prevail in the thermodynamic limit of large site or particle number.  This could be probed employing ultracold bosonic atoms in optical lattices as quantum simulators.

\item[(iv)] {\em Entropies, entanglement and encounters --} The use of quantum information concepts in the framework of MB systems has lead to deep insights into the mechanisms of equilibration, thermalization and the role of quantum coherence~\cite{Eisert15,Rigol08,  Abanin19, Streltsov17}. In this context, RMT has rather successfully been applied to various universal aspects of particular information measures~\cite{Keating04, Chan18}. It has been of further utility for studying weakly connected bipartite chaotic systems as well~\cite{Bandyopadhyay02, Lakshminarayan19, Pulikkottil20, Pulikkottil22}. In contrast, the use of many-body semiclassical methods face severe technical difficulties and only limited work has been accomplished. The origin of the problem is the extreme nonlinear way the propagator appears when calculating such measures, typically involving functions with dependencies such as $\log K K^{*}$, lacking the usual structure of multiple sums over paths where further analysis based on action correlations is possible. Only the so-called purity (a linearized version of the entanglement entropy) has allowed for a semiclassical study in first-quantized systems as carried out in \cite{Jacquod09,Bonanca11}. Properties of entanglement of two (non)interacting particles in the quantum chaotic Chirikov standard map were very recently numerically considered in Ref.~\cite{Ermann22}.

\item[(v)] 
{\em Dual-unitary dynamics --} 
In Refs.~\cite{Gutkin16,Akila16} a correspondence between unitary propagation in time and a non-unitary evolution in particle number in terms of an operator dual to the time evolution operator was established.  The case where the dual operator is also unitary, referred to as “self-dual” \cite{Bertini18}, is of special interest: corresponding (disordered) spin chains have been shown to exhibit RMT behavior~\cite{Kos18}, whereas deviations from self-duality can lead to many-body localization~\cite{Braun20}.
 Correspondingly, so-called dual-unitary circuits consist of networks of special unitary gates leading to interrelated unitary evolution in space and time directions~\cite{Bertini19,Gopalakrishnan19}; for a recent review see~\cite{Prozen21}. The locality of the gates, together with dual-unitary propagation implies correlations to be possible only along a “ray” $|x| = |t|$ in “space time”. Such dual-unitary circuits are special and particularly appealing as they represent a class of exactly solvable quantum systems that exhibit RMT behavior. It would be interesting to devise dual-unitary quantum dynamical systems with a semiclassical limit. This could open up the possibility to extend and apply the semiclassical tools outlined in this review to a conceptually simple case that still carries quantum chaotic features.

\item[(vi)] {\em Quantum chaos meets quantum gravity --}
During the last two decades, quantum chaos concepts have entered the realm of research towards possible quantization of gravitational degrees of freedom. Early suggestions pointing to the chaotic character of black holes~\cite{Sekino08} were finally made precise through the study of scrambling of (quantum) information and the proposal of black holes as systems where such scrambling is maximal, see \cite{Maldacena16} and references therein. This connection between toy models of quantum gravity where, unlike the realistic scenario describing the universe, a full solution is available and quantum chaos was made precise in the cornerstone paper \cite{Saad19}. There, the authors showed the dual relation, by means of the equivalence of correlation functions, between quantized Jackiw-Teitelboim gravity, a solvable model of gravity coupled with a dilaton field in 1+1 dimensions, and a suitably double-scaled theory of random matrices. This finding has triggered lots of attempts towards understanding the origins of this duality with prospective links to supersymmetry and  semiclassical analysis in quantum chaos~\cite{Altland21}.

\end{itemize}
Just these few speculations illustrate the richness of possibilities from following various semiclassical paths to interesting future theoretical challenges in MB quantum chaos.


\section{Acknowledgments}

This review is dedicated to the memory of our dear colleagues, Fritz Haake and Petr Braun. We are very thankful to both of them for many many inspiring conversations about semiclassical physics, quantum chaos and beyond science during the last 30 years and numerous encounters at various places all over the world. Our close connections started with the Symposium {\em Quantum Aspects of Nonlinear Systems}, organized by Fritz Haake and Robert Graham in 1990, in retrospective playing a role for quantum chaos similar to the role of the 1911 Solvay conference for quantum physics. K.R. is further indebted to Petr Braun for various conversations also about Rydberg atomic physics, and for his kind hospitality during a longer stay in St.~Petersburg in 1990.
 
Part of the results on MB semiclassics presented here arose out of two PhD theses~\cite{Engl15a,Rammensee19} conducted in Regensburg. Hence we particularly thank Th. Engl and J. Rammensee for their important work and for many related conversations.
We further thank P. Schlagheck and D. Ullmo, as knowlegable discussion partners on advanced topics of many-body semiclassical methods over many years.
We would also like to thank numerous colleagues for conversations on topics that entered the many-body part of this review, including A. Altland, A. Buchleitner, R. Dubertrand, B. Geiger, T. Guhr, B. Gutkin, Q. Hummel, R. Jalabert, D. Vierl, D. Waltner and D. Wisniacki.

 We acknowledge financial support from the Deutsche Forschungsgemeinschaft (German Research Foundation) through Projects Ri681/14-1, through Ri681/15-1 within the Reinhart-Koselleck Programme, as well as funding through Vielberth Foundation in Regensburg.

\section{Literature}

\bibliographystyle{unsrt}
\bibliography{mean-dos,general_ref,classicalchaos,extreme,furtherones,manybody,molecular,nano,oceanacoustics,quantumchaos,rmtmodify}
\end{document}